\documentclass[10pt]{iopart}
\usepackage{iopams}
\usepackage{graphicx}
\usepackage[graphicx]{realboxes}
\usepackage{hyperref}
\hypersetup{colorlinks=true,
            pdfstartview=FitV,
            linkcolor=DarkOrchid,
            citecolor=DarkSlateGrey,
            urlcolor=DarkMagenta}

\expandafter\let\csname equation*\endcsname\relax
\expandafter\let\csname endequation*\endcsname\relax
\usepackage{mathtools}  

\usepackage[titletoc]{appendix}

\usepackage{longtable}
\usepackage{rotating}
\usepackage{fancyhdr}
\usepackage[comma,numbers,sort&compress]{natbib}
\usepackage{dcolumn}
\usepackage{multirow}
\usepackage{lscape}
\usepackage{times}
\usepackage[varg]{txfonts}
\usepackage{booktabs}
\usepackage{messwithfonts}
\usepackage{tablefootnote}
\usepackage[perpage]{footmisc}
\usepackage{subcaption}
\usepackage[textfont=footnotesize,labelfont=bf,format=hang]{caption}

\usepackage[svgnames]{xcolor} 
\usepackage{ifthen}

\newcommand{\msun}{{\mathrm M}_{\odot}}
\def\gw#1{gravitational wave#1 (GW#1)\gdef\gw{GW}}

\newcommand{\refsection}[1]{Section~\ref{#1}}

\newcommand{\voy}{LIGO Voyager}

\def\TODO#1 { {\color{red} {#1}}}
\def\sci#1#2{#1\times10^{#2}}

\begin{document}
\title{Astrophysical science metrics for next-generation
       gravitational-wave detectors}


\author{R.~X~Adhikari$^1$,
P.~Ajith$^2$,
Y.~Chen$^3$,
J.~A.~Clark$^4$,
V.~Dergachev$^5$,
N.~V.~Fotopoulos$^1$,
S.~E.~Gossan$^1$,
I.~Mandel$^{6,7,8}$,
M.~Okounkova$^3$,
V.~Raymond$^9$,
J.~S.~Read$^{10}$}

\ead{rana@caltech.edu}

\address{$^1$ LIGO Laboratory, California Institute of Technology, Pasadena, CA 91125 USA}
\address{$^2$ International Centre for Theoretical Sciences, Tata Institute of Fundamental Research, Bangalore~560089, India}
\address{$^3$ TAPIR, California Institute of Technology, Pasadena, CA 91125 USA}
\address{$^4$ Georgia Tech, Atlanta, GA USA}
\address{$^5$ Max-Planck-Institut  f\"ur Gravitationphysik, Callinstrasse 38, 30167, Hannover, Germany}
\address{$^6$ Institute for Gravitational Wave Astronomy and School of Physics and Astronomy, University of Birmingham, Edgbaston, Birmingham B15 2TT, United Kingdom}
\address{$^7$ Monash Centre for Astrophysics, School of Physics and Astronomy, Monash University, Clayton, Victoria 3800, Australia}
\address{$^8$ OzGrav, Australian Research Council Centre of Excellence for Gravitational Wave Discovery}
\address{$^9$ Cardiff School of Physics and Astronomy, Cardiff University, Queens Buildings, The Parade, Cardiff CF24 3AA, UK}
\address{$^{10}$ California State University, Fullerton, CA USA}


\begin{abstract}
The second generation of gravitational-wave detectors are being built and
tuned all over the world.
The detection of signals from binary black holes is beginning to
fulfil the promise of gravitational-wave astronomy.
In this work, we examine several possible configurations for
third-generation laser interferometers in existing km-scale
facilities.
We propose a set of astrophysically motivated metrics to evaluate
detector performance.
We measure the impact of detector design choices against these metrics, providing a quantitative cost-benefit analyses of the resulting scientific payoffs.
\end{abstract}

\tableofcontents



\section{Introduction}
\label{s:intro}
The recent detections of gravitational-wave (GW) signals from merging binary black holes and neutron stars are beginning to fulfil the promise of GW astronomy. 
The Advanced LIGO detectors observed GW from a coalescences 
of two $\sim 30 M_\odot$ binary black holes on 14 September, 2015~\citep{GW150914}.   
Further observations of binary black mergers \cite{BBH:O1O2} and a binary neutron star merger \cite{GW170817} followed during the first two observing runs.  The two Advanced LIGO detectors in the U.S. and the Virgo detector in Italy are gradually approaching their design sensitivity, the KAGRA detector in Japan is coming on-line, and the third LIGO interferometer in India is expected to join the world-wide network around 2025.

Within the LIGO Scientific Collaboration (LSC), the Interferometer 
Working Groups have identified a set of design concepts for the 
next generation of interferometer (known as \voy{}) in 
publicly available technical documents~\cite{Straw,ISWP:2016}
and a manuscript in preparation~\cite{Voyager:Inst}.
It is expected that the following decades would see the
development of new facilities supporting the proposed
Einstein Telescope~\cite{3G:Science} and/or 
Cosmic Explorer~\cite{abbott2017exploring} observatories.

In this work, we use the recently generated sensitivity curves 
to make quantitative estimates of the
scientific potential of \voy{}. In particular, we construct a
Jacobian (cf. Tables \ref{tabl:Jacobian1} \& \ref{tabl:Jacobian2}),
which relates the scientific outputs with changes to the
interferometer's parameters. This Jacobian will help to make design
trade-offs as the \voy{} design moves forward.

In \refsection{s:IFO}, the design parameters
of the interferometer are explained and the sensitivity
curves shown.
In \refsection{s:sources}, the astrophysical sources behind the
Jacobian are introduced and the scientific
metrics which form the rows of the Jacobian are described. The rest of
the article describes the scientific targets from the various astrophysical
sources.

\section{Interferometer Design}
\label{s:IFO}
The output of a workshop in January 2012 was a set of \textit{Strawman} designs for
a third generation LIGO~\cite{Straw} which have been iterated as new
understanding arises.
The three (Red, Green, and Blue) design teams worked to come up with separate
designs for an interferometer that could
be installed in the existing LIGO facilities without major facility
modifications (i.e. keeping the same arm lengths, no
modification of the 4\,km beam tubes, etc.). Extra vacuum chambers, tubes,
cryogenic equipment, and other vacuum
equipment modifications were allowed for the purpose of this exercise.

\begin{figure}[h]
  \centering
      \includegraphics[width=\columnwidth, clip, trim=0 0 -3cm 0]{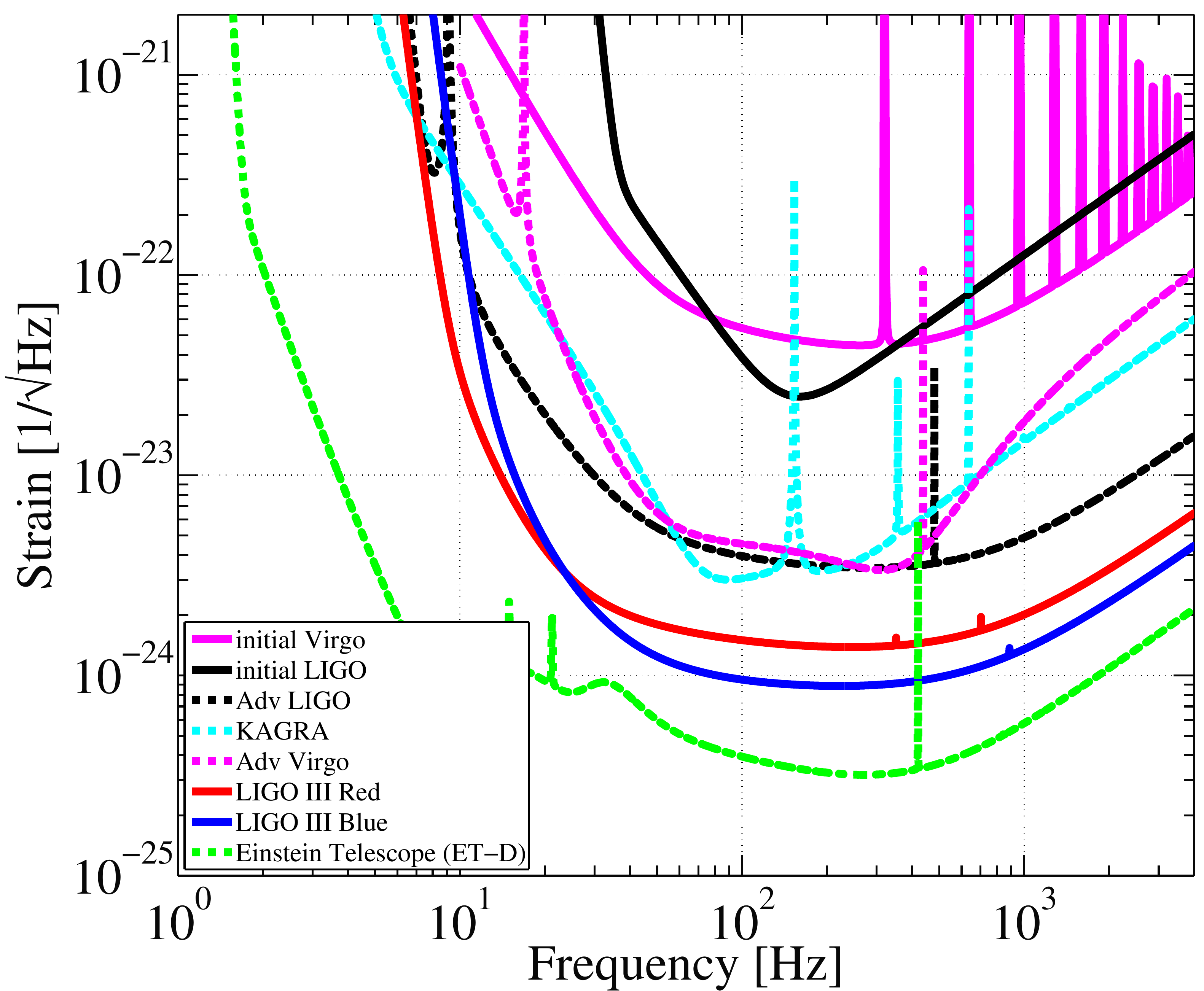}
      \caption[Strain Spectra of ground based interferometers]
      {Shown are the strain noise spectral density estimates for first,
      second, and third generation detectors. The Red and Blue
      \voy{} designs are shown here as well as the ET-D sensitivity estimate.}
    \label{fig:IFOnoises}
\end{figure}

Within most of this document we only consider the Red and Blue designs.
In particular, for the Jacobian table, we use the Blue design as the point
of departure. However, as can be seen from Figure~\ref{fig:IFOnoises},
the two designs are similar enough that this does not change the result
too much. Throughout \refsection{s:SciencePot}, all three design concepts are
used to estimate scientific potential.

\subsection{Sensitivity Limits of Second Generation Detectors}
The second generation interferometers (Advanced LIGO, Advanced Virgo and KAGRA) all have
similar noise limits.
Figure~\ref{fig:IFOnoises} shows the estimated strain spectral density curves.
The differences between the LIGO, Virgo, and KAGRA curves below 40\,Hz arise from some
uncertainty in the estimation of the true suspension thermal noise
\cite{SUS:FEA2009, SUS:2012} as well inherent differences in the baseline
interferometer configurations chosen for the Virgo and KAGRA curves: the detuning
leads to better sensitivity around 100\,Hz, but worse quantum noise performance
at lower and higher frequencies.
Note that these are just calculated noise curves and the true noise performance~\cite{Den:PRL2016} of all
detectors is likely to exceed these optimistic estimates in a few frequency bands.

\begin{figure}[tb]
  \centering
      \includegraphics[width=\columnwidth]{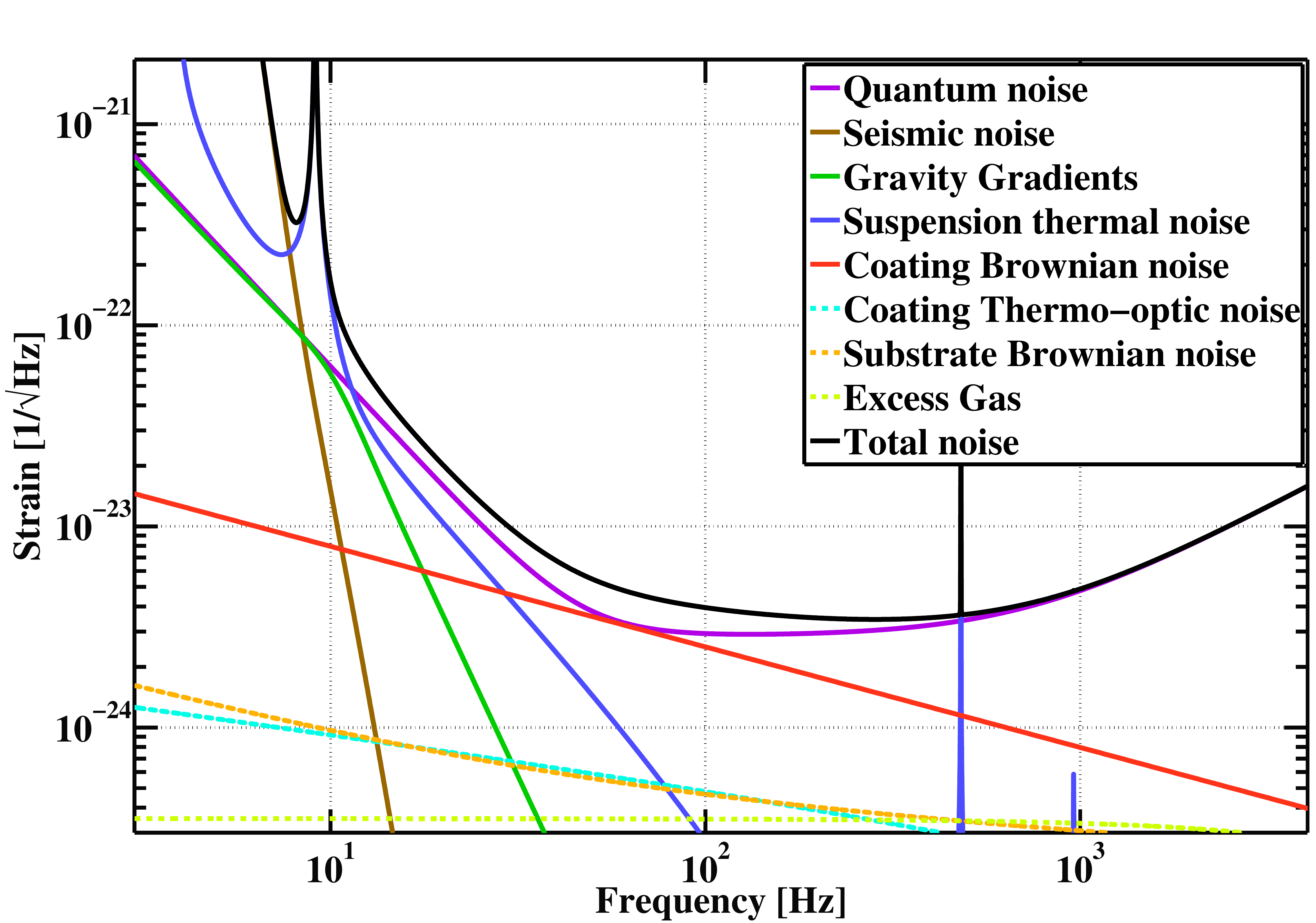}
      \caption[The aLIGO GWINC Noise Budget]
              {The Advanced LIGO\cite{Den:PRL2016} noise budget, computed using GWINC (120418)}
    \label{fig:gwinc}
\end{figure}


\subsection{Design Choices}
In order to facilitate quantitative design choices, we have constructed
the astrophysics and cosmology detector Jacobian tables and discuss the
impact of various design changes on astrophysical and cosmological
science goals in \refsection{s:sources}.
The columns in the Jacobian tables correspond to different
interferometer design choices, while
rows corresponds to different astrophysical figures of merit.

In all cases, the major cost is not monetary, rather it is the time
spent in the installation and commissioning of these upgrades which
must be considered when making the cost-benefit analysis of making the changes.

Here, we briefly describe the various columns in the detector Jacobians:

\begin{figure}[ht]
  \centering
      \includegraphics[width=\columnwidth]{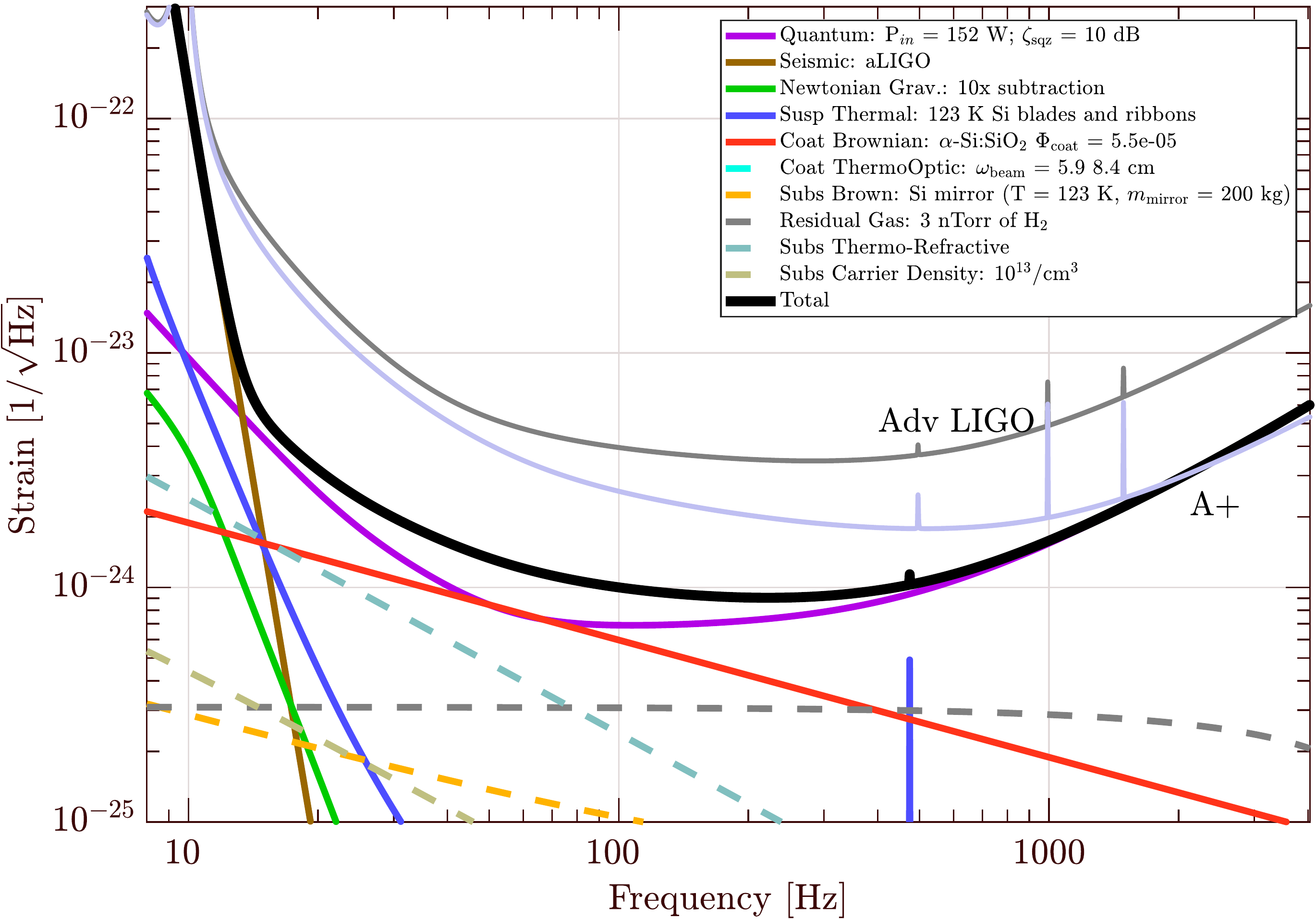}
      \caption[LIGO (Blue) Voyager Noise Budget]{LIGO (Blue) Voyager
        Strain noise budget. Also shown are the Adv. LIGO design\cite{Den:PRL2016} and LIGO upgrade (a.k.a. A+)\cite{PhysRevD.91.062005} noise curves.
      Comparison of interferometer physical parameters given
      in Appendix~\ref{s:IFOparams}.}
    \label{fig:LIGO3gwinc}
\end{figure}

\begin{itemize}
\item {\bf NN}: Fluctuations in the local Newtonian gravity field produce
  accelerations of the test mass which mask low frequency GW
  signals~\cite{Harms:LR2015}. Reduction of this noise will require
  the expansion and improvement of the seismometer array used to
  estimate and subtract this noise offline.

\item {\bf SEI}: The motion of mirrors due to seismic disturbances in
  the 5\,--\,50\,Hz band may be due to vibrations of the ground or the internal
  noise of the active seismic isolation systems~\cite{aLIGO:SEI:2015}.
  Factors of a few reduction are possible using some few years of work
  on improving seismic sensors or redesigning the suspension configuration.

\item {\bf SUS}: The Brownian thermal noise of the mirror suspenion fibers
  is as significant at low frequencies as the seismic and gravity noise.
  Further development in suspension design~\cite{silicon:SUS} and
  materials science may lead to incremental progress in this band.

\item {\bf SPOT/CTN}: The limit to the interferometer sensitivity near
  100\,Hz is the Brownian thermal noise in the mirror
  coatings~\cite{reid2016development} for several different
  interferometer concepts.
  The power spectrum of this noise scales
  inversely with the mechanical quality factor $Q$ of the coating;
  it also scales inversely with the laser beam diameter.

\item {\bf SQZ}: Increasing the amount of squeezing delivered to the
  interferometer results in a broadband reduction of the quantum
  noise (both radiation pressure and shot noise). For technical
  reasons~\cite{Haixing:CC2012}, this may have some frequency dependence
  but for the purposes of this analysis, we assume the naive
  broadband improvement.

\item {\bf POW}: Increasing the laser power mainly increases the low
  frequency radiation pressure noise and reduces the high frequency
  shot noise. When changing this parameter for the Jacobian table,
  we do not consider the thermal wavefront distortion effects due
  to the increased heat load on the mirrors.

\item {\bf FCL}: Increasing the length of the optical filter
  cavity~\cite{FC:2013}, which is used to
  rotate the squeezing quadrature, reduces the degradation of the
  squeezed light which is injected into the interferometer dark
  port, and thereby improves the broadband sensitivity.

\item {\bf MASS}: A relatively simple way to reduce the quantum backaction noise
  (i.e. quantum radiation pressure) is to increase the mirror mass. In this
  column of the Jacobian, we do not consider the effects of increased mass on the
  suspension thermal noise.
\end{itemize}

\section{Astrophysical Metrics}
\label{s:sources}
The main aim of this paper is to introduce a number of metrics to quantify the ability of the LIGO Voyager detectors to perform various astrophysical measurements, and study the variation of these figures of merit with respect to changes in different design parameters of the detector. Here we provide a brief overview of the astrophysical science that can be potentially performed by these detectors and to discuss figures of metrics related to these astrophysical measurements.

\subsection{Astrophysics with Compact Binary Coalescence}
\label{s:cbc}

Binaries of compact objects (black holes or neutron stars) can be
produced in a number of astrophysical scenarios 
(cf.~\cite{MandelOShaughnessy:2010,GW150914:astro,MandelFarmer:2018, Mapelli:2018} for overviews).  
Once formed, they will radiate 
GWs, gradually shrinking the orbit through an inspiral that ends with
the objects merging,
then settling down into a spinning compact object. GWs
emitted during the coalescence of binaries consisting of neutrons stars
(NSs) and/or black holes (BHs, in the mass range from a few to $10^3 M_\odot$)
can be detected by ground-based GW observatories.

NS-NS and NS-BH binaries are typically expected to form through the
evolution of isolated field binaries.  An isolated binary composed of
two main-sequence stars undergoes an evolution that involves 
several mass-transfer phases, possibly a common-envelope phase, and
two core-collapse events of the binary's components.  A review of the
process, including the possible orderings of this sequence, is given
in~\cite{2007PhR...442...75K,PostnovYungelson:2014,Tauris:2017}. 
Occasionally, this process leaves
behind a binary that is sufficiently compact to merge in a Hubble time
through radiation reaction from GW emission 
(cf. \cite{Belczynski:2008,Pfahl:2005,VossTauris:2003,Dominik:2014,VignaGomez:2018}).

BH-BH binaries such as those responsible for the GW150914 event
can evolve through isolated binary evolution as
described above \citep[e.g.,][]{Belczynski:2016,EldridgeStanway:2016,Lipunov:2016,Stevenson:2017}, 
but could also be formed via chemically homogeneous 
evolution~\citep{MandeldeMink:2016,Marchant:2016}, from primordial 
BHs~\citep{Bird:2016,Sasaki:2016}, or dynamically in dense stellar
environments, such as globular clusters or nuclear clusters in
galaxies~\citep[e.g.,][]{Rodriguez:2016,Stone:2016,Bartos:2016}.
There, a combination of three-body and four-body interactions, direct
two-body capture, Kozai resonance and other dynamical effects can lead
to the formation of coalescing binary BHs. These dynamical
capture mechanisms could also drive mergers involving
intermediate-mass BHs (IMBHs), which are discussed in more
detail in \refsection{sec:IMBH}.
The predicted merger rate for compact object binaries was uncertain, with plausible
ranges spanning three orders of magnitude before the first GW detections~\citep{ratesdoc}. LIGO observations have made it possible to constrain the binary black
hole merger rate to a range spanning a factor of $\sim 10$ \citep{BBH:O1O2}.  

\begin{figure}[h]
  \centering
  \includegraphics[width=\columnwidth]{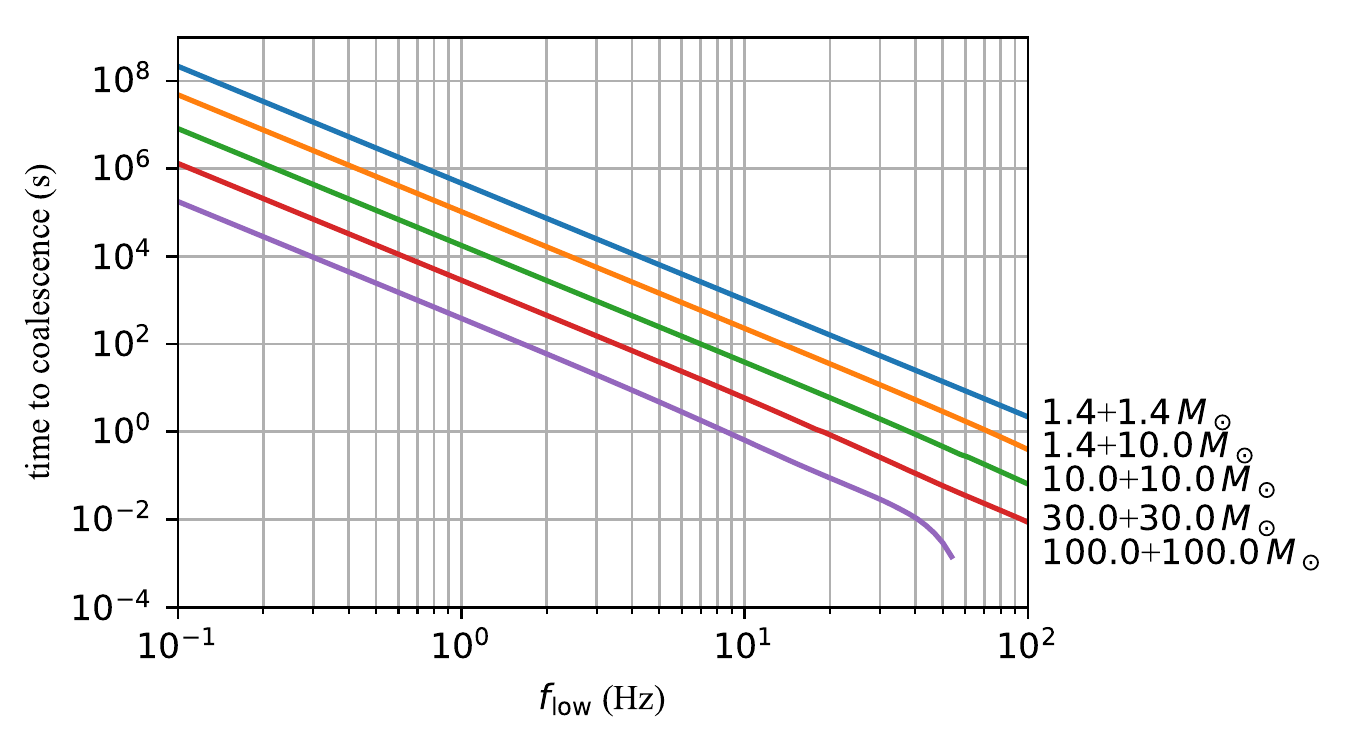}
  \caption[Time to Merger]{Time until merger, ``chirp length'',
    vs start frequency for NS/NS, NS/BH, and BH/BH systems.}
   \label{fig:chirplen}
\end{figure}

The inspiral enters the detector's band when the orbital decay
progresses to the point that the GW frequency (twice the orbital
frequency for a circular system) is above the low frequency seismic ``wall''.  
As shown in Fig.~\ref{fig:chirplen}, NS-NS waveforms will
remain in band for many minutes with a low-frequency cut-off at 10\,Hz,
though most of the signal-to-noise ratio (SNR) and bandwidth, which enable detection and
parameter estimation, lie in the final seconds before 
merger (see \refsection{sec:CBCsensitivity} for a more detailed
discussion).
Binaries with a neutron star contain matter that can be ejected and fuel
an electromagnetic counterpart, though potentially BH-BH binaries might also
excite ambient material to become luminous.  The discovery of an electromagnetic counterpart to the double neutron star merger GW170817 \citep{GW170817:MMA} led to an unprecedented observing campaign spanning all wavelengths, answering a number of long-standing questions, and posing a host of new ones.


\begin{figure}[t]
  \centering
       \includegraphics[width=\columnwidth]{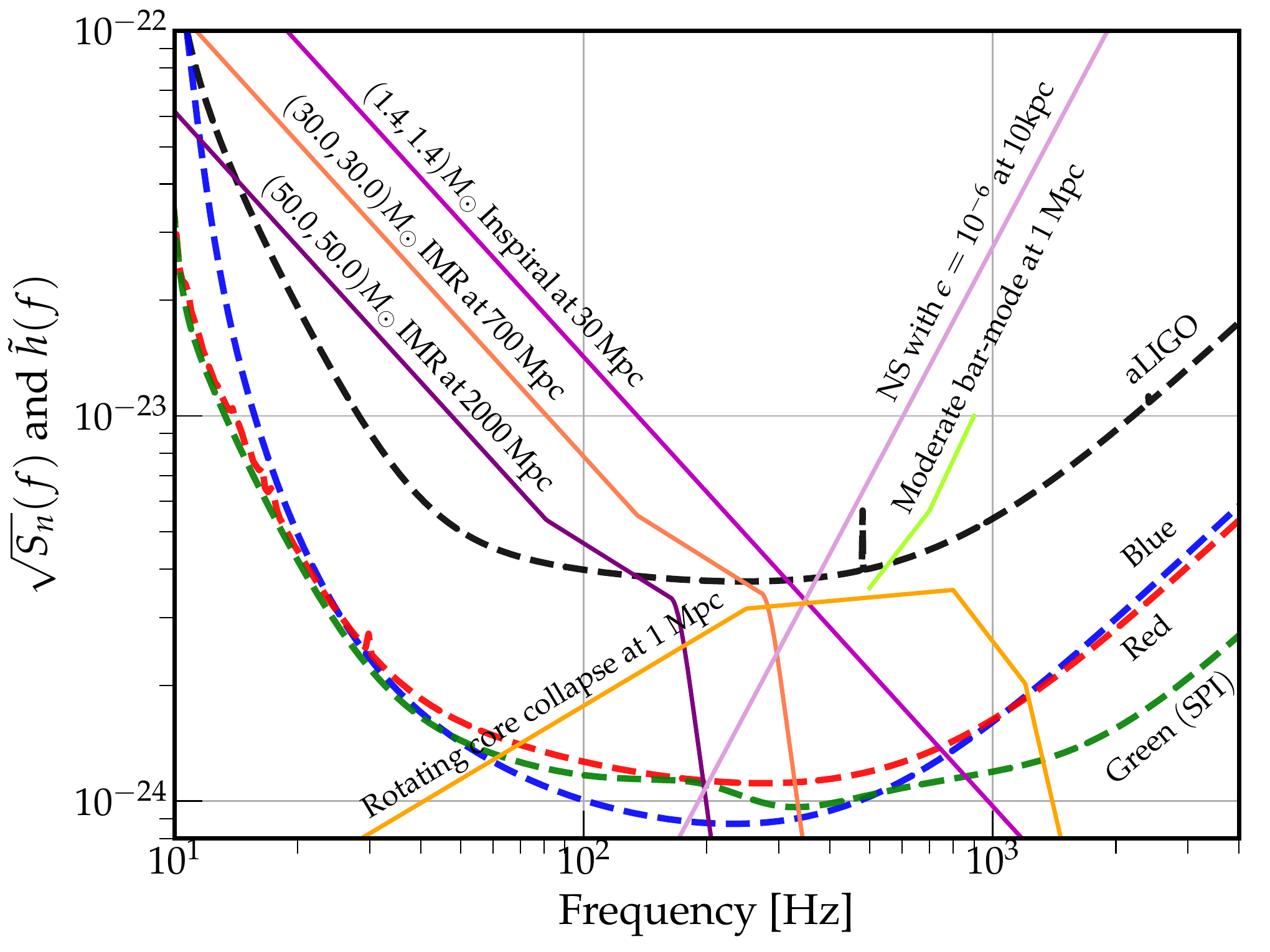}
       \caption[Cutler-Thorne style plot]
       {Baseline detector noise spectra compared with selected astrophysical sources.}
    \label{fig:SourcesandNoiseCurves}
\end{figure}


\begin{sidewaystable}
\vspace{4in}
\resizebox{\textwidth}{!}{%
\begin{tabular}{r l c c c c c}
\toprule
 & \textbf{Science Goals}   &  \textbf{Metric} & aLIGO & {\color{blue} \textbf{Blue}} &{\color{red} \textbf{Red}} &{\color{green} \textbf{Green}} \\
\toprule
\refsection{sec:PulsarSpinDownMountains}
& CW blind search volume & [kpc$^3$] & 4.4 & 158 & 158 & 556 \\
& $\epsilon$ limit (HF targ search)  & equatorial ellipticity & $1.4 \times 10^{-8}$ & $4.3 \times 10^{-9}$ & $4.3 \times 10^{-9}$ & $2.9 10^{-9}$ \\
& $\epsilon$ limit (LF targ search)  & equatorial ellipticity & $3.3 \times 10^{-5}$ & $8.2 \times 10^{-6}$ & $7.9 \times 10^{-6}$ & $1.8 10^{-5}$ \\
\hline
\refsection{sec:CBCsensitivity}
 & NS-NS ($1.4,1.4 M_\odot$)  population & Horizon redshift  & 0.10 & 0.46 & 0.35 & 0.28 \\
 & BH-BH ($10,10 M_\odot$) population & Horizon redshift &  0.52 & 3.4 & 2.3 & 1.7 \\
 & BH-BH ($30,10 M_\odot$) population & Horizon redshift & 1.4 & 7.2 & 6.5 & 4.8 \\
 & NS-NS merger early warning & Early warning time with SNR 8 [s] & & 390 & 500 & 140\\
\hline
\refsection{sec:DensMatterEoS}
 & NS-NS post-merger wave~\tablefootnote{Here we give the range of values as different equations of states are used.}
  &SNR	& 1.49\,--\,3.13	    & 5.24\,--\,11.15      &  4.71\,--\,9.91   & 9.00\,--\,17.26  \\
 & tidal deformability from NS-NS & $\lambda$ Distinguishability $D_\mathrm{eff}$[Mpc]    & 344  & 1394    & 1146 & 1451      \\
 &  tidal deformability from NS-BH 		& $\lambda$ Distinguishability $D_\mathrm{eff}$[Mpc]    & 412 & 1673 & 1369     &  1741     \\
\hline
\refsection{s:collapse_scipot}
& Convection \& SASI (typical SN) & SNR & 3.45 & 13.70 & 12.18 & 11.47\\
& Rapidly rotating core collapse (long GRB)& SNR & 74.38 & 287.87 & 249.94 & 279.00 \\
\hline
\refsection{s:cosmo}
 & Standard Sirens 		&  $\delta H$ / $H$  &   $0.16$   &  $0.01$   &   $0.01$ & $0.02$    \\
\hline
\refsection{sec:TestGRProbGWs}
 & Speed of GW (with EM+GW) 		&  Horizon Distance for 1.4-1.4 $M_\odot$ BNS (Gpc) & 0.44 &   1.7 &    1.5 & 1.2 \\
 & Graviton mass (with EM+GW) 		&  Horizon Distance for 1.4-1.4 $M_\odot$ BNS (Gpc) & 0.44 &   1.7 &    1.5 & 1.2 \\
 & Graviton mass (CBC) 		&  Graviton Compton Wavelength from 30-30 $M_\odot$ BBH (pc) & 1.4 &	2.8	& 2.5	& 2.3 \\
 & Decay of GWs 		& Horizon Distance for 1.4-1.4 $M_\odot$ BNS (Gpc) & 0.44 &   1.7 &    1.5 & 1.2
  \\
 \hline
\refsection{sec:TestGRSourcePhysics}
 & Parametrized deviations from PN theory &  frac. error in 3.5\,PN phasing term from 1.4-1.4 $M_\odot$ BNS & 0.14	& 0.04 &	0.04 &	0.04 \\
 & Tails 					&  frac. error in 1.5\,PN phasing term from 1.4-1.4 $M_\odot$ BNS&  $4.9 \times 10^{-3}$	& $1.6 \times 10^{-3}$ &	$1.2 \times 10^{-3}$ &	$1.8 \times 10^{-3}$ \\
 & Memory effect 			&  SNR for $h^{\rm mem}$ in Eq.~\eqref{hmem} (500Mpc) & 2.4 &  9.5   &  7.9 & 6.7     \\
 & No hair theorem (IMRI)    &  SNR for 100-1.4 $M_\odot$ BHNS &  6.7  &  25.3  &  24.8 &   16.7 \\
 & No hair theorem (ringdown) &  Ringdown SNR for 30-30 $M_\odot$ BBH & 6.6 & 29.5 & 22.2  &    19.1 \\
 & Unaccounted loss of $E ~ \& ~ J$ & SNR for 30-30 $M_\odot$ BBH & 75.9 &	 300.9 &	 256.7 &	 202.6 \\
\bottomrule
\end{tabular}}
\caption[Baseline Astrophysical Scores for the interferometer configurations]
{Tabulated comparison of next generation interferometer configurations. A representative (but not exhaustive) list of astrophysical sources and their respective baseline 'scores' are listed here for each of the configurations.}
\label{tabl:sources_baseline}
\end{sidewaystable}


\begin{table}
\resizebox{\textwidth}{!}{%
\centering
\begin{tabular}{r l c c c c}
\toprule
 & \textbf{Science Goals}           &  \textbf{NN} & \textbf{SEI} & \textbf{SUS} & \textbf{SPOT/CTN} \\
\toprule
\refsection{sec:PulsarSpinDownMountains}
& CW blind search volume & $ 0 $ & $ 0 $ & $ 0 $ & $ 0.04 $      \\
& $\epsilon$ limit (HF targ search) & $ 0 $ & $ 0 $ & $ 0 $ & $ -0.01 $      \\
& $\epsilon$ limit (LF targ search)  & $ -0.05 $ & $ 0 $ & $ -0.13 $ & $ -0.14 $      \\
\hline
\refsection{sec:CBCsensitivity}
 & NS-NS horizon & $0.01$ & $0$ & $0.02$ & $0.21$ \\
 & BH-BH (10+10) horizon & $0.01$ & $0$ & $0.03$ & $0.34$ \\
 & BH-BH (30+30) horizon & $0.02$ & $0$ & $0.04$ & $0.09$ \\
 & CBC early warning & $0.11$ & $0.09$ & $0.22$ & $0$ \\
\hline
\refsection{sec:DensMatterEoS}
 & NS-NS post-merger SNR~\tablefootnote{The range of values correspond to different equations of states.}	 	&  0   & 0     & 0 & 0 -- 0.01 \\
 & tidal deformability from NS-NS      &   0  &  $0$    &  0  &  0.06     \\
 & tidal deformability from NS-BH     & 0   & $0$      & 0    & 0.01      \\
 & NS $f$-mode 1590\,Hz (SGR) 	& 0 & 0 & 0 & 0     \\
 & NS mode 100\,--\,200\,Hz (SGR) 	& 0 & 0 & 0 & $0.30 $     \\
\hline
\refsection{s:collapse_scipot}
 & Convection \& SASI (typical SN) & 0 & 0 & 0 & 0.16 \\
 & Rapidly rotating core collapse (long GRB) & 0 & 0 & 0 & 0.07 \\
\hline
\refsection{s:cosmo}
 & Standard Sirens 		&  $0.01$ &  $0.01$    &  $0.03$   &    $0.53$   \\
\hline
\refsection{sec:TestGRProbGWs}
 & Speed of GW (EM+GW) 		& 0.01 & 0.00 & 0.01 & 0.35 \\
 & Graviton mass (EM+GW) 		& 0.01 & 0.00 & 0.01 & 0.35 \\
 & Graviton mass (CBC) 		& 0.00 &	0.00 &	0.01 &	0.11 \\
 & Decay of GWs 				& 0.01 & 0.00 & 0.01 & 0.35 \\
 \hline
\refsection{sec:TestGRSourcePhysics}
 & Parametrized deviations from PN theory &  0.02 &	0.03	& 0.03	& 0.13 \\
 & Tails 					& 0.04 &	0.06 &  0.06 &	0.15 \\
 & Memory effect (BBH 30+30)	& $0.01$ & 0 & $0.01$ & $0.34$ \\
 & No hair theorem (IMRI) 	&  0.02 & 0  & 0.05 & 0.27 \\
 & No hair theorem (ringdown) 	& 0 & 0 & 0 & 0.18 \\
 & Unaccounted loss of $E$ \& $J$ & 0.01 & 0.00 & 0.02  & 0.35 \\
\bottomrule
\end{tabular}}
\caption[CBC Science Jacobian I]
  {Jacobian of Science goals as a function of interferometer upgrade technology:
    Each column corresponds to a configurable parameter or noise source:  Newtonian noise (NN), seismic noise (SEI), suspension thermal noise at 10\,Hz (SUS), coating thermal noise (CTN), and arm cavity laser beam spot size (SPOT).
    (Note that the sensitivity changes exactly the same way with respect to CTN and SPOT. Hence these quantities are shown in the same column).
	Each row corresponds to a particular science goal $\mathcal{S}$.
	Each element of the matrix is the logarithmic partial derivative $\partial \log \mathcal{S}/\partial x$
	of the metric for a science goal $\mathcal{S}$ with respect to
	a parameter $x$ that represents a change in a particular interferometer component (e.g. laser power)
	or a component noise source (e.g. seismic noise). Both $\mathcal{S}$ and $x$ are normalized such that
	$\mathcal{S} = x = 1$ for the `baseline' detector
        configuration. \emph{Positive} Jacobian elements correspond to
        detector changes which \emph{increase} the SNR / decrease measurement errors.}
\label{tabl:Jacobian1}
\end{table}

\begin{table}
\resizebox{\textwidth}{!}{%
\centering
\begin{tabular}{r l c c c c}
\toprule
 & \textbf{Science Goals}  & \textbf{SQZ} & \textbf{POW} &  \textbf{FCL} & \textbf{MASS} \\
\toprule
\refsection{sec:PulsarSpinDownMountains}
& CW blind search volume & $ 1 $ & $ 0.3 $ &  0 & $ 0 $      \\
\hline
& $\epsilon$ limit (HF targ search) & $ -0.24 $ & $ -0.08 $ & 0 & 0      \\
& $\epsilon$ limit (LF targ search)  & $ -0.14 $ & $ -0.05 $ & $ 0.01 $ & $ -0.06 $      \\
\hline
\refsection{sec:CBCsensitivity}
 & NS-NS horizon & $0.48$ & $0.27$ & $-0.05$ & $0$ \\
 & BH-BH (10+10) horizon & $0.75$ & $0.42$ & $-0.07$ & $0.01$ \\
 & BH-BH (30+30) horizon & $0.29$ & $0.17$ & $-0.07$ & $0.01$ \\
 & CBC early warning & $0.13$ & $-0.07$ & $0.15$ & $0.12$ \\
\hline
\refsection{sec:DensMatterEoS}
 & NS-NS post-merger SNR &  0.81   &  0.45      & 0 & 0 \\
 & tidal deformability from NS-NS &  $0.7$  & $0.39$      & 0     & 0     \\
 &  tidal deformability from NS-BH  & $0.62$    &  $0.34$   & $0.01$     & 0      \\
 & NS $f$-mode 1590\,Hz (SGR) 	& $1.1 $ & $0.37$ & 0.37 & 0.37   \\
 & NS mode 100\,--\,200\,Hz (SGR) 	& $ 0.08$ & 0.07 & 0.12 & 0.11     \\
\hline
\refsection{s:collapse_scipot}
& Convection \& SASI (typical SN) & $0.58$ & $0.32$ & 0 & $-0.01$\\
& Rapidly Rotating Core Collapse (long GRB) & $0.70$ & $0.38$ & 0 &  0 \\
\hline
\refsection{s:cosmo}
 & Standard Sirens		 	&  $0.52$   &  $0.35$    &  $-0.05$   &  $-0.38$     \\
\hline
\refsection{sec:TestGRProbGWs}
 & Speed of GW (EM+GW) 		  &  0.34 &  0.23 & -0.03 & -0.21 \\
 & Graviton mass (EM+GW) 		  &  0.34 &  0.23 & -0.03 & -0.21 \\
 & Graviton mass (CBC) 		  & 0.22 &	0.13	& 0.00	& -0.03 \\
 & Decay of GWs 				  & 0.34 &  0.23 & -0.03 & -0.21 \\
\hline
\refsection{sec:TestGRSourcePhysics}
 & Parametrized deviations from PN theory &  0.48	& 0.29 &	-0.01	& -0.04	\\
 & Tails 					& 0.38 &	0.23 &	-0.02	& -0.04 \\
 & Memory effect (BBH 30+30)	& $0.38$ & $0.24$ & $-0.02$ & $-0.19$ \\
 & No hair theorem (IMRI)	&  0.32 &    0.21 &   -0.05 &  -0.10 \\
 & No hair theorem (ringdown) 	& 0.66 &   0.36 &   $0$  & -0.01 \\
 & Unaccounted loss of $E$ \& $J$ & 0.35 &  0.23 & -0.03  & -0.19 \\
\bottomrule
\end{tabular}}
\caption[CBC Science Jacobian II]{Jacobian of Science goals as a function of interferometer upgrade technology (Table \ref{tabl:Jacobian1} continued):
    Each column corresponds to a configurable parameter or noise source:
	squeezing factor (SQZ), arm cavity stored power (POW), loss of the filter cavity for
	squeezing angle rotation (FCL), and mirror mass (MASS).
        Each row corresponds to a particular science goal $\mathcal{S}$.
        Each element of the matrix is the logarithmic partial derivative $\partial \log \mathcal{S}/\partial x$
        of the metric for a science goal $\mathcal{S}$ with respect to
	a parameter $x$ that represents a change in a particular interferometer component (e.g. laser power)
	or a component noise source (e.g. seismic noise). Both $\mathcal{S}$ and $x$ are normalized such that
	$\mathcal{S} = x = 1$ for the `baseline' detector
        configuration.
        \emph{Positive} Jacobian elements correspond to detector
        changes which \emph{increase} the SNR / decrease measurement errors.}
\label{tabl:Jacobian2}
\end{table}

\subsubsection{Binary Parameters and Populations.}
\label{s:SciencePot}
\label{sec:CBCsensitivity}
GW observations of compact binary systems provide a great deal
of information about the component objects as well as the populations of
NSs and BHs in the Universe.
The analysis of individual detections yields the masses and spins
of the compact objects involved~\citep{GW150914:PE,Veitch:2014,VitaleEvans:2017}.  
The distribution of
these parameters in the population, along with the overall merger
rates, will give critical insights into the processes that govern binary
evolution. These include mass transfer in progenitors of compact
binaries, supernova kicks, the efficiency of common-envelope ejection, and the
dominant processes governing dynamical binary formation in dense
stellar environments~\citep[e.g.,][and references therein]{MandelOShaughnessy:2010,Rodriguez:2016big,PostnovYungelson:2014}.

Extracting this information will require a solution to the inverse
problem of GW astrophysics: reconstructing the astrophysics from a
collection of GW detections of coalescing binaries. This can be
accomplished by a combination of two techniques: comparing observed
distributions against modeled distributions predicted by
population-synthesis simulations under a variety of 
assumptions~\citep[e.g.,][]{Stevenson:2015,Stevenson:2017spin,Barrett:2017FIM}; or
model-independent efforts to distinguish subpopulations, such as those
of dynamically formed black-hole binaries and evolved isolated 
binaries~\citep[e.g.,][]{Mandel:2015,Vitale:2015,Mandel:2016cluster,Farr:2017}.

For either approach, both source statistics and accurate estimates 
of the parameters of individual systems will be required.  Here, we 
use the maximum (horizon) redshift at which a single detector could 
observe an optimally oriented, overhead source as a proxy for the 
amount of astrophysical knowledge that can be gained.  Greater horizon 
redshifts increase the overall number of detections and make it 
possible to probe the evolution of merger rates and mass distributions 
with redshift; they also imply greater local sensitivity, increasing 
the number of high-SNR detections, where inference on component masses 
and spins will be most accurate.

\voy{} in any incarnation will have sensitivity to binary mergers at
cosmological distances. For NS-NS binaries, the horizon redshift approaches
$z\approx0.5$, rising to $z\approx7$ for BH-BH binaries with $30\,M_\odot$ components. 
Hundreds or thousands of detections will allow us to
precisely characterize the long-sought NS equation of
state~\cite{Read:2009} and NS and BH mass and spin
distributions in merging binaries.  The equation of state measurement
requires high-frequency sensitivity individually, but benefits also
from a greater number of detections (cf. \refsection{s:NS}).

As mentioned earlier, multi-messenger observations are particularly exciting. 
These associate GWs from binary
coalescences with coincident observations of the electromagnetic
merger signatures, including detections with radio, optical, x-ray and
gamma-ray telescopes, and low- and high-energy neutrino detectors.  For example,
observations accompanying GW170817 have firmly established the relationship between
gamma ray bursts and compact binary coalescences involving NSs~\cite{GW170817:GRB}.

\begin{figure}
  \centering
     \includegraphics[height=4.18cm]{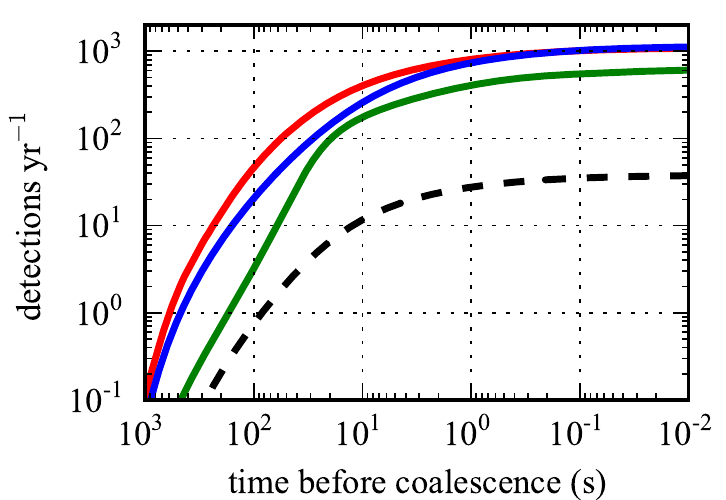}
     \includegraphics[height=4.18cm]{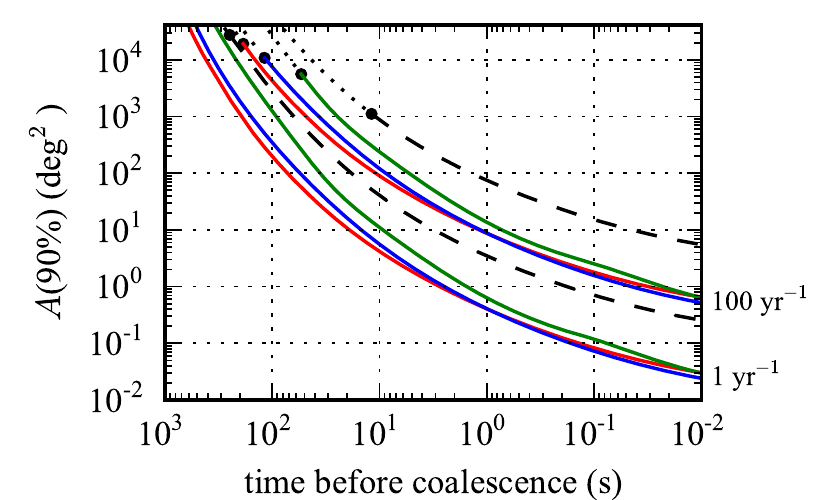}
  \caption[NS-NS early warning performance]
  {NS-NS early warning performance.
    Left: NS-NS detection rate
    per year (SNR = 8) at times before merger,
    for an arbitrary reference merger rate density.  
  Right: The evolving sky
  localization estimate at times before merger. For each detector,
  there are curves for two systems whose final amplitudes are set by
  their detection rates, labeled on the right. This plot assumes HLV
  as sites, but identical (LIGO) detectors. The thick dots in the
  upper left indicate when $\textrm{SNR} = 8$ is accumulated.
  The red, green, and blue curves are the respective \voy{} designs
  while the dashed line is aLIGO.}
\label{fig:early}
\end{figure}

The intriguing prospect of early-warning detection~\cite{2012ApJ...748..136C}
is primarily
dependent on low-frequency sensitivity to accumulate SNR in the early
part of the inspiral. We characterize early-warning detection with
$t_{\mbox{early}}$, the time before coalescence at which a once-a-year
event will accumulate SNR of 8 (optimally oriented binary, single detector).
The SNR and sky localization accuracy as functions of time before merger are
depicted in Fig.~\ref{fig:early}. Cosmological volume corrections have
been taken into account, but the NS-NS merger rate is held
constant at a fiducial value.

\subsubsection{Intermediate Mass Black Holes}
\label{sec:IMBH}

The evidence for the existence of intermediate-mass black holes
(IMBHs) in the $10^2$\,--\,$10^4\,\msun$ mass range is still inconclusive
at present.  Attempts to look for electromagnetic signatures are hampered by the small dynamical footprint of low-mass IMBHs and
the difficulty of associating phenomena such as ultraluminous x-ray sources specifically with
IMBHs (see \cite{MillerColbert:2004} for a review).  On the other
hand, a handful of promising sources have been observed
(e.g. \cite{Farrell:2009}), and multiple formation scenarios have been
proposed---though none without problems (see the introduction
of~\cite{Collaboration:S5HighMass} for a brief review).  Thus, GW
observations of compact objects in this mass range, which would be
enabled by a future detector with good low-frequency sensitivity,
could yield the first definitive proof of IMBH existence at the low
end of the IMBH mass range \cite{Veitch:2015}.  Such measurements could also answer
outstanding questions about the dynamics of globular clusters and
about the formation history of today's massive black holes~\cite{Gair:2009ETrev}.

In general, three types of GW signatures of binary coalescences
involving IMBHs can be detected by upgraded detectors:

\begin{itemize}
\item \emph{Intermediate-mass-ratio inspirals:} If a reasonable fraction of
globular clusters host IMBHs in the 100\,--\,1000\,$\msun$ range, there
is a good possibility of being able to detect intermediate-mass-ratio
inspirals (IMRIs) of stellar-mass compact objects (NSs or BHs) into
IMBHs.  The dominant mechanism is likely to be the successive
hardening of a binary involving an IMBH and a compact object by
three-body interactions, until it can merge through GW 
radiation reaction\,\cite{Mandel:2008,Macleod:2016,Haster:2016}.  Although IMBH occupation
fractions in globular clusters are highly uncertain, we could assume
that 10\% of globular clusters host an IMBH in a suitable mass range.
Globular clusters have a comoving density of $0.3\,\text{Mpc}^{-3}$.  We use
the fiducial values of 0.5\,$\msun$ for interloper mass in 3-body
interactions, $10^{5.5}\,\text{pc}^{-3}$ for the stellar density and
$10\,\text{km}\,\text{s}^{-1}$ for the velocity dispersion.
The merger rate depends on the
component masses; it is $1/T_\text{m}$, where the merger time scale is
\begin{equation*}
T_\text{m} \approx 3  \times 10^8 \left(\frac{m}{M_\odot}\right)^{-0.2} \left(\frac{M} {100
  M_\odot}\right)^{-0.4}\,\text{yr},
\end{equation*}
$M$ is the IMBH mass, and $m$ is the
inspiraling compact object mass.  We can assume that $M$ is
distributed from 100\,--\,1000~$\msun$ with $p(M) \propto
M^{-2}$, and $m$ is uniformly distributed from 1.2\,--\,12\,$\msun$.

\item {\it IMBH binaries in stellar clusters:} If a young dense stellar
cluster has a sufficiently high binary fraction, and the deep core
collapse timescale is sufficiently short, an IMBH-IMBH binary could
form via the collisional runaway scenario~\cite{Fregeau:2006}.  IMBH
binaries could also form through the collision of two globular
clusters, each containing an IMBH~\cite{AmaroSeoaneSantamaria:2009}.
Estimates of the rates and mass distributions of these processes is
highly uncertain, but a plausible framework for doing so is provided
in Section~3.2 of \cite{Gair:2009ETrev}.

\item {\it Low-mass MBH seeds:} According to hierarchical models of massive
black hole formation~\cite{MadauRees:2001}, today's massive black
holes are the product of multiple mergers and accretion episodes,
starting with light seeds of $\sim$100 or a few hundred solar masses,
possibly arising from the direct collapse of population III stars.  If
so, it may be possible to directly detect the first mergers of these
low-mass seeds at redshifts of $10$ or $15$, thereby testing these
predictions~\cite{Sesana:2009ET,Gair:2009ET}.

\end{itemize}



\subsubsection{Testing General Relativity.}
\label{sec:TestGRProbGWs}
\label{sec:TestGRSourcePhysics}

GWs from compact-binary mergers will provide a unique probe of
strong-field dynamics~\cite{Will:2001, Yunes:2011}.
LIGO's first observations of GWs from binary black holes have already allowed made it possible 
perform the first tests of general relativity (GR) in the highly relativistic 
strong-field regime~\cite{GW150914:GR}. LIGO Voyager instruments will allow us to significantly 
improve the precision of such tests.

\paragraph{Speed of propagation of GWs from joint GW-EM observations:}

According to GR, GWs travel at the speed of light, $c$. In other
theories, the speed $v_g$ of propagation of GWs could be
different~\cite{lrr-2006-3}. Coincident observation of electromagnetic
(EM) and GW signals from astrophysical sources such as GRBs or
core-collapse supernovae make it possible to measure the time-delay
$\Delta t_a$ between the EM and GW signals, and thus to constrain the
speed of GWs. For the case of a source located at a distance $D$,
\begin{equation}
  1-\frac{v_g}{c} \simeq \frac{c \Delta t}{D}; ~~~~ \Delta t
  = \Delta t_a - \left[(1+z) \Delta t_s \right]
\label{eq:GWspeedbounds}
\end{equation}
where $\Delta t_s$ is the time-delay between the GW and EM emissions
\emph{at the source} and $z$ the cosmological red shift.

The most promising astrophysical sources for this test are short-hard
GRBs \cite{GW170817:GRB}.
It can be seen from Eq.(\ref{eq:GWspeedbounds}) that the sensitivity of this test is
proportional to the distance to the source, and the best bound is
provided by sources located at the horizon distance of the detector.
The precise bound that we can place on $v_g$ depends on the the time 
delay $\Delta t_s$  at the source, which is currently uncertain.
Hence we use the horizon distance at which a double neutron star inspiral
$(m_1 = m_2 = 1.4 M_\odot)$ can be detected with an optimal SNR of 8
as a figure of merit for this measurement.

\paragraph{Mass of the graviton from joint GW-EM observations:}
One particular scenario in which the speed of GWs $v_g$ could differ from
$c$ is in the case of graviton having a non-zero rest mass. This is
characterized by the dispersion relation ${v_g^2}/{c^2} = 1 -
m_g^2\,c^4/{E_g^2}$, where $m_g$ is the rest mass and $E_g \equiv 2 \pi \hbar
f_{\mathrm{GW}}$ the rest energy of the graviton with frequency
$f_{\mathrm{GW}}$, $\hbar$ being the reduced Planck constant~\cite{Will98}.
From this dispersion relation and using Eq.(\ref{eq:GWspeedbounds}), a lower
bound on the Compton wavelength $\lambda_g = 2\pi\hbar/m_g c$ (or, an upper
bound on the mass $m_g$ of the graviton) can be inferred from joint GW-EM
measurements:

\begin{equation}
\lambda_g \gtrsim \left[\frac{Dc}{2\Delta t f_\mathrm{GW}^2}\right]^{1/2}
\end{equation}
It can be seen that the best bound is provided by distant sources.
Here we also use the horizon distance to a double neutron star inspiral
$(m_1 = m_2 = 1.4 M_\odot)$ as a figure of merit for this measurement.

\paragraph{Mass of the graviton from GW observations of CBCs:}
CBC observations also make it possible to estimate the mass of the graviton even
in the absence of an EM counterpart. In the case of CBCs, the GW
frequency sweeps from lower to higher frequencies. If the graviton is
massive, different frequency components travel with different speeds,
causing a distortion in the observed waveform~\cite{Will98}. In
particular, the observed GW phase $\Psi(f)$ in the frequency domain
will be deviated from the phase $\Psi_\mathrm{GR}(f)$ predicted by GR:
\begin{equation}
\Psi(f) = \Psi_\mathrm{GR}(f) - \frac{\pi D}{\lambda_g^2 (1+z)} \, f^{-1},
\end{equation}
where $\lambda_g \equiv h/m_gc$ is the Compton wavelength of the
graviton. LIGO's first observation of a binary black hole system has provided
one of the best lower bound on 
$\lambda_g \sim 10^{13}~\mathrm{km}$~\cite{PhysRevLett.116.221101}. Here we
use the expected lower bound on $\lambda_g$ from the observation of a binary
black hole system with parameters similar to the first LIGO event ($m_1 = m_2 = 30 M_\odot$,
located at a distance of 500 Mpc) as the figure of merit. The bounds reported
in Table~\ref{tabl:sources_baseline} are computed using the Fisher 
matrix formalism, outlined in~\cite{Keppel:2010qu}.

\paragraph{Decay of GWs during propagation:}
If GWs decay during propagation (apart from the expected $1/r$ falloff; e.g. due to dissipation),
distant sources would appear to be systematically dimmer. The detection of this requires a
population of coincident GW+EM observations with redshift $z$ estimation
(say, from the merger binary neutron stars).
Then we could look for a systematic suppression of GW amplitude for higher-$z$ sources.
The sensitivity of this test would be proportional to the distance traveled by the
GWs. Assuming that the redshift can be accurately estimated for sources located
at arbitrary distances, the relevant figure of merit for GW detectors is simply
the horizon distance. We take the horizon distance (SNR of 8) to non-spinning binary neutron
star inspirals with $m_1 = m_2 = 1.4 M_\odot$ as the figure of merit for this test.

\paragraph{Parametrized deviations from post-Newtonian theory:}
Here we introduce parametrized deviations from GR in the inspiral
waveforms computed using the post-Newtonian (PN) approxmation to GR,
and examine our ability to constrain these deviations from the data.
Consistency with their GR values is a null-hypothesis test of
relativity~\cite{Arun:2006a,Mishra:2010,Li:2012}.

The frequency domain phase of the PN waveforms can be written as:
\begin{equation}
\Psi(f) = 2\pi f t_0 + \phi_0 + \sum_{k=0}^{7} (\psi_k + \psi^\mathrm{L}_k \ln f ) f^{(k-5)/3},
\end{equation}
where $\psi_k$ and  $\psi^\mathrm{L}_k$ are the PN coefficients of the
phase at order $k/2$ PN.
We introduce deviations in the PN coefficients in the following way:
\begin{equation}
\psi_k \rightarrow \psi_k + \Delta \chi_k, ~~~ \psi_k^\mathrm{L} \rightarrow \psi^\mathrm{L}_k  + \Delta \chi^\mathrm{L}_k,
\end{equation}
where $\Delta \chi_k$ and $\Delta \chi^\mathrm{L}_k$ are zero in GR. For a 
non-spinning binary, the waveform $h(f) = A(f) \exp [i \ \Psi(f)]$ depends on 
the intrinsic binary parameters $m_1, m_2$, the extrinsic parameters $t_0, \phi_0, D$ 
and the deviation parameters $\Delta \chi_k, \Delta \chi^\mathrm{L}_k$. We deform 
the GR waveform by introducing one deviation parameter at a time and compute 
the expected constraint on this from the Cram\'er-Rao bound, from an 
archetypal binary neutron star system with $m_1 = m_2 = 1.4 M_\odot$ 
located at a distance of 500\,Mpc.
The $1\sigma$ errors in estimating the deviation parameter
$\Delta \chi_7$ at 3.5PN order, as a fraction of the known value
$\psi_7$ of the 3.5PN term in GR, are given in
Table~\ref{tabl:sources_baseline}.

\paragraph{Detecting the tails:}
The backscattering of gravitational waves emitted by a binary are by the binary's own gravitational field\cite{Blanchet1993,blanchet2017first} are known as \emph{tails}.
In a post-Newtonian expansion, this effect appears first in the 1.5\,PN phasing coefficient. The figure of merit will be our estimation error for this coefficient (that is, $\Delta \chi_3$ as defined in the previous section) from a double neutron star system ($m_1 = m_2 = 1.4 M_\odot$) located at 500\,Mpc.


\paragraph{Detecting the memory effect:}
The {\it memory wave} can be viewed as the gravitational effect of the
stress-energy carried by previously emitted waves
(Christodoulou, or nonlinear memory) with a contribution from the
final momentum distribution of the binary
(linear memory, e.g., due to kick)~\cite{Favata:2009,favata:10,Pollney:2011,PhysRevD.98.064031}.
The ramping up of the memory wave may be a measurable~\cite{Lasky2016,Yu2018,PhysRevLett.121.071102} test of GR. Let us provide a simple estimate of this effect, by writing
\begin{equation}
  \label{hmem}
  h^{\rm mem}(t) \propto \frac{R^2}{M} \int_{-\infty}^t \dot h^2(t') dt'
\end{equation}
Here $h$ is the leading gravitational waveform, while $h^{\rm mem}$ is the 
nonlinear memory effect generated by the stress-energy associated with $h$.  
We note that $h^{\rm mem}$ is mostly a slowly increasing function of time, 
except during the merger process. We can use the SNR for $h^{\rm mem}$ as the 
FOM for detecting the memory effect. We choose a fiducial BBH of 
$(30+30)\,\msun$ at 500\,Mpc as representative for this FOM.

\paragraph{Testing the no-hair theorem with inspirals}
According to the no-hair theorem of General Relativity, the spacetime around 
a singularity fully enclosed by an event horizon, with no closed timelike 
curves outside the horizon, and subject to several additional conditions, must 
be described by the Kerr metric.
In particular, the full structure of the 
spacetime can be described by a multipole moment decomposition with only 
two free parameters, mass and spin, and all higher-order moments given by 
the mass and spin.

Ryan proved that the full multipole moment structure of the spacetime
can in principle be measured through observations of
gravitational-wave emission~\cite{Ryan:1995}.  
Therefore, gravitational waves can be used to test deviations from the null 
hypothesis that massive compact objects are black holes by checking for consistency 
between higher-order multipole moments and their values as predicted by the Kerr metric.  
The mass quadrupole moment is the first measurable term beyond the mass and spin, 
entering the post-Newtonian expansion at the second post-Newtonian order in the phase.  
Therefore, its measurement is likely to be the focus of no-hair theorem tests, 
though other tests (e.g., through consistency of ringdown 
modes~\cite[e.g.,][and see below]{Meidam:2014}) are possible.

Intermediate-mass-ratio inspirals (IMRIs) of low-mass compact objects into intermediate 
mass black holes provide a particularly promising tool for constraining the mass 
quadrupole moment and null hypothesis deviations\,\cite{Brown:2007,Rodriguez:2012}.  
Given the uncertainty in existing waveform families in the intermediate-mass-ratio 
regime\,\cite[e.g.,][]{SmithIMRI:2013}, the exact precision of the measurement is 
at present compromised by systematic waveform uncertainty.  Therefore, we will use the SNR of an intermediate-mass-ratio inspiral of a $1.4$ solar mass neutron star 
into a $100\,\msun$ black hole at a luminosity distance of 1\,Gpc 
(consistent with an astrophysically plausible rate of such 
inspirals in globular clusters\,\cite{Mandel:2008}) as a proxy for the 
detectability of IMRIs and the measurability of the mass quadrupole moment.

\paragraph{Testing the No-Hair theorem with Ringdowns}
GW signals from the ringdown phase of a BBH coalescence are expected to be dominated 
by a spectrum of quasi-normal modes (QNMs). According to the no-hair theorem, the 
frequencies $\omega_{\ell m}^\mathrm{GR}$ and damping times $\tau_{\ell m}^\mathrm{GR}$ 
of these QNMs are unique functions of the mass $M_f$ and spin $a_f$ of the final 
Kerr black hole~\cite{Berti:2009}. Thus, in principle, the mass and spin of the 
final black hole can be extracted from different QNMs, which have to be consistent 
with each other. In practice, one can introduce parameters $\Delta \omega_{\ell m}$ 
and $\Delta \tau_{\ell m}$ that describe deviations from the GR prediction of the 
frequencies and damping times of the QNMs
\begin{equation}
\omega_{\ell m} = \omega_{\ell m}^\mathrm{GR}(M_f, a_f) ~ (1 + \Delta \omega_{\ell m}), ~~~
\tau_{\ell m} = \tau_{\ell m}^\mathrm{GR}(M_f, a_f) ~ (1 + \Delta \tau_{\ell m}).
\end{equation}
and constrain those deviations~\cite{Meidam:2014}.
We use the SNR of the ringdown phase of the expected signal from a BBH system with 
$m_1 = m_2 = 30\,\msun$, located at a distance of 500 Mpc as a simple FOM for our 
ability to constrain these deviation parameters.

\paragraph{Unaccounted loss of energy and angular momentum:}
If a GW signal from a BBH coalescence is observed with sufficient SNR,
the masses and spins of the black holes can be estimated from just the
inspiral part of the signal.
Using these estimates of the initial parameters of the binary,
the mass and spin of the final black hole can be uniquely predicted by
making use of numerical relativity simulations.
In addition, the mass and spin of the final black hole can be independently 
estimated from the ringdown part of the signal~\cite{Hughes:2004vw,Ghosh:2016qgn}. 
Any inconsistency between these two estimates will point to an
unaccounted loss in the energy / angular momentum from the system.

This consistency test requires binaries with the right masses and
spins such that the inspiral, merger and ringdown parts of the signal
are all observed with sufficiently large SNRs.
As a simple FOM, we use the SNR of a non-spinning binary with 
$m_1 = m_2 = 30 M_\odot$, located at 500\,Mpc.

\subsubsection{Dense Matter Equation of State from the Tidal Deformation of Neutron Stars}
\label{sec:DensMatterEoS}
The inspiral and merger of BH/NS or NS/NS binaries can
provide a wealth of information about the NS Equation of State (EOS).
This may come about through observing the phase evolution of a NS/NS or BH/NS in
the late inspiral to merger, the pulsations of a hypermassive NS (or
newly-born stable NS), which may form during a NS/NS merger, and the frequency of tidal disruption of the NS  in a BH/NS inspiral; see \cite{GW170817:EOS} for inference on the EOS from the gravitational-wave signal of the NS/NS merger GW170817.

\begin{figure}
    \centering
    \includegraphics[width=\columnwidth]{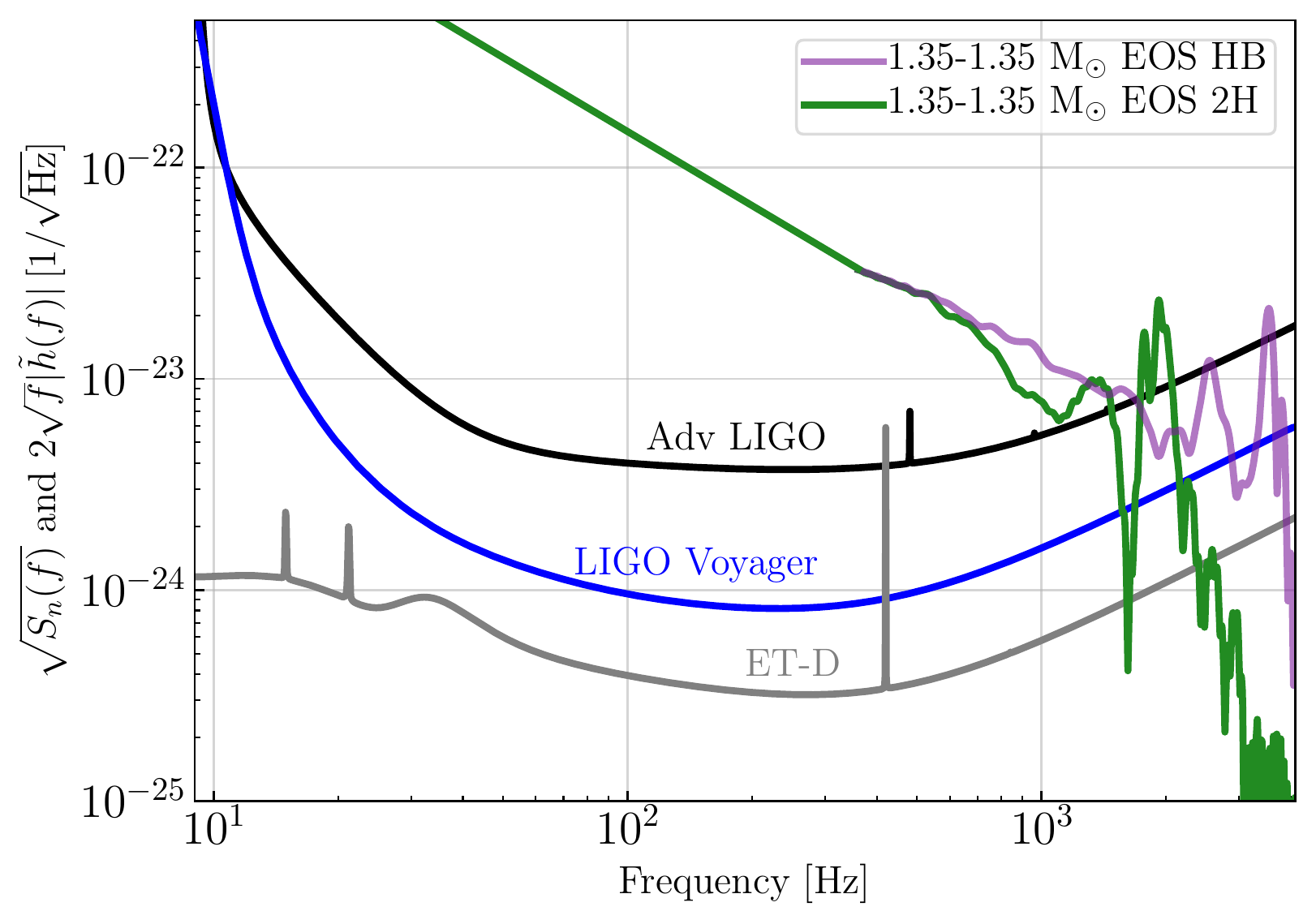}
    \caption[The Amplitudes of BNS waveforms for different EoS]
    {The amplitudes of NS/NS waveforms,
compared to noise spectra of various GW detectors, showing the frequency range
where various EoS effects may be seen. Hybrid waveforms for merging binary
neutron stars with two different EoS from~\cite{2009PhRvD..79l4033R} are shown.
EoS HB produces a neutron-star radius of 11.6\,km and EoS 2H a radius of
15.2\,km.
Both waveforms are from 1.35\,--\,1.35\,M$_\odot$
binaries at an effective distance of 100\,Mpc.  Differences in waveform
amplitudes due to strong tidal interactions are seen before the merger
(up to  $\sim\,1000$\,Hz for EoS 2H, and up to $\sim\,2000$\,Hz for EoS HB.) The
effects of the EoS on the phase evolution are not visible in the amplitudes
below $\sim\,500$\,Hz but would be measurable in Advanced LIGO for extreme
EoS like 2H~\cite{hinderer:10}.  Post-merger oscillations of a hypermassive
neutron star formed by the merger produce peaks in the spectrum near $2000$\,Hz
for EoS 2H, and near $3500$\,Hz for EoS HB.}
    \label{fig:NS_EoS_Strains}
\end{figure}

For most of the NS/NS or BH/NS inspiral, NSs are well approximated as
point particles; detections and rate estimates can be made with templates that
ignore finite size effects. However, as the size of the orbit becomes comparable
to the size of the neutron stars, the EoS will begin to modify the phase
evolution as tidal effects deform the neutron stars.  These modifications in the high-frequency portions of the waveforms are shown in Figure~\ref{fig:NS_EoS_Strains} for two representative equations of state.

The leading order effect of the EoS on the GWs is characterized
by a parameter that relates the size of the quadrupole deformation induced in
the star to the strength of the external tidal field. A dimensionless form of
this parameter is
\begin{equation}
\Lambda = \frac{2}{3} k_2 \left(\frac{R}{M}\right)^5,
\end{equation}
which depends on the Love number $k_2$, the radius of the star $R$, and the
mass $M$ of the star. Both $k_2$ and $R$ are determined by the EoS. Tidal effects
contribute to the waveform formally at 5th and higher post-Newtonian (PN) orders
\cite{hinderer:08,vines:11}.
The PN model will break down at high frequency, as the stars
interact more strongly, and eventually as the stars collide at a frequency which
depends on the EoS. However, the magnitude of the waveform effects predicted in
PN models is approximately equal to the magnitude of effects seen in both EOB
calculations\,\cite{Damour:2012yf} and estimates from hybrid waveforms using
numerical simulations~\cite{read:11c}.  In this paper, we therefore
use the leading order tidal contribution in TaylorF2 post-Newtonian models to
estimate the measurability of EoS effects. In practice, a more accurate waveform
model will be required to measure this parameter without significant systematic
errors~\cite{2014PhRvL.112j1101F, 2014PhRvD..89j3012W, GW170817:EOS}.

The figure of merit for neutron star tidal deformability is related to how well we can differentiate between different values of $\Lambda$ (and hence different equations of state) for various noise curves. We estimate how well we can differentiate between two waveforms using the distinguishability
criteria $\delta h$, as discussed in \cite{read:11c}. We find $\delta h$ for a given system at 100\,Mpc, and then
compute 100\,Mpc $\times \delta h$ to find the effective distance at which $\delta h = 1$ (the minimum distance for distinguishability).

\begin{figure}
  \centering
       \includegraphics[width=0.7\columnwidth]{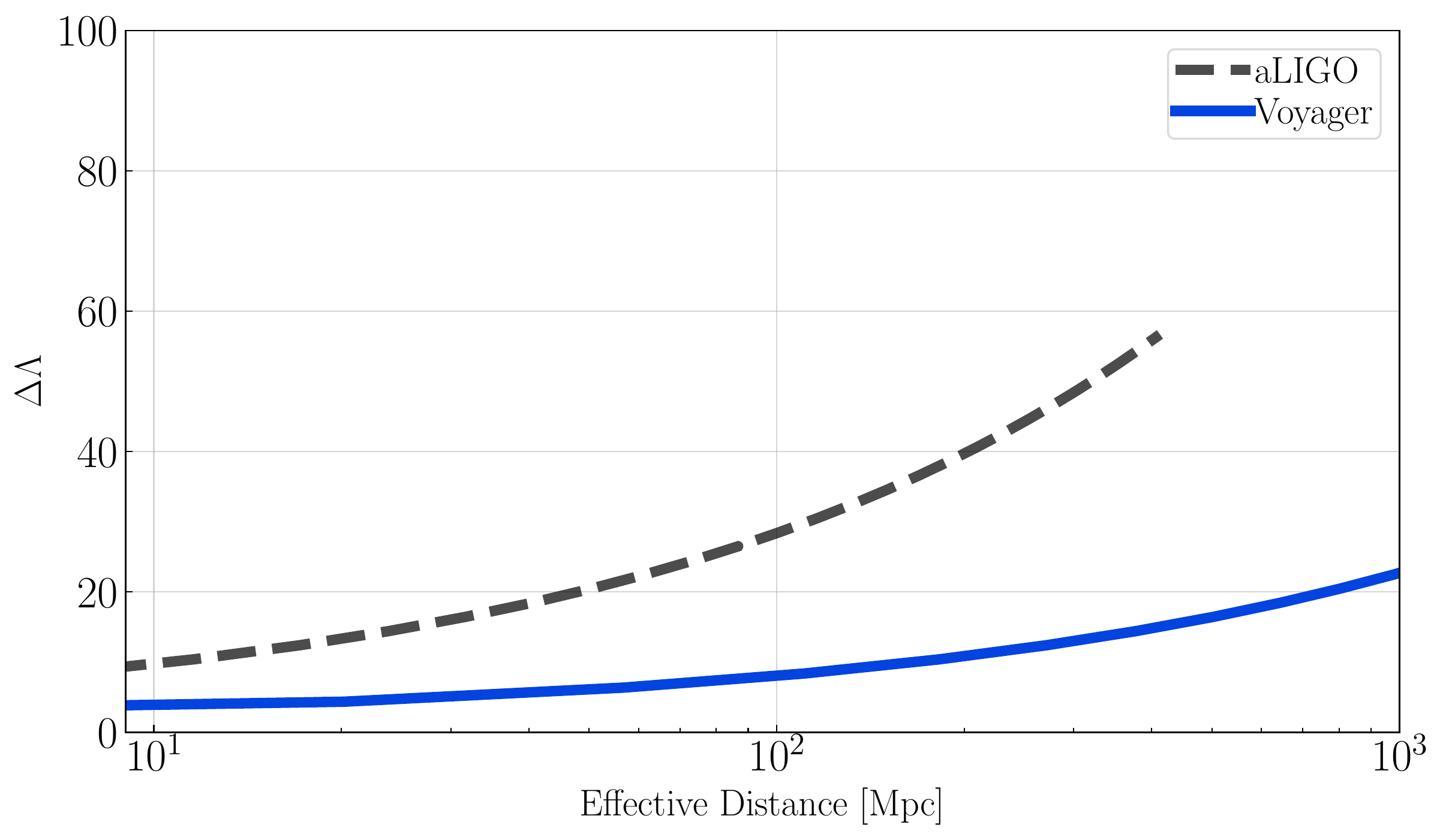}
       \caption[Tidal deformability figure of merit]
       {Tidal deformability figure of merit. $\Lambda$ is the parameter
       characterizing the tidal deformability of various equations of state.
       We show the maximum effective distance at which a given difference in 
       $\Lambda$ is distinguishable. This figure considers a NS-NS, $1.4 - 1.4$\,M$_\odot$ binary, with waveforms
       generated using Taylor T4, with tidal terms parametrized by $\Lambda$. Note that to distinguish between the SLY and AP4 equations of state,
       for example, one would need to find $\Delta \Lambda \approx 50$.}
    \label{fig:tidal_deformability}
\end{figure}

We consider a $1.4 - 1.4$\,$\msun$ NS-NS binary, and find how well we can
recover a given $\Lambda$ by computing overlaps of Taylor T4 waveforms.
For our figure of merit, we consider the maximum effective
distance at which the SLY~\cite{Haensel:2002cia} and 
AP4~\cite{Lattimer:2012nd} equations of state can be distinguished, where $\Lambda_\mathrm{SLY} = 323$
and $\Lambda_\mathrm{AP4} = 270$ for the given binary system. The accuracy in the recovery of the tidal deformability parameter $\Lambda$ is plotted in Figure~\ref{fig:tidal_deformability}.  For NS/BH binaries,
we consider the maximum effective distance where these two equations of state would be 
distinguishable in a $1.4 - 10.0$\,M$_\odot$ binary. The Jacobian quantities
using this FOM are then computed by the same prescription as in the rest of this
study.

\subsubsection{Measurement of the Cosmological Expansion}
\label{s:cosmo}

In the era of precision cosmology, measurements of cosmological parameters
are based primarily on two types of observations: standard candles and
large-scale structure.
Standard candle measurements rely on measuring data points in the
distance--redshift Hubble diagram by considering sources with a known intrinsic
luminosity, such as type IA supernovae, allowing the distance to be extracted
from their apparent luminosity.  Such observations famously led to the discovery
of the acceleration of the cosmological expansion of the
Universe~\cite{Perlmutter:1999}.
Together with measurements of large-scale structure, and particularly of the
cosmic microwave background radiation, these observations have led to modern
concordance cosmology: a flat Universe, with a Hubble constant of
approximately 70 km/s/Mpc, dominated by dark energy
($\Omega_\Lambda \approx 0.7$) and cold dark matter, with a dark energy
equation of state ($p = w \rho$) parameter $w = -1$ within
$\sim$15\%~\cite{2011ApJS..192...18K}.

Despite the success of these observations, they may suffer from systematic
errors, such as any redshift dependence in type IA SNe instrinsic luminosity.
There is also mild contention between the WMAP and Planck data regarding the
value of the Hubble constant~\cite{Planck:2015}.  Gravitational waves may therefore provide a compelling alternative measurement of
cosmography, since they are sensitive to different sources and subject to
different systematics~\cite{Sathya:Cosmo2010,TaylorGair:2012}.  The
promise of gravitational-wave cosmology was pointed out by
Schutz~\cite{Schutz:1986}.  Although gravitational-wave observations of binary
inspirals can be used to directly measure the distance to the source, making them
standard sirens~\cite{HolzHughes:2005}, the redshift is generally
degenerate with the source mass, since the mass provides the only timescale
in the evolution of the binary.

Several possibilities exist for breaking this degeneracy and measuring the
binary's redshift.  If the binary is associated with an electromagnetic
counterpart that allows accurate localization to a unique host galaxy, the redshift can be measured directly.  This was done for GW170817, with a resulting measurement of the Hubble constant at the 15\% level \cite{GW170817:H0}.  Otherwise, in the presence of a galaxy catalog and under
the assumption that binaries merge in galaxies, a statistical association is
possible~\cite{Schutz:1986,DelPozzo:2012}.   If one of the binary components is a neutron
star, matter effects (e.g., tides) provide an alternative scale in the
problem -- the physical size of the neutron star -- making it possible to
break the mass-redshift degeneracy~\cite{MessengerRead:2012,DelPozzo:2017}.  
Finally, if the intrinsic distribution
of the source masses is sufficiently narrow, the degeneracy can again be
broken~\cite{Taylor:2012}.

Although all of the above approaches are promising, in this study we will use
the accuracy of the measurement of the Hubble constant based on the
intrinsically narrow distribution of neutron star masses in binary neutron
star systems as the figure of merit.
A useful scaling law for the fractional Hubble constant measurement
accuracy\,\cite{Taylor:2012} is:
\begin{equation}
\frac{\delta H}{H} \approx \frac{1}{\sqrt N} \left[
\frac{\sigma_\mathcal{M}}{z~\mathcal{M}} + \frac{\delta {d_L}}{d_L} \right]\, ,
\end{equation}
where the fractional spread in the binary neutron star chirp mass is
${\sigma_\mathcal{M}}/{\mathcal{M}} \approx (0.06 M_\odot) / (1.2 M_\odot)  = 0.05$ \cite{Kiziltan:2013},
$z$ is the maximum redshift at which a neutron star binary can be detected, $N$ is the total number of detections, and ${\delta {d_L}}/{d_L}~\approx~0.3$
is the fractional uncertainty on the distance measurement for the most
distant source~\cite{Berry:2015}.


\subsection{Astrophysics with Stellar Collapse}
\label{s:supernova}

The core collapse of massive stars has long been considered an interesting 
source of GWs~\cite{mtw}. While the intricacies of the core-collapse supernova (CCSN) explosion mechanism 
are not well understood, state-of-the-art 3D simulations (see, e.g.,
\cite{roberts:2016,kuroda:2016,takiwaki:2016,radice:2016,abdikamalov:2015,lentz:2015,melson:2015-strangeq,bmueller:2015,moesta:2015,couch:2014,takiwaki:2014,murphy:2013,hanke:2013,couch:2013,dolence:2013,ott:2013,emueller:2012,takiwaki:2012,hanke:2012,burrows:2012,kotake:2011,scheidegger:2010}) suggest 
that rapid rotation, turbulent convection, and instabilities of the stalled 
accretion shock play important roles in re-energising the shock 
and aiding stellar explosion. GW emission from the initial core collapse and 
subsequent explosion is strongly influenced by the physical processes driving 
the explosion. For this reason, GW observations can be used to directly probe 
the CCSN central engine and gain insight into the explosion
mechanism~\cite{logue:12,powell:2016}. 

The angular momentum and degree of differential rotation of the precollapse stellar 
core are thought to strongly influence the dynamics of the initial collapse, the subsequent 
explosion, and the compact remnant formed (see, 
e.g.~\cite{macfadyen:1999,woosley:2006,yoon:2006,georgy:2009}). Observations of the evolving pulsar population suggest a broad distribution of 
moderately rotating NSs at birth with initial spin periods around 
10\,--\,100\,ms~\cite{faucher-giguere:2006,popov:2010,gullon:2014}. Wave-driven 
angular momentum transport in massive stars during the late stages of shell 
burning may strongly impact the pre-collapse rotation rate, predicting 
a distribution of initial NS periods consistent with
observations~\cite{fuller:2015}. Binary interactions are also expected to have a
marked effect on the rotation of massive stars~\cite{demink:2013,Zaldarriaga:2017,Qin:2019}.

For stellar cores with pre-collapse periods exceeding a few tens of seconds, 
delayed revival of the stalled shock is thought to be driven by
the neutrino mechanism. In this scenario, some small fraction of the binding
energy released in neutrinos is absorbed in a layer between the stalled shock
and proto-NS. Increased pressure behind the shock from neutrino heating and 
multi-dimensional hydrodynamic instabilities drive it outwards and aid explosion. 

State-of-the-art 3D simulations suggest that turbulent
convection and the standing accretion-shock instability (SASI) are expected to 
dominate the explosion dynamics~\cite{andresen:etal:2017,roberts:2016,radice:2016,abdikamalov:2015,lentz:2015,melson:2015-strangeq,bmueller:2015,yakunin:2015-gw}. 
Extensive research on the GW signature from slowly rotating core collapse has been done in 2D 
(see, e.g.,\cite{yakunin:2015-gw,bmueller:2013}, for recent studies) and 3D 
(see, e.g.,~\cite{yakunin:2017,kuroda:2016,andresen:etal:2017,ott:2013,emueller:2012} 
for recent studies).

Proto-neutron star (PNS) oscillation modes can source appreciable GW emission from 
$\sim(100-200)\,\mathrm{ms}$ after core bounce, following a short quiescent
period, for up to $\sim1\,\mathrm{s}$. The GW frequency naturally follows the 
dominant PNS surface g-mode frequency, increasing linearly with time from 
$\sim(100-200)\,\mathrm{Hz}$ to over $1\,\mathrm{kHz}$ as the PNS 
evolves~\cite{murphy:2009,marek:2009,bmueller:2013,yakunin:2015-gw,yakunin:2017}.
Strong fluid downflows associated with the SASI can modify the accretion rate 
at the PNS, inducing quadrupolar oscillations at $\sim(100-200)\,\mathrm{Hz}$ at later
times~\cite{murphy:2009,marek:2009,kuroda:2016}.

GW memory, a non-oscillatory contribution to the GW amplitude at leading
quadrupole order (see, e.g.~\cite{braginskii:87,favata:10}), may be 
created by anisotropic neutrino emission
\cite{epstein:78,mueller:04,ott:09,kotake:11,marek:09b,yakunin:10,
  muellere:12,mueller:13gw,yakunin:2015-gw,yakunin:2017} and aspherical 
explosive outflows (e.g., in jets or if the shock acceleration is not spherically symmetric)
of matter and magnetic stresses
\cite{obergaulinger:06a,obergaulinger:06b,takiwaki:11,ott:09}. For
anisotropic neutrino emission, the GW memory effect causes emission at less than
10\,Hz~\cite{yakunin:10,marek:09b,ott:09,yakunin:2015-gw,yakunin:2017}.

Rapidly rotating stars, which are expected to make up
$\sim$\,1\,--\,10\% of the massive star population\,\cite{ott:06spin,woosley:06,heger:05},
could explode via a bipolar magnetohydrodynamic explosion leading to 
large explosion asymmetries and relativistic
outflows~\cite{burrows:07b,takiwaki:09,scheidegger:2010,kuroda:2014,moesta:14b,moesta:2015,masada:2015,takiwaki:2016,richers:2017}.
The inner cores in rotating progenitor stars become centrifugally deformed in the late
stages of collapse. This results in a large quadrupole moment, which
changes rapidly at core bounce, leading to a strong and pronounced
peak in the GW signal that is followed by ring-down oscillations of
the PNS~\cite{dimmelmeier:08,richers:2017,abdikamalov:14,kuroda:2014}.

PNSs with strongly differentially rotating profiles are often 
subject to a rotational shear instability that drives the development
of nonaxisymmetric dynamics in the PNS 
core~\cite{scheidegger:10b,ott:07prl,ott:09,fu:11,muhlberger:14}. Also known as 
the co-rotational (or low $T/|W|$) instability, typical GW energy emitted can be 
as high as $10^{-7}M_\odot c^2$. 

A number of energetic SN explosions have been seen in coincidence with
nearby long gamma-ray bursts (GRBs), providing an
observationally robust connection between long GRBs and stellar
collapse \cite{hjorth:11,modjaz:11}. The central engine in a long GRB is 
thought to be either a nascent black hole surrounded by a fallback accretion 
disk (a collapsar~\cite{woosley:93,macfadyen:01,wb:06}) or millisecond 
proto-magnetar~\cite{wheeler:02,metzger:11}. In systems with accretion disks 
from fallback material, various instabilities may develop and lead to GW 
emission (e.g., \cite{piro:07,korobkin:11,kiuchi:11pp}). Classical dynamical 
instabilities are unlikely to occur in regular core collapse events, but may be 
relevant in extreme cases that lead to long GRBs and/or black hole formation
\cite{fryer:02,ott:09,ott:11a,oconnor:11,corsi:09,piro:11}. 

BH formation as a consequence of fallback accretion onto the PNS
is thought to be the formation channel for most stellar
BHs~\cite{oconnor:11,ugliano:12}. The timescale on which this occurs is
dependent on the accretion rate (directly influenced by
the properties of the progenitor star), the angular momentum of the PNS, and the
nuclear matter equation of state (EOS). In most systems, this happens
$\sim0.5-3\,\mathrm{s}$ after core bounce, and is characterised by a short GW
burst with a broad spectrum peaking at
$2.5$--$3.5\,\mathrm{kHz}$~\cite{cerda-duran:2013,oconnor:11,ott:2011}.

Rapid rotation is also expected to be present in massive accreting
white dwarfs and in the cores of white dwarf merger remnants (e.g.,
\cite{yoon:05b}). Such massive degenerate cores are expected to
collapse to neutron stars rather than explode as thermonuclear
supernovae. Such ``accretion-induced collapse'' (AIC) events are
expected to give off a strong burst of gravitational waves~\cite{ott:09, abdikamalov:10}.

\subsubsection{Detection Prospects.}
The galactic rate of core collapse events is low and estimates vary
from $\sim$1 in 40 years to 1 per century~\cite{vdb:91,mannucci:05,keane:08};
including the Large and Small
Magellanic Clouds and the Andromeda galaxy (M31) roughly doubles the
rate~\cite{vdb:91}. A significant increase in the event rate occurs
only outside of the Local Group of Galaxies with the M81 group of
galaxies at $\sim$3\,--\,5\,Mpc~\cite{ando:05, kistler:11}. There,
$\sim$\,0.5 core-collapse supernovae are discovered per year, suggesting
a rate of around 1 event per year, when assuming that $\sim$50\% of
the events remain undiscovered due to obscuration or weak/absent EM
emission. The integrated event rate out to 10\,Mpc is
likely around $\gtrsim$2 events per year~\cite{ando:05, kistler:11}.

While the search for CCSNe with the first generation of ground-based GW 
detectors yielded no observations, constraints on energy emitted in GWs by 
CCSNe were made for the first time~\cite{abbott:etal:2016:1gSNsearch}. With 
the second-generation instruments, we don't expect to see the typical 
CCSN beyond a few kpc for slowly rotating progenitors, while rapidly rotating 
progenitors might yield GW emission observable throughout the galaxy and 
Magellanic clouds~\cite{gossan:etal:2016,hayama:etal:2015,nakamura:etal:2016}.

\subsubsection{Probing Core-Collapse Supernova Physics.}
\label{s:collapse_scipot}
The physics that may be learned from a detection of 
GWs from stellar collapse goes far beyond constraining GW emission
and explosion mechanisms. CCSNe and related
phenomena are cosmic laboratories for high-energy-density
gravitational, plasma, nuclear, and particle physics. In particular,
it may be possible to extract information on the nuclear EOS directly
from GW observations~\cite{richers:2017,abdikamalov:2014,roever:09,murphy:2009,marek:09b,kuroda:2016}.
Electromagnetic (EM) observations can tell us little directly about the distribution of rotation rates of
pre-collapse iron cores, such that existing constraints come primarily from 
observations of the resultant young compact objects. For rapidly rotating
pre-collapse cores, GW observations can be used to directly infer the 
angular momentum distribution~\cite{abdikamalov:14,engels:2014,edwards:2014}, aided by  multimessenger observations with neutrinos~\cite{yokozawa:2015}.
Coincident GW and neutrino observations will be of extreme importance
if the next Galactic core collapse event leads to black hole formation
(without electromagnetic display). MeV-energy neutrinos from any Galactic or 
nearby extragalatic core collapse event will be observed by current and future neutrino
detectors (e.g., \cite{halzen:09,ikeda:07,icecube:11sn,scholberg:11,hyperkamiokande:11}). 

In the absence of a more specific FOM covering GWs from both slowly and 
rapidly rotating core collapse, we consider the signal SNR as the astrophysical
FOM. 
\voy{}, in the baseline design considered in this report, yields SNRs
for core-collapse supernova waveforms (see Table~\ref{tab:changeSNR} and
\refsection{s:supernova}) that are a factor of 4\,--\,5 greater than
aLIGO, which means that robust core-collapse supernova model selection
may be possible out to distances of $\sim$\,8\,--\,10\,kpc,
providing coverage virtually throughout the Galaxy. There will 
likely be at most one core collapse event in the Milky Way in the
lifetime of LIGO, so extending our reach throughout the Galaxy is
crucial.

According to the current understanding of core-collapse supernova
theory~\cite{janka:07}, the most likely and most robust
GW emission mechanism is turbulent neutrino-driven convection in the
context of the ``neutrino mechanism''~\cite{ott:09,mueller:13gw}. This
leads to broadband GW emission with most power at
100\,--\,1000\,Hz. Table~\ref{tab:changeSNR} shows that most
improvement above the baseline \voy{} design will come from reducing the
shot noise either through more squeezing or more laser power.
%
%
Note that the vast majority of stellar collapse events lead to standard-energy
type-II supernovae and are unlikely to be strong GW
emitters.  Even \voy{} will not be able to observe such events to
distances greater than $\sim 100\,\mathrm{kpc}$, which covers the
Milky Way, and the Magellanic Clouds.

A number of theoretical models (\refsection{s:supernova}) predict strong
GW emission connected with hyper-energetic
core-collapse supernovae and/or long-duration GRBs.  A potential
candidate emission process relies on long lasting non-axisymmetric bar-like
deformations of an extremely rapidly spinning PNS (or
``protomagnetar'') due to a rotational instability
\cite{ott:09,ott:10dcc,piro:11,fryer:02}. The GW
emission is expected to be narrow-band and at high frequency
($\sim$\,400\,--\,2000\,Hz) and Table~\ref{tab:changeSNR} lists
results for a range of potential waveforms, generated using the ad-hoc
bar model of \cite{ott:10dcc}. Even the weakest signal considered here
may be detectable by aLIGO at a distance of a few hundred
$\mathrm{kpc}$. \voy{} in its baseline configuration could be able to
detect this signal out to 5\,--\,10\,Mpc. At this distance,
$\gtrsim$$2$ core collapse events occur per year. It would thus be
possible to put strong constraint on the presence of such strong
emission models.

\subsubsection{Gravitational-Wave Memory in Stellar Collapse in the Milky Way}
GW memory may be left behind by most stellar collapse events, even
those that do not result in an explosion. The typical growth timescale
of the memory is of order $\gtrsim$$0.1\,\mathrm{s}$, which makes it
the only known low-frequency GW emission process in stellar
collapse. Detecting the GW memory from a galactic
event with aLIGO may be a difficult task even if the full projected
low-frequency sensitivity is reached, but the baseline \voy{} design
would allow detection. Searches for GW memory would most
benefit from improvements of the coating thermal noise or the arm
cavity spot size (see Table~\ref{tab:changeSNR}).

\begin{table}[t]
\resizebox{\textwidth}{!}{%
\centering
\begin{tabular}{lcccccccccc}
\textbf{Waveform Type} & \textbf{aLIGO SNR} & \textbf{\voy{} SNR} & \textbf{NN} & \textbf{SEI} & \textbf{SUS} & \textbf{SPOT/CTN} & \textbf{SQZ} & \textbf{POW} & \textbf{FCL} & \textbf{MASS}\\
& \textbf{@ 10 kpc} & \textbf{baseline @ 10 kpc} \\
\bottomrule
\textbf{$\nu$ GW Memory}\\
B12-WH07-$\nu$ (2D) \cite{yakunin:2015-gw} & 19.70 & 75.91 & $4.22\times10^{-3}$ & 
$3.93\times10^{-4}$ & $1.00\times10^{-2}$ & $3.14\times10^{-1}$ & 
$-2.18\times10^{-2}$ & $3.40\times10^{-1}$ & $2.20\times10^{-1}$ &
$-1.86\times10^{-1}$\\
B20-WH07-$\nu$ (2D) \cite{yakunin:2015-gw} & 16.54 & 63.73 & $4.18\times10^{-3}$ &
$3.89\times10^{-4}$ & $1.00e-02\times10^{-2}$ & $3.14\times10^{-1}$ & 
$-2.19\times10^{-2}$ & $3.40\times10^{-1}$ & $2.20\times10^{-1}$ & 
$-1.86\times10^{-1}$\\
\hline
\textbf{Rapidly Rot. Core Collapse}\\
R1E1CA$_\mathrm{L}$ (3D) \cite{scheidegger:2010} & 1.46 & 5.81 & 
$1.37\times10^{-6}$ & $5.67\times10^{-8}$ &
$5.70\times10^{-6}$ & $1.29\times10^{-1}$ & $6.22\times10^{-1}$ &
$3.39\times10^{-1}$ & $-1.62\times10^{-3}$ & $-6.65\times10^{-3}$\\
R3E1AC$_\mathrm{L}$ (3D) \cite{scheidegger:2010}& 74.38 & 287.87 & 
$2.43\times10^{-7}$ & $1.14\times10^{-8}$ &
$1.09\times10^{-6}$ & $7.49\times10^{-2}$ & $7.03\times10^{-1}$ &
$3.81\times10^{-1}$ & $-8.73\times10^{-4}$ & $-1.84\times10^{-3}$\\
R4E1FC$_\mathrm{L}$ (3D) \cite{scheidegger:2010}& 77.12 & 287.53 & 
$4.42\times10^{-7}$ & $1.40\times10^{-8}$ &
$2.22\times10^{-6}$ & $5.99\times10^{-2}$ & $7.38\times10^{-1}$ &
$3.99\times10^{-1}$ & $-7.69\times10^{-4}$ & $-2.65\times10^{-3}$ \\
R3 (3D) \cite{kuroda:2014} & 
30.48 & 123.72 & $1.98\times10^{-5}$ & $4.95\times10^{-7}$ &
$8.17\times10^{-5}$ & $2.21\times10^{-1}$ & $5.04\times10^{-1}$ &
$2.82\times10^{-1}$ & $-4.10\times10^{-3}$ & $-4.28\times10^{-2}$\\
\hline
\textbf{Convection \& SASI}\\
L15-3 (3D) \cite{mueller:12b} & 3.74 & 15.39 & $4.92\times10^{-4}$ &
$4.19\times10^{-5}$ & $1.23\times10^{-3}$ & $3.28\times10^{-1}$ &
$3.58\times10^{-1}$ & $2.14\times10^{-1}$ & $-8.59\times10^{-3}$ &
$-9.46\times10^{-2}$\\
s27$f_\mathrm{heat}1.05$ (3D) \cite{ott:13a} & 3.45 & 13.70 & 
$1.28\times10^{-7}$ & $4.30\times10^{-9}$ & $3.37\times10^{-6}$ &
$1.63\times10^{-1}$ & $5.84\times10^{-1}$ & $3.20\times10^{-1}$ &
$-2.31\times10^{-3}$ & $-1.48\times10^{-2}$\\
s20s (3D) \cite{andresen:etal:2017} & 5.58 & 21.95 & $1.85\times10^{-4}$ &
$9.08\times10^{-6}$ & $6.14\times10^{-4}$ & $1.88\times10^{-1}$ &
$5.52\times10^{-1}$ & $3.09\times10^{-1}$ & $-5.15\times10^{-3}$ &
$-5.31\times10^{-2}$\\
TM1 (3D) \cite{kuroda:2016} & 8.43 & 32.14 & $6.94\times10^{-7}$ & 
$1.33\times10^{-8}$ & $2.16\times10^{-5}$ & $1.33\times10^{-1}$ & 
$6.39\times10^{-1}$ & $3.51\times10^{-1}$ & $-2.67\times10^{-3}$ & 
$-3.31\times10^{-2}$\\
SFHx (3D) \cite{kuroda:2016} & 12.48 & 47.75 & $7.20\times10^{-7}$ &
$1.76\times10^{-8}$ & $2.75\times10^{-5}$ & $1.79\times10^{-1}$ &
$5.83\times10^{-1}$ & $3.24\times10^{-1}$ & $-3.87\times10^{-3}$ & 
$-5.21\times10^{-2}$ \\
\hline
\textbf{BH Formation in Collapsars}\\
u75rot1 (3D) \cite{ott:2011} & 11.77 & 43.84 & $1.81\times10^{-7}$ & 
$3.55\times10^{-9}$ & $1.76\times10^{-6}$ & $7.33\times10^{-2}$ &
$7.18\times10^{-1}$ & $3.89\times10^{-1}$ & $-9.13\times10^{-4}$ & 
$-3.20\times10^{-3}$\\
u75rot2 (3D) \cite{ott:2011} & 30.91 & 114.84 & $5.70\times10^{-7}$ & 
$1.32\times10^{-8}$ & $2.98\times10^{-6}$ & $7.08\times10^{-2}$ & 
$7.22\times10^{-1}$ & $3.91\times10^{-1}$ & $-8.76\times10^{-4}$ &
$-2.82\times10^{-3}$\\
\hline
\end{tabular}}
\caption{Jacobian of Science Return as a function of interferometer 
  upgrade technology for various stellar collapse waveform families 
  introduced in \refsection{s:supernova}. The baseline FOM is the 
  angle-averaged SNR $<\rho>$.}
\label{tab:changeSNR}
\end{table}

\subsection{Astrophysics with Neutron Stars}
\label{s:NS}

\subsubsection{Bursting Magnetars, Glitching Pulsars.}

Magnetars are neutron stars powered by extreme magnetic fields 
($\sim\,10^{15}$\,G)\,\citep{duncan92}. They are thought to be the
progenitors for the soft gamma repeaters (SGRs) and the anomalous
X-ray pulsars (AXPs), compact X-ray sources which give steeply rising
bursts of soft gamma rays typically lasting less than a second and
with total isotropic burst energies rarely exceeding $10^{42}$\,erg (for
a review see\,\cite{mereghetti08}).  Only a few dozen SGRs and
AXPs are known.  Three extraordinarily giant flares have been observed
in $\sim$30 years from magnetars in our Galaxy and the Large
Magellanic Cloud with observed energies of between
$\sim\sci{1.2}{44}d_\mathrm{55}^2$\,erg\,\citep{mazets79} and
$\sim\sci{5}{46}d_\mathrm{15}^2$\,erg\,\citep{terasawa05} where $d_n =
d/({n\,\mathrm{kpc}})$.  Some short gamma ray bursts (GRBs) might be
extragalactic giant flares.  GRB 070201 might have been a giant flare
located in the Andromeda galaxy with an isotropic energy of
$\sci{1.5}{45}$\,erg\,\citep{mazets08, S5GRB070201}; and GRB 051103
might have been a giant flare in M81 with an energy of
$\sci{7.5}{46}$\,erg\,\citep{frederiks07b}.

GW signals from magnetars would allow us to probe NS physics and
structure.  However, it is not clear when we might begin to expect a
detection.  Recent quantitative predictions or constraints on the
amplitude of GW emission associated with magnetar bursts are
uncertain, and while the most recent are pessimistic 
(see e.g.\,\cite{ioka01, owen05, horowitz09, corsi11, kashiyama11,levin11}) 
these sources are still poorly understood.  Furthermore,
the closest is only about a kpc from Earth and precise sky locations
and trigger times from electromagnetic (EM) bursts allow us to reduce
the false-alarm rate and increase sensitivity relative to all-sky,
all-time searches.  GW might be emitted by damping of non-radial
pulsational NS modes excited by a sudden localized energy release
caused by untwisting of the global interior magnetic field and
associated cracking of the solid NS crust\,\citep{thompson95}, or
global reconfiguration of the internal magnetic field and associated
deformation of the NS hydrostatic equilibrium\,\citep{ioka01,corsi11}.

The $f$-mode is damped principally via GW emission and would ring down
with a predicted damping time of 100\,--\,400\,ms and with a frequency in
the 1\,--\,3\,kHz range depending on the NS model\, \citep{benhar04}.  If
the magnetar outburst dynamics are confined to surface layer modes,
torsional oscillations in the crust might emit GWs at frequencies of
$\sim$10\,--\,2000\,Hz~\citep{mcdermott88}.  It is possible that no
neutron star mode will be excited at a sufficient level to emit
detectable GWs; however, the lack of theoretical understanding implies
the continued relevance of improving observational constraints on GW
emission.

The physical mechanisms behind pulsar glitches are another possible
route for excitation of GW-producing NS modes.  Like the magnetar
burst mechanism, the mechanism underlying pulsar glitches is also
poorly understood.  However, they might be caused by
starquakes\,\cite{middleditch06} or the transfer of angular momentum
from a differentially rotating superfluid core to the solid star
crust\,\cite{anderson75}.

\subsubsection{Continuous Sources of GWs: Spinning Deformed Neutron Stars.}
\label{sec:PulsarSpinDownMountains}
There are an estimated $10^9$ neutron stars in the galaxy 
(e.g., \cite{Andersson2010ETReview}), but only of order $3000$ have
been identified as radio/X-ray/$\gamma$-ray sources. The detection of
low-amplitude continuous GW emission in an all-sky search could mean
the discovery of this large unseen neutron star population.
Small nonaxisymmetric structures (``mountains'') on neutron stars are
the primary sources of continuous GWs that have been
targeted by GW searches. Radio, X-ray, and
$\gamma$-ray pulsars and spinning neutron stars in low-mass/high-mass X-ray binaries
are primary potential sources. All-sky searches
(which are presently computationally limited) for continuous signals
could discover unknown, radio/X-ray/$\gamma$-ray quiet neutron stars. Radio sources with precise timings can be followed up with long term integration that gains more than a magnitude in sensitivity.

The degree of nonaxisymmetry is typically quantified by the
ellipticity $\epsilon = (I_{xx} - I_{yy})/I_{zz}$
where $I_{ij}$ are components of the neutron star quadrupole
moment. 
The fiducial gravitational strain from such a deformation
will then be 
\begin{equation}
h_0 \sim \frac{1}{D} \frac{G}{c^4} (2\pi f)^2 I_{zz} \epsilon\,\,,
\label{eq:h0cont}
\end{equation}
where $f = 2 / P$ ($P$ being the spin period of the neutron star) and
$D$ is the distance to the detector. The signal will be sinusoidal
with long-period modulations due to intrinsic properties of the source,
binary orbital motion and motion of the detector relative to the source.
Since emission is strongest for rapidly spinning sources, the
detection of a low-frequency continuous wave source would be highly
unexpected, but scientifically extremely rewarding.

Theory suggests an upper bound \cite{ushomirsky:00} on $\epsilon$ of
\begin{equation}
\epsilon \lesssim 2\times 10^{-5} \left(\frac{\sigma_\mathrm{break}}{0.1}\right)\,\,,
\end{equation}
where $\sigma_\mathrm{break}$ is the breaking strain of the neutron
star crust, which may be as large as $\sim
0.1$ \cite{horowitz09}. However, regular radio pulsars are expected to
have significantly smaller ellipticities of order $10^{-9} - 10^{-8}$
\cite{Kitiashvili2006ellipticity}, but young magnetars may exhibit
significantly greater deformations
(e.g., \cite{colaiuda:08,Andersson2010ETReview}). Given these
constraints on $\epsilon$ and Eq.~(\ref{eq:h0cont}) it is obvious that
the expected GW strains will be minute even for nearby
sources and can be found only by computationally intense long-term
integrations. Constraining the ellipticity of known pulsars (e.g., Vela \cite{abbott:2010s5pulsar}) can constrain neutron star structure, crust physics, and pulsar models.
Detection of continuous GWs from spinning neutron stars will provide 
interesting probes to the true nature of GWs. 
Having just one continuous wave source with moderately stable
timing will place a very strong constraint on the speed of
GWs, due to the large impact of Doppler corrections
from detector movement.

\begin{figure}
\centering
  \includegraphics[width=\columnwidth]{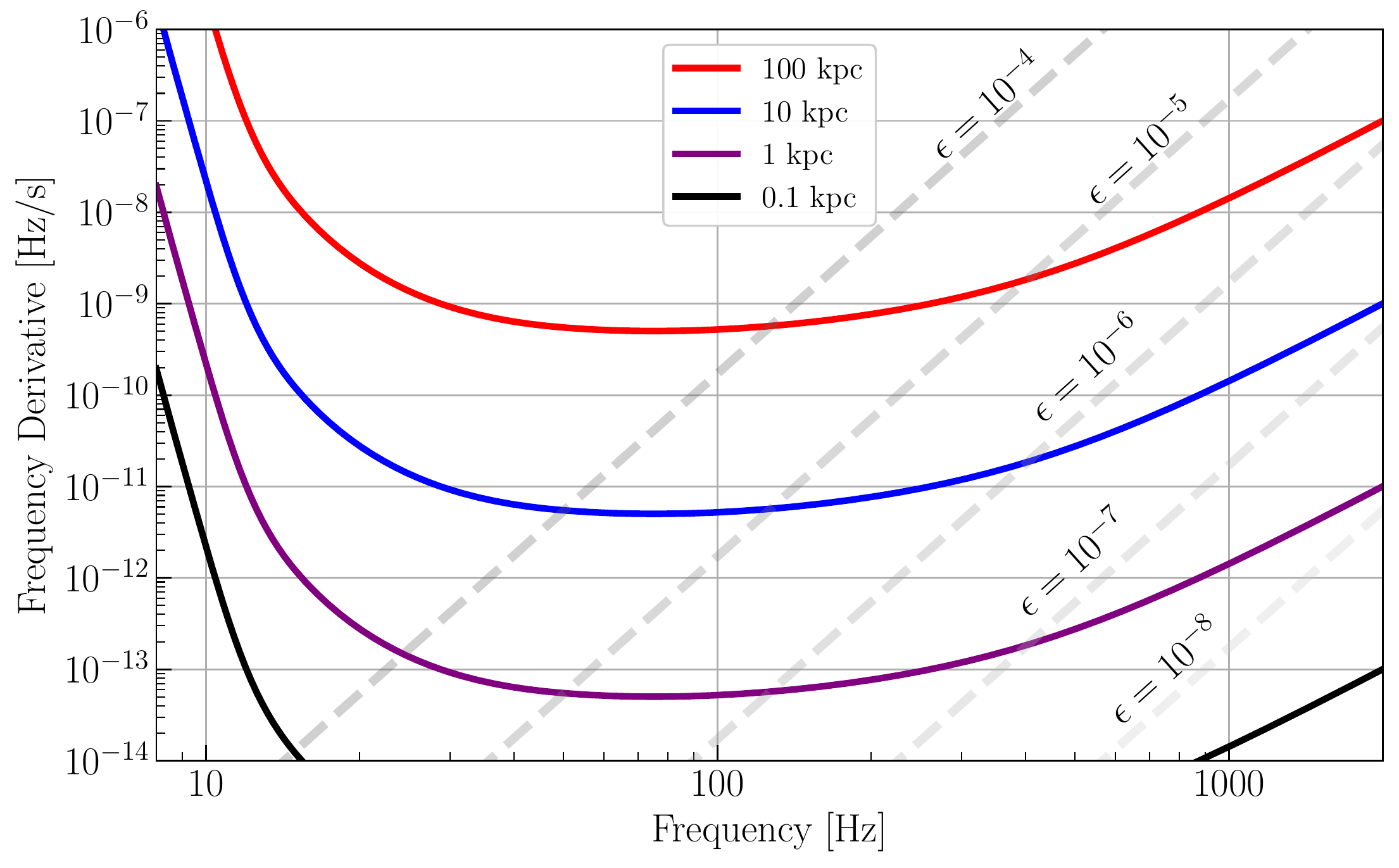}
 \caption[Pulsar upper limits]
 {Range of the PowerFlux search for continuous gravitational
   waves from neutron stars spinning down solely due to gravitational
   radiation, assuming the baseline \voy~detector.  This is a
   superposition of two contour plots.  The solid lines are
   contours of the maximum distance at which a neutron star could be
   detected as a function of gravitational-wave frequency $f$ and its
   derivative $\dot{f}$.  The dashed lines are contours of the
   corresponding ellipticity $\epsilon(f,\dot{f})$.
   }
\label{fig:spindown_range_standard}
\end{figure}

For the purpose of this study, we assume that the continuous wave
search pipeline \emph{PowerFlux}~\cite{Abbot2008S4IncoherentPaper} is
being run on data from an ideal \voy~detector. In
Fig~\ref{fig:spindown_range_standard} we provide estimates for the
astrophysics range of the \voy~baseline design. The neutron star
spindown is assumed to be solely due to GW emission,
which is the most optimistic, astrophysically unrealistic case. For
example, at the frequency of $1000$\,Hz, assuming a modest ellipticity
of $\mathrm{few}\,\times 10^{-7}$ (which is well under the maximum limit in
\cite{horowitz09}) \voy's reach will be almost
$10\,\mathrm{kpc}$. One should expect a gain of around another factor of
10 for a directed/targeted search.



The sensitivity of searches scales as $t^{1/2}$ for
a coherent search down to $t^{1/4}$ for semi-coherent
searches, where $t$ is the integration time. Thus, if one could make a change to the instrument that
improved the sensitivity at frequencies above 1\,kHz by a factor of 3 and
kept this running for 3 months it would be equivalent to running for
more than 2\,years for a coherent search (which is not practical for
blind searches at all) and more than 10\,years for a semi-coherent
search.



There are two main methods of discovery of continuous wave sources:
a search for unknown sources and a followup of pulsars
discovered by radio/X-ray/$\gamma$-ray surveys. Since the
GW strain at fixed ellipticity increases with the
square of the spin frequency, high frequency sources are easier to detect.
On the other hand we know many more radio pulsars with periods of 
$1$~s and higher than millisecond pulsars, but this sample is known 
to be highly biased.

For evaluating variations on the \voy~baseline design, we adopt three
figures of merit:
\begin{enumerate}
 \item  The integrated search volume of a PowerFlux-like blind search, where
  we consider the frequency space up to 1500\,Hz. This effectively assumes a
  flat prior on pulsar frequency. Keeping in mind that we do not know the
  distribution of pulsars with large ellipticity values this is not unreasonable.
  The volume is computed assuming the source has ellipticity $10^{-7}$ with emission frequency between $20$ and $1500$~Hz.
\item An ellipticity bound placed by a targeted search for a high frequency pulsar with
  parameters similar to J1023+0038 (frequency $2\times 592$~Hz, distance
  $900$~pc). Smaller values of this figure of merit are better.
\item An ellipticity bound placed by a targeted search for a low frequency pulsar with
  parameters similar to Vela (frequency $2\times 11.195$~Hz, distance
  $294$~pc). Smaller values of this figure of merit are better.
\end{enumerate}

The computation of these figures of merit uses a nominal constant to convert from design sensitivity to strain upper limits that could be obtained in actual search. The figure was derived by comparing initial LIGO design sensitivity to results of S5 run. This multiplicative factor is common to all numbers in Table \ref{tabl:sources_baseline} and cancels out for entries in Tables~\ref{tabl:Jacobian1} and \ref{tabl:Jacobian2}.

The results for these FOMs under variations of the baseline design are
summarized in Tables~\ref{tabl:Jacobian1} and \ref{tabl:Jacobian2}.
For first two FOMs we find the strongest improvement from noise reduction at
high frequencies, such as obtained with increased squeezing or laser power. Table \ref{tabl:sources_baseline} shows that high-frequency targeted search can place very interesting bounds on pulsar signals - or make a detection.
The third FOM benefits from any improvements to low frequency sensitivity, however it only approaches a region of interesting ellipticities.

\subsubsection{Postmerger Oscillation Signal}
\label{sec:pmns}

The most probable postmerger scenario following binary neutron star
coalescence is the formation of a massive ($M > 2\,\msun$), differentially
rotating neutron star~\cite{shibata:06bns, giacomazzo:11,
  hotokezaka:11, bauswein:12}.  The stability of this postmerger
neutron star (PMNS) against gravitational collapse depends on its mass
and on the details of the nuclear EOS.  Less
massive systems whose component masses add up to less than the maximum
mass that can be supported by the EOS in combination with uniform
rotation (the supramassive limit, e.g., \cite{cook:94c,kaplan:14}) ,
result in long-lived stable PMNSs. For more massive systems,
strong differential rotation temporarily supports the remnant and it
eventually undergoes gravitational collapse due to redistribution of
angular momentum via viscous processes and radiation of GWs. The
role of thermal pressure support is secondary, because at the
densities involved, the temperature-insensitive pressure due to the
nuclear force dominates the EOS~\cite{kaplan:14}. Sufficiently
high-mass systems that cannot be supported even by extreme
differential rotation will result in prompt collapse to a black hole
(BH) upon merger or very shortly after merger, emitting a
high-frequency ring-down \gw{} signal at $\sim$\,6\,--\,7\,kHz (e.g.,
\cite{shibata:06bns}).

Transient non-axisymmetric deformations in the surviving postmerger
remnant lead to a short duration ($\sim$\,10\,--\,100\,ms) burst
of \gw{s} with rich high frequency content, dominated by emission from
$f$-mode oscillations at $\sim$\,2\,--\,4\,kHz and generally
lower-frequency sub-dominant peaks from nonlinear couplings between
certain oscillation modes~\cite{stergioulas:11}.  The general
morphology of the \gw{} signal thus emitted resembles an
amplitude-modulated damped sinusoid, the phase and amplitude evolution
of which are not yet well modeled by numerical simulations of neutron
star coalescence and postmerger evolution.  However, the spectral
properties of this signal carry a particularly distinct signature of
the EOS.

\begin{table}
  \centering
\begin{tabular}{l l l l l l}
\toprule
EoS     & $M_{\mathrm{max}}$ & $R_{\mathrm{max}}$ &  $R_{1.35}$  & $\rho_{\mathrm{c}}/\rho_0$ & $f_{\mathrm{peak}}$\\
        & $[\msun]$      & [km]               & [km]         &   &  [Hz] \\ \hline
APR~\cite{1998PhRvC..58.1804A} (approx)                     & 2.19 &  9.90  & 11.33  & 10.4 &  3405 \\
DD2~\cite{2010PhRvC..81a5803T,2010NuPhA.837..210H} (full)   & 2.42 & 11.90  & 13.21  & 7.2  &  2589 \\
Shen~\cite{1998NuPhA.637..435S}  (full)                     & 2.22 & 13.12  & 14.56  & 6.7  &  2263 \\
NL3~\cite{1997PhRvC..55..540L,2010NuPhA.837..210H} (full)   & 2.79 & 13.43  & 14.75  & 5.6  &  2157 \\
\bottomrule
\end{tabular}
\caption[EoS comparisons]{The nuclear EOS employed in this study.
  References are provided in the first column. EOS indicated by
  ``approx'' refer to models which rely on an approximate treatment of
  thermal effects, whereas ``full'' marks EOS which provide the full
  temperature dependence. $M_{\mathrm{max}}$, $R_{\mathrm{max}}$, and
  $\rho_{\mathrm{c}}$ are the gravitational mass, circumferential
  radius, and central energy density of the maximum-mass
  Tolman-Oppenheimer-Volkoff configurations. We list
  $\rho_{\mathrm{c}}$ in units of the nuclear saturation density
  $\rho_0=2.7\times 10^{14}~\mathrm{g\,cm}^{-3}$. $R_{1.35}$ is the
  circumferential radius of a $1.35\,M_\odot$ NS.  The final column
  $f_{\mathrm{peak}}$ gives the dominant postmerger oscillation
  frequency.}
  \label{tab:eos}
\end{table}

A number of studies~\cite{bauswein:12,hotokezaka:13,bauswein:14} have
identified and confirmed a correlation between the dominant postmerger
oscillation frequency (i.e., half the peak \gw{} emission frequency) and the
radius of a fiducial cold, non-rotating neutron star.  For example,
in~\cite{bauswein:12}, the authors perform a Fisher matrix analysis and find
that it may be possible to use aLIGO observations of postmerger signals to
measure the dominant postmerger oscillation frequency to an accuracy of
$\sim$\,40\,Hz and thus determine the radius of a fiducial
$1.6$\,M$_\odot$ NS to an accuracy of 100\,--\,200\,m with an expected
detection horizon of approximately 15\,Mpc\footnote{Assuming an optimal SNR
detection threshold of 5, justified below.}, corresponding to a detection rate
of just under 1 per century, assuming the ``realistic'' rate in~\cite{ratesdoc}.
Furthermore, a systematic Monte-Carlo analysis using a variety of postmerger
waveforms corresponding to different component masses and EOS and data from the
initial-LIGO/Virgo instruments, recolored to the nominal advanced detector
design sensitivities was presented in~\cite{clark:14}.  There, the results
of~\cite{bauswein:12} are essentially confirmed, albeit at a reduced expected
detection range of $\sim\,5$\,Mpc, where this is now the angle-averaged range
since a network of detectors was used in that analysis and so the notion of
horizon distance is not well-defined. This corresponds to an expected detection
rate of $\sim\,0.5$ events per century. The reduction in sensitivity arises
since a generic burst search was used in the absence of accurate templates.
Fisher analysis, by contrast, assumes that an optimal filtering strategy is
feasible.

\begin{table}
\centering
\resizebox{\textwidth}{!}{%
\begin{tabular}{cccccccccccccc}
\toprule
& \multicolumn{3}{c}{aLIGO} & \multicolumn{3}{c}{Blue \voy{} Baseline} & \multicolumn{3}{c}{Red \voy{} Baseline} & \multicolumn{3}{c}{Green \voy{} Baseline} & \multicolumn{1}{c}{Frequency Recovery}  \\
\cmidrule(lr{0.2em}){2-4} \cmidrule(lr{0.2em}){5-7} \cmidrule(lr{0.2em}){8-10}\cmidrule(lr{0.2em}){11-13}\cmidrule(lr{0.2em}){14-14}
& $\langle \rho \rangle$ & $D_\text{H}$ & $\dot{\mathcal{N}}$ &  $\langle \rho \rangle$ & $D_\text{H}$ & $\dot{\mathcal{N}}$ & $\langle \rho \rangle$ & $D_\text{H}$ & $\dot{\mathcal{N}}$ & $\langle \rho \rangle$ & $D_\text{H}$ & $\dot{\mathcal{N}}$ & $\tilde{\delta f}\pm\mathrm{IQR}$\\
EoS & @ 10 Mpc & [Mpc] & [yr$^{-1}$] &  @ 10 Mpc & [Mpc] & [yr$^{-1}$] & @ 10 Mpc & [Mpc] & [yr$^{-1}$] & @ 10 Mpc & [Mpc] & [yr$^{-1}$] & [Hz] \\
\cmidrule(lr{0.2em}){1-1} \cmidrule(lr{0.2em}){2-4} \cmidrule(lr{0.2em}){5-7} \cmidrule(lr{0.2em}){8-10} \cmidrule(lr{0.2em}){11-13}\cmidrule(lr{0.2em}){14-14}
APR  & $1.49$  & $8.3$  & $3.0\times10^{-3}$ & $5.24$  & $39.63$ & $0.09$ & 4.71 & 33.20 & 0.05 & 9.00  & 52.41  & 0.31 & $10\pm54$ \\
DD2  & $2.55$  & $14.1$ & $6.6\times10^{-3}$ & $9.03$  & $53.26$ & $0.23$ & 8.07 & 52.10 & 0.22 & 14.76 & 94.42  & 0.74 & $-2\pm14$ \\
Shen & $2.64$  & $13.8$ & $6.4\times10^{-3}$ & $9.37$  & $70.19$ & $0.43$ & 8.34 & 54.08 & 0.24 & 14.73 & 94.81  & 0.74 & $8\pm22$  \\
NL3  & $3.13$  & $15.7$ & $7.7\times10^{-3}$ & $11.15$ & $73.84$ & $0.49$ & 9.91 & 67.04 & 0.37 & 17.26 & 115.03 & 0.74 & $5\pm10$  \\
\bottomrule
\end{tabular}}
\caption[BNS postmerger detection parameters v. Interferometer]
  {Detectability of the post-merger oscillation in binary neutron star
  mergers. Angle-averaged
  SNR $\langle\rho\rangle$, horizon distances
  $D_\text{H}$, and expected annual detection rate $\dot{\mathcal{N}}$
  (at SNR = 5).  Local merger rates are based on ``realistic'' rates~\cite{ratesdoc},
  but are uncertain within three orders of magnitude.
  Note that the true SNR recovered by a
  burst search will be factors of a few below the values given
  here (which are based on matched filtering).
  We also estimate the accuracy in the determination of the
  dominant postmerger oscillation frequency, based on a Monte-Carlo
  analysis (see text for details).}
\label{tab:pmns_estimates}
\end{table}

Here, we estimate the detectability of the postmerger signal from a surviving
PMNS for four different EOS using three figures of merit: the angle-averaged
signal-to-noise ratio, described in \ref{s:snr_estimates}; the horizon
distance, the distance at which an optimally-oriented face-on source yields an
optimal matched-filter SNR of some fiducial value and the expected annual
detection rate based on the ``realistic'' rates from~\cite{ratesdoc}\footnote{As
with the estimates for core-collapse supernovae, these FOMs assume that the
waveform is sufficiently well-modeled to permit a matched filtering detection
strategy.}.  Our interest here is in the detectability of the merger/postmerger
signal arising from surviving PMNS. We therefore choose to window the waveforms
prior to computing their optimal and angle-averaged SNR such that the
time-domain amplitude of the waveform is zero at times prior to the merger
(taken as the point of maximum amplitude).  This helps to ensure that there is
essentially no contribution to the SNR (and hence, our detectability estimates)
from the pre-merger inspiral signal.  Finally, the SNR is computed for
frequencies of 1\,kHz and above.

Since the postmerger signal is likely to only be observable for relatively
nearby, rare events, it is reasonable to assume that its time is known to
extremely high accuracy from the time of coalescence measured from the much
higher SNR inspiral precursor.  We therefore adopt a relatively low nominal SNR
threshold of $5$ in computing the horizon distance and detection rate.  The EOS
used for these estimates are a subset of those used in the more extensive study
of~\cite{clark:14} and range from the rather soft APR (high frequency \gw{}
signal) to the somewhat softer NL3 (lower frequency \gw{} signal).
Table~\ref{tab:pmns_estimates} reports the FOMs described above for each
waveform, for both the aLIGO noise curve and that of the red, blue, and green
\voy{} baseline designs.

While the mere detection of the postmerger signal will itself have
significant consequences for our understanding of the neutron star EOS
by excluding those EOS unable to support a long-lived postmerger
object, one of the most useful, and simple, measurements that will be
possible is the identification of the dominant postmerger oscillation
frequency.  It is, therefore, informative to also estimate the
accuracy with which the postmerger frequency may be determined.
To that end, we have performed a Monte-Carlo study wherein a
population of postmerger waveforms with a uniform distribution in SNR
have been injected into Gaussian noise with the noise spectral density
of the blue \voy{} baseline design.  The signals are recovered using a Bayesian
nested sampling algorithm with a simple sine-Gaussian template and the peak
frequency of the postmerger signal is recovered from the maximum-likelihood
estimate (ML) of the sine-Gaussian template.  This is a reasonable approximation
to a realistic burst search; the sine-Gaussian yields a 70\% match, or better,
with the postmerger waveforms and allows for accurate recovery of the peak
frequency. We quantify the accuracy of the frequency estimation via the error
$\delta f = f_{\mathrm{True}} - f_{\mathrm{ML}}$.
Table~\ref{tab:pmns_estimates} reports the median value and the interquartile
range\footnote{the difference between the 25$^{\mathrm{th}}$ and 75$^{\mathrm{th}}$
percentiles, representing a robust measure of the spread in the frequency error.}
of the frequency error for each waveform, for all injections recovered with a
sine-Gaussian SNR $>6$.  This higher threshold is chosen to reflect the fact
that this burst-like analysis will recover a factor of order unity less of the
SNR that an optimal matched-filter strategy would.  Since this frequency
estimate is essentially only SNR-dependent (i.e., the differences in the shapes
of the aLIGO and \voy{} baseline design noise curves at such high frequencies do
not affect this analysis), the frequency error is common to both noise curves.

In conclusion, we find that the blue \voy{} baseline design will provide a
realistic chance of the detection, and characterization, of the \gw{} signal
associated with postmerger oscillations following binary NS coalescence.  A
search for these signals with data from \voy{} would still require some level of
optimism and a relatively nearby (i.e., 10\,--\,50\,Mpc) event, but, given the
uncertainties in the expected rate of binary NS coalescence in the local
Universe~\cite{ratesdoc}, the uncertainties
in the numerical modeling of the postmerger signal and the science possible with
the detection and accurate measurement of the dominant postmerger oscillation
frequency ($\delta f \sim 10$\,Hz), this constitutes an important high-frequency
source for the next generation \gw{} observatories.



\begin{table}
\renewcommand{\arraystretch}{1.2}
\centering
\label{table:pmns_jacobian}
\resizebox{\textwidth}{!}{%
\begin{tabular}{>{\tiny} l >{\tiny}c >{\tiny}c >{\tiny}c >{\tiny}c >{\tiny}c >{\tiny} c >{\tiny} c >{\tiny} c}
\toprule
\textbf{Waveform} &\textbf{NN} & \textbf{Sei} & \textbf{SUS} & \textbf{SPOT/CTN} & \textbf{SQZ} & \textbf{POW} & \textbf{FCL} & \textbf{MASS}\\

\toprule
APR  & 0 & 0 & $4.50\times10^{-7}$ & $4.89\times 10^{-3}$ & 0.81 & 0.45 & $-2.73\times 10^{-5}$ & $-2.79\times 10^{-5}$ \\
DD2  & 0 & 0 & $1.62\times10^{-7}$ & $7.89\times 10^{-3}$ & 0.81 & 0.45 & $-4.14\times 10^{-5}$ & $-4.26 \times 10^{-5}$ \\
Shen & 0 & 0 & $1.46\times10^{-7}$ & 0.01 & 0.81 & 0.45 & $-5.18\times 10^{-5}$ & $-5.31\times 10^{-5}$ \\
NL3  & 0 & 0 & $1.83\times10^{-8}$ & 0.01 & 0.81 & 0.45 & $-5.76\times 10^{-5}$ & $-5.93\times 10^{-5}$ \\
\hline
\bottomrule
\end{tabular}}
\caption[Jacobian of Science Return for various postmerger waveforms]
{Jacobian of science goals as a function of interferometer upgrade technology.
The baseline FOM is the ideal optimally-oriented single-detector
matched filtering SNR for various postmerger waveforms (cf. \refsection{sec:pmns}).
The true SNR recovered by a burst search will be factors of order unity
below the values given for aLIGO and the baseline \voy{} design.}
\label{tab:changeSNRpmns}
\end{table}

\subsection{Stochastic Background and Unanticipated Discoveries}
The \voy{} detector could be used to look for a few different types
of stochastic background radiation: spatially resolved regions of
space with quasi-periodic or quasi-continuous radiation of a random
nature, an unresolved foreground of astrophysical sources, and a
background of cosmological origin. This has already been previously
explored within the context of the Einstein Telescope Design
study~\cite{ET:study}.

For the purposes of optimizing the design of this detector, we do not
include stochastic sources in the cost function.  The expected cosmological
background from inflation is either too weak to
detect~\cite{Bruce:1988} or is non-existent~\cite{BoSt2005, Steinhardt:2011}. 
Other cosmological stochastic background sources, such as early-Universe strongly first-order phase transitions, also suffer from uncertain amplitudes and peak frequencies \cite{GiblinThrane:2014}.
Estimates of the astrophysical foregrounds~\cite{PhysRevLett.118.121101}
have significant error bars at present, and work is still ongoing to quantify the scientific benefits of observing these foregrounds \cite{Callister:2016}.
Therefore, stochastic backgrounds do not influence the detector design in any 
quantitative way at present.  
Qualitatively, we remain excited about the prospect of
making serendipitous discoveries of heretofore unforeseen stochastic
sources.\\




Whenever a new vista onto the cosmos has been exposed in the past, it
has revolutionized our understanding of the Universe and its denizens.
We anticipate a similarly dramatic upheaval as the gravitational wave
Universe reveals itself.
A detector with broadband sensitivity is best suited for exploring the full range of serendipitous discoveries.


\section*{Conclusions}
\label{s:conclusion}
We have shown that a number of significant quantitative improvements can be achieved relative to a
wide array of known astrophysical targets by upgrading the LIGO interferometers within the existing facilities.
Precision tests of extreme spacetime curvatures can be made with
these improved instruments, perhaps even shedding light on what really happens at the black hole horizons.
In order to aid with making design tradeoffs for the LIGO Voyager
detector, we have numerically computed derivatives for these targets,
indicating how much scientific value there is in incremental improvements in the interferometers.
It is clear from the Jacobian tables that there are significant astrophysical gains to be made for modest investments in the reduction of technical noise in the audio band (40\,--\,8000\,Hz).
To make improvements for the low frequency (10\,--\,40\,Hz) science targets (e.g. GW memory or mergers of higher mass black holes) would require order-of-magnitude improvements in the seismic isolation, suspension thermal noise, and Newtonian gravity noise. 

This work should serve as a guide in making these detector design choices as well a starting point for more exhaustive evaluation of other science targets.

\section*{Acknowledgements}
RXA, PA and IM performed part of this work at the Aspen Center for Physics, which
is supported by National Science Foundation grant PHY-1066293.
JAC acknowledges support under NSF PHYS-1505824 and PHYS-1505524.
JSR acknowledges support from NSF PHYS-1307545 and the Research Corporation for Science Advancement.
RXA, PA and YC acknowledge support from the Indo-US Centre for the
Exploration of Extreme Gravity funded by the Indo-US Science and
Technology Forum (IUSSTF/JC-029/2016).
In addition, P.~A.'s research was supported by a Ramanujan Fellowship from the Science and Engineering Research Board (SERB), India, and by the Max Planck Society through a Max Planck Partner Group at ICTS.

\clearpage

\begin{appendices}

\section{Estimating Signal-To-Noise Ratios}
\label{s:snr_estimates}

The gravitational-wave signal $h(t)$ in an interferometric
detector is given by
\begin{equation}
h(t) = F_+(\theta, \varphi, \Psi)h_+(t,D,\iota,\beta) + F_\times(\theta,\varphi, \Psi) h_\times(t,D,\iota,\beta)\,\,.
\end{equation}

$F_+$ and $F_\times$ are the interferometer beam pattern functions
\cite{thorne:87}. $h_+$ and $h_\times$ are the two independent
polarizations of the gravitational-wave strain amplitude. The source
is located at distance $D$ and position $(\theta,\varphi)$
(spherical polar angles) on the sky. In the reference frame of the
source, Earth is located at a position $(\iota,\beta)$ (spherical
polar angles). $\Psi$ is the angle of the waves' polarization axis on
the plane of the sky relative to the detector's orientation (see
Fig.~9.2 of \cite{thorne:87}).

The optimal  Wiener matched-filter signal-to-noise ratio
(SNR, $\rho$) is
\begin{equation}
\rho^2 = 4 \int_0^\infty \frac{|\tilde{h}(f)|^2}{S(f)} df\,\,,
\end{equation}
where
\begin{equation}
\tilde{h}(f) = \int_{-\infty}^\infty h(t) \exp(2\pi i f t)\, df\,\,,
\end{equation}
and $S(f)$ is the one-sided power spectral density of noise
in the detector.

Since sky location, orientation of the source, and polarization angle
are unknown, it is most meaningful to compute and state SNRs that are
averages over these angles. In this, we follow Flanagan \& Hughes\,\cite{flanhughes:98}
and use their equation (2.30), which gives the
angle-averaged SNR in terms of the spectral energy density of the waves
\begin{equation}
\langle \rho^2 \rangle = \frac{2}{5\pi^2 D^2}\frac{G}{c^3} \int_0^\infty d\ln{f}\,
\frac{1}{f S(f)}  \frac{d E_\mathrm{GW}}{d f}\,\,.
\end{equation}

When comparing GW spectra with LIGO 3 detector design sensitivity
curves (the one-sided amplitude spectral density $\sqrt{S(f)}$), we
plot
\begin{equation}
\bar{h}(f) = \sqrt{\frac{2}{5\pi^2 D} \frac{G}{c^3}\frac{1}{f} \frac{dE_\mathrm{GW}}{df}
}\,\,,
\end{equation}
which has dimension of (frequency)$^{-1/2}$ and thus can be compared
directly with $\sqrt{S(f)}$: the ratio of $\bar{h}$ and $\sqrt{S(f)}$ is
dimensionless.

\clearpage
\section{Interferometer Parameters}
\label{s:IFOparams}


\providecommand\DefGwincVal[2]{%
  \expandafter\newcommand\csname gwincval-#1\endcsname{#2}%
}
\providecommand{\GwincVal}[1]{\csname gwincval-#1\endcsname}
\DefGwincVal{aLIGO.Infrastructure.Length}{3995}
\DefGwincVal{aLIGO.Infrastructure.ResidualGas.pressure}{4e-07}
\DefGwincVal{aLIGO.Infrastructure.ResidualGas.mass}{3.35e-27}
\DefGwincVal{aLIGO.Infrastructure.ResidualGas.polarizability}{7.8e-31}
\DefGwincVal{aLIGO.Constants.E0}{8.8542e-12}
\DefGwincVal{aLIGO.Constants.hbar}{1.0546e-34}
\DefGwincVal{aLIGO.Constants.c}{299792458}
\DefGwincVal{aLIGO.Constants.G}{6.6726e-11}
\DefGwincVal{aLIGO.Constants.kB}{1.3807e-23}
\DefGwincVal{aLIGO.Constants.h}{6.6261e-34}
\DefGwincVal{aLIGO.Constants.R}{8.3145}
\DefGwincVal{aLIGO.Constants.Temp}{290}
\DefGwincVal{aLIGO.Constants.yr}{31556926.08}
\DefGwincVal{aLIGO.Constants.Mpc}{3.084132270281887e+22}
\DefGwincVal{aLIGO.Constants.MSol}{1476.6848}
\DefGwincVal{aLIGO.Constants.g}{9.81}
\DefGwincVal{aLIGO.Constants.fs}{16384}
\DefGwincVal{aLIGO.Constants.fInspiralMin}{3}
\DefGwincVal{aLIGO.TCS.s_cc}{7.024}
\DefGwincVal{aLIGO.TCS.s_cs}{7.321}
\DefGwincVal{aLIGO.TCS.s_ss}{7.631}
\DefGwincVal{aLIGO.TCS.SRCloss}{0}
\DefGwincVal{aLIGO.Seismic.Site}{LHO}
\DefGwincVal{aLIGO.Seismic.KneeFrequency}{10}
\DefGwincVal{aLIGO.Seismic.LowFrequencyLevel}{1e-09}
\DefGwincVal{aLIGO.Seismic.Gamma}{0.8}
\DefGwincVal{aLIGO.Seismic.Rho}{1800}
\DefGwincVal{aLIGO.Seismic.Beta}{0.5}
\DefGwincVal{aLIGO.Seismic.Omicron}{1}
\DefGwincVal{aLIGO.Suspension.BreakStress}{750000000}
\DefGwincVal{aLIGO.Suspension.Temp}{300}
\DefGwincVal{aLIGO.Suspension.VHCoupling.theta}{0.001}
\DefGwincVal{aLIGO.Suspension.Silicon.Rho}{2330}
\DefGwincVal{aLIGO.Suspension.Silicon.C}{772}
\DefGwincVal{aLIGO.Suspension.Silicon.K}{4980}
\DefGwincVal{aLIGO.Suspension.Silicon.Alpha}{1e-09}
\DefGwincVal{aLIGO.Suspension.Silicon.dlnEdT}{2.5e-10}
\DefGwincVal{aLIGO.Suspension.Silicon.Phi}{2e-09}
\DefGwincVal{aLIGO.Suspension.Silicon.Y}{150000000000}
\DefGwincVal{aLIGO.Suspension.Silicon.Dissdepth}{0.0015}
\DefGwincVal{aLIGO.Suspension.FiberType}{0}
\DefGwincVal{aLIGO.Suspension.Silica.Rho}{2200}
\DefGwincVal{aLIGO.Suspension.Silica.C}{772}
\DefGwincVal{aLIGO.Suspension.Silica.K}{1.38}
\DefGwincVal{aLIGO.Suspension.Silica.Alpha}{3.9e-07}
\DefGwincVal{aLIGO.Suspension.Silica.dlnEdT}{0.000152}
\DefGwincVal{aLIGO.Suspension.Silica.Phi}{4.1e-10}
\DefGwincVal{aLIGO.Suspension.Silica.Y}{72000000000}
\DefGwincVal{aLIGO.Suspension.Silica.Dissdepth}{0.015}
\DefGwincVal{aLIGO.Suspension.C70Steel.Rho}{7800}
\DefGwincVal{aLIGO.Suspension.C70Steel.C}{486}
\DefGwincVal{aLIGO.Suspension.C70Steel.K}{49}
\DefGwincVal{aLIGO.Suspension.C70Steel.Alpha}{1.2e-05}
\DefGwincVal{aLIGO.Suspension.C70Steel.dlnEdT}{-0.00025}
\DefGwincVal{aLIGO.Suspension.C70Steel.Phi}{0.0002}
\DefGwincVal{aLIGO.Suspension.C70Steel.Y}{212000000000}
\DefGwincVal{aLIGO.Suspension.MaragingSteel.Rho}{7800}
\DefGwincVal{aLIGO.Suspension.MaragingSteel.C}{460}
\DefGwincVal{aLIGO.Suspension.MaragingSteel.K}{20}
\DefGwincVal{aLIGO.Suspension.MaragingSteel.Alpha}{1.1e-05}
\DefGwincVal{aLIGO.Suspension.MaragingSteel.dlnEdT}{0}
\DefGwincVal{aLIGO.Suspension.MaragingSteel.Phi}{0.0001}
\DefGwincVal{aLIGO.Suspension.MaragingSteel.Y}{187000000000}
\DefGwincVal{aLIGO.Suspension.Type}{Quad}
\DefGwincVal{aLIGO.Suspension.Stage(1).Mass}{39.6}
\DefGwincVal{aLIGO.Suspension.Stage(1).Length}{0.602}
\DefGwincVal{aLIGO.Suspension.Stage(1).Dilution}{NaN}
\DefGwincVal{aLIGO.Suspension.Stage(1).K}{NaN}
\DefGwincVal{aLIGO.Suspension.Stage(1).WireRadius}{NaN}
\DefGwincVal{aLIGO.Suspension.Stage(1).Blade}{NaN}
\DefGwincVal{aLIGO.Suspension.Stage(1).NWires}{4}
\DefGwincVal{aLIGO.Suspension.Stage(2).Mass}{39.6}
\DefGwincVal{aLIGO.Suspension.Stage(2).Length}{0.341}
\DefGwincVal{aLIGO.Suspension.Stage(2).Dilution}{106}
\DefGwincVal{aLIGO.Suspension.Stage(2).K}{5200}
\DefGwincVal{aLIGO.Suspension.Stage(2).WireRadius}{0.00031}
\DefGwincVal{aLIGO.Suspension.Stage(2).Blade}{0.0042}
\DefGwincVal{aLIGO.Suspension.Stage(2).NWires}{4}
\DefGwincVal{aLIGO.Suspension.Stage(3).Mass}{21.8}
\DefGwincVal{aLIGO.Suspension.Stage(3).Length}{0.277}
\DefGwincVal{aLIGO.Suspension.Stage(3).Dilution}{80}
\DefGwincVal{aLIGO.Suspension.Stage(3).K}{3900}
\DefGwincVal{aLIGO.Suspension.Stage(3).WireRadius}{0.00035}
\DefGwincVal{aLIGO.Suspension.Stage(3).Blade}{0.0046}
\DefGwincVal{aLIGO.Suspension.Stage(3).NWires}{4}
\DefGwincVal{aLIGO.Suspension.Stage(4).Mass}{22.1}
\DefGwincVal{aLIGO.Suspension.Stage(4).Length}{0.416}
\DefGwincVal{aLIGO.Suspension.Stage(4).Dilution}{87}
\DefGwincVal{aLIGO.Suspension.Stage(4).K}{3400}
\DefGwincVal{aLIGO.Suspension.Stage(4).WireRadius}{0.00052}
\DefGwincVal{aLIGO.Suspension.Stage(4).Blade}{0.0043}
\DefGwincVal{aLIGO.Suspension.Stage(4).NWires}{2}
\DefGwincVal{aLIGO.Suspension.Ribbon.Thickness}{0.000115}
\DefGwincVal{aLIGO.Suspension.Ribbon.Width}{0.00115}
\DefGwincVal{aLIGO.Suspension.Fiber.Radius}{0.000205}
\DefGwincVal{aLIGO.Materials.Coating.Yhighn}{140000000000}
\DefGwincVal{aLIGO.Materials.Coating.Sigmahighn}{0.23}
\DefGwincVal{aLIGO.Materials.Coating.CVhighn}{2100000}
\DefGwincVal{aLIGO.Materials.Coating.Alphahighn}{3.6e-06}
\DefGwincVal{aLIGO.Materials.Coating.Betahighn}{1.4e-05}
\DefGwincVal{aLIGO.Materials.Coating.ThermalDiffusivityhighn}{33}
\DefGwincVal{aLIGO.Materials.Coating.Phihighn}{0.00023}
\DefGwincVal{aLIGO.Materials.Coating.Indexhighn}{2.0654}
\DefGwincVal{aLIGO.Materials.Coating.Ylown}{72000000000}
\DefGwincVal{aLIGO.Materials.Coating.Sigmalown}{0.17}
\DefGwincVal{aLIGO.Materials.Coating.CVlown}{1641200}
\DefGwincVal{aLIGO.Materials.Coating.Alphalown}{5.1e-07}
\DefGwincVal{aLIGO.Materials.Coating.Betalown}{8e-06}
\DefGwincVal{aLIGO.Materials.Coating.ThermalDiffusivitylown}{1.38}
\DefGwincVal{aLIGO.Materials.Coating.Philown}{4e-05}
\DefGwincVal{aLIGO.Materials.Coating.Indexlown}{1.45}
\DefGwincVal{aLIGO.Materials.Substrate.c2}{7.6e-12}
\DefGwincVal{aLIGO.Materials.Substrate.MechanicalLossExponent}{0.77}
\DefGwincVal{aLIGO.Materials.Substrate.Alphas}{5.2e-12}
\DefGwincVal{aLIGO.Materials.Substrate.MirrorY}{72700000000}
\DefGwincVal{aLIGO.Materials.Substrate.MirrorSigma}{0.167}
\DefGwincVal{aLIGO.Materials.Substrate.MassDensity}{2200}
\DefGwincVal{aLIGO.Materials.Substrate.MassAlpha}{3.9e-07}
\DefGwincVal{aLIGO.Materials.Substrate.MassCM}{739}
\DefGwincVal{aLIGO.Materials.Substrate.MassKappa}{1.38}
\DefGwincVal{aLIGO.Materials.Substrate.RefractiveIndex}{1.45}
\DefGwincVal{aLIGO.Materials.MassRadius}{0.17}
\DefGwincVal{aLIGO.Materials.MassThickness}{0.2}
\DefGwincVal{aLIGO.Laser.Wavelength}{1.064e-06}
\DefGwincVal{aLIGO.Laser.Power}{125}
\DefGwincVal{aLIGO.Optics.Type}{SignalRecycled}
\DefGwincVal{aLIGO.Optics.SRM.CavityLength}{55}
\DefGwincVal{aLIGO.Optics.SRM.Transmittance}{0.2}
\DefGwincVal{aLIGO.Optics.SRM.Tunephase}{0}
\DefGwincVal{aLIGO.Optics.PhotoDetectorEfficiency}{0.95}
\DefGwincVal{aLIGO.Optics.Loss}{3.75e-05}
\DefGwincVal{aLIGO.Optics.BSLoss}{0.0005}
\DefGwincVal{aLIGO.Optics.coupling}{1}
\DefGwincVal{aLIGO.Optics.Curvature.ITM}{1970}
\DefGwincVal{aLIGO.Optics.Curvature.ETM}{2192}
\DefGwincVal{aLIGO.Optics.SubstrateAbsorption}{5e-05}
\DefGwincVal{aLIGO.Optics.pcrit}{10}
\DefGwincVal{aLIGO.Optics.ITM.BeamRadius}{0.055}
\DefGwincVal{aLIGO.Optics.ITM.CoatingAbsorption}{5e-07}
\DefGwincVal{aLIGO.Optics.ITM.Transmittance}{0.014}
\DefGwincVal{aLIGO.Optics.ITM.CoatingThicknessLown}{0.308}
\DefGwincVal{aLIGO.Optics.ITM.CoatingThicknessCap}{0.5}
\DefGwincVal{aLIGO.Optics.ETM.BeamRadius}{0.062}
\DefGwincVal{aLIGO.Optics.ETM.Transmittance}{5e-06}
\DefGwincVal{aLIGO.Optics.ETM.CoatingThicknessLown}{0.27}
\DefGwincVal{aLIGO.Optics.ETM.CoatingThicknessCap}{0.5}
\DefGwincVal{aLIGO.Optics.PRM.Transmittance}{0.03}
\DefGwincVal{aLIGO.Optics.Quadrature.dc}{1.5708}
\DefGwincVal{aLIGO.Squeezer.Type}{None}
\DefGwincVal{aLIGO.Squeezer.AmplitudedB}{10}
\DefGwincVal{aLIGO.Squeezer.InjectionLoss}{0.05}
\DefGwincVal{aLIGO.Squeezer.SQZAngle}{0}
\DefGwincVal{aLIGO.Squeezer.FilterCavity.fdetune}{-14.5}
\DefGwincVal{aLIGO.Squeezer.FilterCavity.L}{100}
\DefGwincVal{aLIGO.Squeezer.FilterCavity.Ti}{0.00012}
\DefGwincVal{aLIGO.Squeezer.FilterCavity.Te}{0}
\DefGwincVal{aLIGO.Squeezer.FilterCavity.Lrt}{0.0001}
\DefGwincVal{aLIGO.Squeezer.FilterCavity.Rot}{0}
\DefGwincVal{aLIGO.OutputFilter.Type}{None}
\DefGwincVal{aLIGO.OutputFilter.FilterCavity.fdetune}{-30}
\DefGwincVal{aLIGO.OutputFilter.FilterCavity.L}{4000}
\DefGwincVal{aLIGO.OutputFilter.FilterCavity.Ti}{0.01}
\DefGwincVal{aLIGO.OutputFilter.FilterCavity.Te}{0}
\DefGwincVal{aLIGO.OutputFilter.FilterCavity.Lrt}{0.0001}
\DefGwincVal{aLIGO.OutputFilter.FilterCavity.Rot}{0}
\DefGwincVal{aLIGO.Laser.Wavelength_nm}{1064}
\DefGwincVal{aLIGO.Materials.MassRadius_cm}{17}
\DefGwincVal{aLIGO.Materials.MassThickness_cm}{20}
\DefGwincVal{aLIGO.Optics.ITM.BeamRadius_cm}{5.5}
\DefGwincVal{aLIGO.Optics.ETM.BeamRadius_cm}{6.2}
\DefGwincVal{aLIGO.Substance}{fused silica}
\DefGwincVal{aLIGO.Suspension.FiberType_str}{fiber}
\DefGwincVal{aLIGO.Squeezer.FilterCavity.Lrt_ppm}{100}

\providecommand\DefGwincVal[2]{%
  \expandafter\newcommand\csname gwincval-#1\endcsname{#2}%
}
\providecommand{\GwincVal}[1]{\csname gwincval-#1\endcsname}
\DefGwincVal{Red.Infrastructure.Length}{3995}
\DefGwincVal{Red.Infrastructure.ResidualGas.pressure}{4e-07}
\DefGwincVal{Red.Infrastructure.ResidualGas.mass}{3.35e-27}
\DefGwincVal{Red.Infrastructure.ResidualGas.polarizability}{7.8e-31}
\DefGwincVal{Red.Constants.E0}{8.8542e-12}
\DefGwincVal{Red.Constants.hbar}{1.0546e-34}
\DefGwincVal{Red.Constants.c}{299792458}
\DefGwincVal{Red.Constants.G}{6.6726e-11}
\DefGwincVal{Red.Constants.kB}{1.3807e-23}
\DefGwincVal{Red.Constants.h}{6.6261e-34}
\DefGwincVal{Red.Constants.R}{8.3145}
\DefGwincVal{Red.Constants.Temp}{290}
\DefGwincVal{Red.Constants.yr}{31556926.08}
\DefGwincVal{Red.Constants.Mpc}{3.084132270281887e+22}
\DefGwincVal{Red.Constants.MSol}{1476.6848}
\DefGwincVal{Red.Constants.g}{9.81}
\DefGwincVal{Red.Constants.fs}{16384}
\DefGwincVal{Red.Constants.fInspiralMin}{3}
\DefGwincVal{Red.TCS.s_cc}{7.024}
\DefGwincVal{Red.TCS.s_cs}{7.321}
\DefGwincVal{Red.TCS.s_ss}{7.631}
\DefGwincVal{Red.TCS.SRCloss}{0}
\DefGwincVal{Red.Seismic.Site}{LLO}
\DefGwincVal{Red.Seismic.KneeFrequency}{10}
\DefGwincVal{Red.Seismic.LowFrequencyLevel}{1e-09}
\DefGwincVal{Red.Seismic.Gamma}{0.8}
\DefGwincVal{Red.Seismic.Rho}{1800}
\DefGwincVal{Red.Seismic.Beta}{0.5}
\DefGwincVal{Red.Seismic.Omicron}{5}
\DefGwincVal{Red.Suspension.BreakStress}{750000000}
\DefGwincVal{Red.Suspension.Temp}{290}
\DefGwincVal{Red.Suspension.VHCoupling.theta}{0.001}
\DefGwincVal{Red.Suspension.Silicon.Rho}{2330}
\DefGwincVal{Red.Suspension.Silicon.C}{772}
\DefGwincVal{Red.Suspension.Silicon.K}{4980}
\DefGwincVal{Red.Suspension.Silicon.Alpha}{1e-09}
\DefGwincVal{Red.Suspension.Silicon.dlnEdT}{2.5e-10}
\DefGwincVal{Red.Suspension.Silicon.Phi}{2e-09}
\DefGwincVal{Red.Suspension.Silicon.Y}{150000000000}
\DefGwincVal{Red.Suspension.Silicon.Dissdepth}{0.0015}
\DefGwincVal{Red.Suspension.FiberType}{0}
\DefGwincVal{Red.Suspension.Silica.Rho}{2200}
\DefGwincVal{Red.Suspension.Silica.C}{3860}
\DefGwincVal{Red.Suspension.Silica.K}{1.38}
\DefGwincVal{Red.Suspension.Silica.Alpha}{3.9e-07}
\DefGwincVal{Red.Suspension.Silica.dlnEdT}{0.000152}
\DefGwincVal{Red.Suspension.Silica.Phi}{4.1e-10}
\DefGwincVal{Red.Suspension.Silica.Y}{72000000000}
\DefGwincVal{Red.Suspension.Silica.Dissdepth}{0.015}
\DefGwincVal{Red.Suspension.C70Steel.Rho}{7800}
\DefGwincVal{Red.Suspension.C70Steel.C}{486}
\DefGwincVal{Red.Suspension.C70Steel.K}{49}
\DefGwincVal{Red.Suspension.C70Steel.Alpha}{1.2e-05}
\DefGwincVal{Red.Suspension.C70Steel.dlnEdT}{-0.00025}
\DefGwincVal{Red.Suspension.C70Steel.Phi}{0.0002}
\DefGwincVal{Red.Suspension.C70Steel.Y}{212000000000}
\DefGwincVal{Red.Suspension.MaragingSteel.Rho}{7800}
\DefGwincVal{Red.Suspension.MaragingSteel.C}{460}
\DefGwincVal{Red.Suspension.MaragingSteel.K}{20}
\DefGwincVal{Red.Suspension.MaragingSteel.Alpha}{1.1e-05}
\DefGwincVal{Red.Suspension.MaragingSteel.dlnEdT}{0}
\DefGwincVal{Red.Suspension.MaragingSteel.Phi}{0.0001}
\DefGwincVal{Red.Suspension.MaragingSteel.Y}{187000000000}
\DefGwincVal{Red.Suspension.Type}{Quad}
\DefGwincVal{Red.Suspension.Stage(1).Mass}{160}
\DefGwincVal{Red.Suspension.Stage(1).Length}{1.2}
\DefGwincVal{Red.Suspension.Stage(1).Dilution}{NaN}
\DefGwincVal{Red.Suspension.Stage(1).K}{NaN}
\DefGwincVal{Red.Suspension.Stage(1).WireRadius}{NaN}
\DefGwincVal{Red.Suspension.Stage(1).Blade}{NaN}
\DefGwincVal{Red.Suspension.Stage(1).NWires}{4}
\DefGwincVal{Red.Suspension.Stage(2).Mass}{120}
\DefGwincVal{Red.Suspension.Stage(2).Length}{0.341}
\DefGwincVal{Red.Suspension.Stage(2).Dilution}{106}
\DefGwincVal{Red.Suspension.Stage(2).K}{5200}
\DefGwincVal{Red.Suspension.Stage(2).WireRadius}{0.00031}
\DefGwincVal{Red.Suspension.Stage(2).Blade}{0.0042}
\DefGwincVal{Red.Suspension.Stage(2).NWires}{4}
\DefGwincVal{Red.Suspension.Stage(3).Mass}{66}
\DefGwincVal{Red.Suspension.Stage(3).Length}{0.277}
\DefGwincVal{Red.Suspension.Stage(3).Dilution}{80}
\DefGwincVal{Red.Suspension.Stage(3).K}{3900}
\DefGwincVal{Red.Suspension.Stage(3).WireRadius}{0.00035}
\DefGwincVal{Red.Suspension.Stage(3).Blade}{0.0046}
\DefGwincVal{Red.Suspension.Stage(3).NWires}{4}
\DefGwincVal{Red.Suspension.Stage(4).Mass}{44}
\DefGwincVal{Red.Suspension.Stage(4).Length}{0.416}
\DefGwincVal{Red.Suspension.Stage(4).Dilution}{87}
\DefGwincVal{Red.Suspension.Stage(4).K}{3400}
\DefGwincVal{Red.Suspension.Stage(4).WireRadius}{0.00052}
\DefGwincVal{Red.Suspension.Stage(4).Blade}{0.0043}
\DefGwincVal{Red.Suspension.Stage(4).NWires}{2}
\DefGwincVal{Red.Suspension.Ribbon.Thickness}{0.000115}
\DefGwincVal{Red.Suspension.Ribbon.Width}{0.00115}
\DefGwincVal{Red.Suspension.Fiber.Radius}{0.000283}
\DefGwincVal{Red.Materials.Coating.Yhighn}{140000000000}
\DefGwincVal{Red.Materials.Coating.Sigmahighn}{0.23}
\DefGwincVal{Red.Materials.Coating.CVhighn}{2100000}
\DefGwincVal{Red.Materials.Coating.Alphahighn}{3.6e-06}
\DefGwincVal{Red.Materials.Coating.Betahighn}{1.4e-05}
\DefGwincVal{Red.Materials.Coating.ThermalDiffusivityhighn}{33}
\DefGwincVal{Red.Materials.Coating.Phihighn}{0.00023}
\DefGwincVal{Red.Materials.Coating.Indexhighn}{2.0654}
\DefGwincVal{Red.Materials.Coating.Ylown}{72000000000}
\DefGwincVal{Red.Materials.Coating.Sigmalown}{0.17}
\DefGwincVal{Red.Materials.Coating.CVlown}{1641200}
\DefGwincVal{Red.Materials.Coating.Alphalown}{5.1e-07}
\DefGwincVal{Red.Materials.Coating.Betalown}{8e-06}
\DefGwincVal{Red.Materials.Coating.ThermalDiffusivitylown}{1.38}
\DefGwincVal{Red.Materials.Coating.Philown}{4e-05}
\DefGwincVal{Red.Materials.Coating.Indexlown}{1.45}
\DefGwincVal{Red.Materials.Substrate.c2}{7.6e-12}
\DefGwincVal{Red.Materials.Substrate.MechanicalLossExponent}{0.77}
\DefGwincVal{Red.Materials.Substrate.Alphas}{5.2e-12}
\DefGwincVal{Red.Materials.Substrate.MirrorY}{72700000000}
\DefGwincVal{Red.Materials.Substrate.MirrorSigma}{0.167}
\DefGwincVal{Red.Materials.Substrate.MassDensity}{2200}
\DefGwincVal{Red.Materials.Substrate.MassAlpha}{3.9e-07}
\DefGwincVal{Red.Materials.Substrate.MassCM}{739}
\DefGwincVal{Red.Materials.Substrate.MassKappa}{1.38}
\DefGwincVal{Red.Materials.Substrate.RefractiveIndex}{1.45}
\DefGwincVal{Red.Materials.Substrate.Temp}{290}
\DefGwincVal{Red.Materials.MassRadius}{0.275}
\DefGwincVal{Red.Materials.MassThickness}{0.306}
\DefGwincVal{Red.Laser.Wavelength}{1.064e-06}
\DefGwincVal{Red.Laser.Power}{125}
\DefGwincVal{Red.Optics.Type}{SignalRecycled}
\DefGwincVal{Red.Optics.SRM.CavityLength}{55}
\DefGwincVal{Red.Optics.SRM.Transmittance}{0.2}
\DefGwincVal{Red.Optics.SRM.Tunephase}{0}
\DefGwincVal{Red.Optics.PhotoDetectorEfficiency}{0.95}
\DefGwincVal{Red.Optics.Loss}{3.75e-05}
\DefGwincVal{Red.Optics.BSLoss}{0.0005}
\DefGwincVal{Red.Optics.coupling}{1}
\DefGwincVal{Red.Optics.Curvature.ITM}{1849}
\DefGwincVal{Red.Optics.Curvature.ETM}{2173}
\DefGwincVal{Red.Optics.SubstrateAbsorption}{5e-05}
\DefGwincVal{Red.Optics.pcrit}{10}
\DefGwincVal{Red.Optics.ITM.BeamRadius}{0.1692}
\DefGwincVal{Red.Optics.ITM.CoatingAbsorption}{5e-07}
\DefGwincVal{Red.Optics.ITM.Transmittance}{0.014}
\DefGwincVal{Red.Optics.ITM.CoatingThicknessLown}{0.308}
\DefGwincVal{Red.Optics.ITM.CoatingThicknessCap}{0.5}
\DefGwincVal{Red.Optics.ETM.BeamRadius}{0.199}
\DefGwincVal{Red.Optics.ETM.Transmittance}{5e-06}
\DefGwincVal{Red.Optics.ETM.CoatingThicknessLown}{0.27}
\DefGwincVal{Red.Optics.ETM.CoatingThicknessCap}{0.5}
\DefGwincVal{Red.Optics.PRM.Transmittance}{0.03}
\DefGwincVal{Red.Optics.Quadrature.dc}{1.5708}
\DefGwincVal{Red.Squeezer.Type}{Freq Dependent}
\DefGwincVal{Red.Squeezer.AmplitudedB}{10}
\DefGwincVal{Red.Squeezer.InjectionLoss}{0.05}
\DefGwincVal{Red.Squeezer.SQZAngle}{0}
\DefGwincVal{Red.Squeezer.FilterCavity.fdetune}{-17.4}
\DefGwincVal{Red.Squeezer.FilterCavity.L}{300}
\DefGwincVal{Red.Squeezer.FilterCavity.Ti}{0.000445}
\DefGwincVal{Red.Squeezer.FilterCavity.Te}{0}
\DefGwincVal{Red.Squeezer.FilterCavity.Lrt}{4e-05}
\DefGwincVal{Red.Squeezer.FilterCavity.Rot}{0}
\DefGwincVal{Red.OutputFilter.Type}{None}
\DefGwincVal{Red.OutputFilter.FilterCavity.fdetune}{-30}
\DefGwincVal{Red.OutputFilter.FilterCavity.L}{4000}
\DefGwincVal{Red.OutputFilter.FilterCavity.Ti}{0.01}
\DefGwincVal{Red.OutputFilter.FilterCavity.Te}{0}
\DefGwincVal{Red.OutputFilter.FilterCavity.Lrt}{0.0001}
\DefGwincVal{Red.OutputFilter.FilterCavity.Rot}{0}
\DefGwincVal{Red.Laser.Wavelength_nm}{1064}
\DefGwincVal{Red.Materials.MassRadius_cm}{27.5}
\DefGwincVal{Red.Materials.MassThickness_cm}{30.6}
\DefGwincVal{Red.Optics.ITM.BeamRadius_cm}{16.9}
\DefGwincVal{Red.Optics.ETM.BeamRadius_cm}{19.9}
\DefGwincVal{Red.Substance}{fused silica}
\DefGwincVal{Red.Suspension.FiberType_str}{fiber}
\DefGwincVal{Red.Squeezer.FilterCavity.Lrt_ppm}{40}

\providecommand\DefGwincVal[2]{%
  \expandafter\newcommand\csname gwincval-#1\endcsname{#2}%
}
\providecommand{\GwincVal}[1]{\csname gwincval-#1\endcsname}
\DefGwincVal{Blue.Infrastructure.Length}{3995}
\DefGwincVal{Blue.Infrastructure.ResidualGas.pressure}{4e-07}
\DefGwincVal{Blue.Infrastructure.ResidualGas.mass}{3.35e-27}
\DefGwincVal{Blue.Infrastructure.ResidualGas.polarizability}{8.1e-31}
\DefGwincVal{Blue.Constants.E0}{8.8542e-12}
\DefGwincVal{Blue.Constants.hbar}{1.0546e-34}
\DefGwincVal{Blue.Constants.kB}{1.3807e-23}
\DefGwincVal{Blue.Constants.h}{6.6261e-34}
\DefGwincVal{Blue.Constants.R}{8.3145}
\DefGwincVal{Blue.Constants.m_e}{9.1094e-31}
\DefGwincVal{Blue.Constants.c}{299792458}
\DefGwincVal{Blue.Constants.Temp}{295}
\DefGwincVal{Blue.Constants.yr}{31556926.08}
\DefGwincVal{Blue.Constants.M_earth}{5.972e+24}
\DefGwincVal{Blue.Constants.R_earth}{6378100}
\DefGwincVal{Blue.Constants.fs}{16384}
\DefGwincVal{Blue.Constants.AU}{149597870700}
\DefGwincVal{Blue.Constants.parsec}{3.085677581491367e+16}
\DefGwincVal{Blue.Constants.Mpc}{3.085677581491367e+22}
\DefGwincVal{Blue.Constants.SolarMassParameter}{1.3271244e+20}
\DefGwincVal{Blue.Constants.G}{6.6741e-11}
\DefGwincVal{Blue.Constants.MSol}{1.988475415338144e+30}
\DefGwincVal{Blue.Constants.g}{9.806}
\DefGwincVal{Blue.Constants.H0}{67110}
\DefGwincVal{Blue.Constants.omegaM}{0.3175}
\DefGwincVal{Blue.Constants.omegaLambda}{0.6825}
\DefGwincVal{Blue.Constants.fInspiralMin}{3}
\DefGwincVal{Blue.Constants.BesselZeros}{[array]}
\DefGwincVal{Blue.TCS.s_cc}{7.024}
\DefGwincVal{Blue.TCS.s_cs}{7.321}
\DefGwincVal{Blue.TCS.s_ss}{7.631}
\DefGwincVal{Blue.TCS.SRCloss}{0}
\DefGwincVal{Blue.Seismic.Site}{LHO}
\DefGwincVal{Blue.Seismic.KneeFrequency}{10}
\DefGwincVal{Blue.Seismic.LowFrequencyLevel}{1e-09}
\DefGwincVal{Blue.Seismic.Gamma}{0.8}
\DefGwincVal{Blue.Seismic.Rho}{1800}
\DefGwincVal{Blue.Seismic.Beta}{0.8}
\DefGwincVal{Blue.Seismic.Omicron}{10}
\DefGwincVal{Blue.Seismic.TestMassHeight}{1.5}
\DefGwincVal{Blue.Seismic.RayleighWaveSpeed}{250}
\DefGwincVal{Blue.Seismic.darmSeiSusFile}{seismic.mat}
\DefGwincVal{Blue.Seismic.darmseis_f}{[array]}
\DefGwincVal{Blue.Seismic.darmseis_x}{[array]}
\DefGwincVal{Blue.Suspension.BreakStress}{750000000}
\DefGwincVal{Blue.Suspension.Temp}{[array]}
\DefGwincVal{Blue.Suspension.VHCoupling.theta}{0.001}
\DefGwincVal{Blue.Suspension.Silicon.Rho}{2329}
\DefGwincVal{Blue.Suspension.Silicon.C}{300}
\DefGwincVal{Blue.Suspension.Silicon.K}{700}
\DefGwincVal{Blue.Suspension.Silicon.Alpha}{1e-10}
\DefGwincVal{Blue.Suspension.Silicon.dlnEdT}{-2e-05}
\DefGwincVal{Blue.Suspension.Silicon.Phi}{2e-09}
\DefGwincVal{Blue.Suspension.Silicon.Y}{155800000000}
\DefGwincVal{Blue.Suspension.Silicon.Dissdepth}{0.00025}
\DefGwincVal{Blue.Suspension.Silicon123K.Rho}{2329}
\DefGwincVal{Blue.Suspension.Silicon123K.C}{300}
\DefGwincVal{Blue.Suspension.Silicon123K.K}{700}
\DefGwincVal{Blue.Suspension.Silicon123K.Alpha}{1e-10}
\DefGwincVal{Blue.Suspension.Silicon123K.dlnEdT}{-2e-05}
\DefGwincVal{Blue.Suspension.Silicon123K.Phi}{2e-09}
\DefGwincVal{Blue.Suspension.Silicon123K.Y}{155800000000}
\DefGwincVal{Blue.Suspension.Silicon123K.Dissdepth}{0.00025}
\DefGwincVal{Blue.Suspension.Silicon120K.Rho}{2329}
\DefGwincVal{Blue.Suspension.Silicon120K.C}{300}
\DefGwincVal{Blue.Suspension.Silicon120K.K}{700}
\DefGwincVal{Blue.Suspension.Silicon120K.Alpha}{1e-10}
\DefGwincVal{Blue.Suspension.Silicon120K.dlnEdT}{-2e-05}
\DefGwincVal{Blue.Suspension.Silicon120K.Phi}{2e-09}
\DefGwincVal{Blue.Suspension.Silicon120K.Y}{155800000000}
\DefGwincVal{Blue.Suspension.Silicon120K.Dissdepth}{0.00025}
\DefGwincVal{Blue.Suspension.Silicon295K.Rho}{2329}
\DefGwincVal{Blue.Suspension.Silicon295K.C}{400}
\DefGwincVal{Blue.Suspension.Silicon295K.K}{148}
\DefGwincVal{Blue.Suspension.Silicon295K.Alpha}{2.6e-06}
\DefGwincVal{Blue.Suspension.Silicon295K.dlnEdT}{-7e-05}
\DefGwincVal{Blue.Suspension.Silicon295K.Y}{166000000000}
\DefGwincVal{Blue.Suspension.Silicon295K.Phi}{2e-08}
\DefGwincVal{Blue.Suspension.Silicon295K.Dissdepth}{0.00025}
\DefGwincVal{Blue.Suspension.Silicon300K.Rho}{2329}
\DefGwincVal{Blue.Suspension.Silicon300K.C}{400}
\DefGwincVal{Blue.Suspension.Silicon300K.K}{148}
\DefGwincVal{Blue.Suspension.Silicon300K.Alpha}{2.6e-06}
\DefGwincVal{Blue.Suspension.Silicon300K.dlnEdT}{-7e-05}
\DefGwincVal{Blue.Suspension.Silicon300K.Y}{166000000000}
\DefGwincVal{Blue.Suspension.Silicon300K.Phi}{2e-08}
\DefGwincVal{Blue.Suspension.Silicon300K.Dissdepth}{0.00025}
\DefGwincVal{Blue.Suspension.FiberType}{1}
\DefGwincVal{Blue.Suspension.Silica.Rho}{2200}
\DefGwincVal{Blue.Suspension.Silica.C}{772}
\DefGwincVal{Blue.Suspension.Silica.K}{1.38}
\DefGwincVal{Blue.Suspension.Silica.Alpha}{3.9e-07}
\DefGwincVal{Blue.Suspension.Silica.dlnEdT}{0.000152}
\DefGwincVal{Blue.Suspension.Silica.Phi}{4.1e-10}
\DefGwincVal{Blue.Suspension.Silica.Y}{72000000000}
\DefGwincVal{Blue.Suspension.Silica.Dissdepth}{0.015}
\DefGwincVal{Blue.Suspension.Silica300K.Rho}{2200}
\DefGwincVal{Blue.Suspension.Silica300K.C}{772}
\DefGwincVal{Blue.Suspension.Silica300K.K}{1.38}
\DefGwincVal{Blue.Suspension.Silica300K.Alpha}{3.9e-07}
\DefGwincVal{Blue.Suspension.Silica300K.dlnEdT}{0.000152}
\DefGwincVal{Blue.Suspension.Silica300K.Phi}{4.1e-10}
\DefGwincVal{Blue.Suspension.Silica300K.Y}{72000000000}
\DefGwincVal{Blue.Suspension.Silica300K.Dissdepth}{0.015}
\DefGwincVal{Blue.Suspension.Silica295K.Rho}{2200}
\DefGwincVal{Blue.Suspension.Silica295K.C}{772}
\DefGwincVal{Blue.Suspension.Silica295K.K}{1.38}
\DefGwincVal{Blue.Suspension.Silica295K.Alpha}{3.9e-07}
\DefGwincVal{Blue.Suspension.Silica295K.dlnEdT}{0.000152}
\DefGwincVal{Blue.Suspension.Silica295K.Phi}{4.1e-10}
\DefGwincVal{Blue.Suspension.Silica295K.Y}{72000000000}
\DefGwincVal{Blue.Suspension.Silica295K.Dissdepth}{0.015}
\DefGwincVal{Blue.Suspension.Silica123K.Rho}{2200}
\DefGwincVal{Blue.Suspension.Silica123K.C}{300}
\DefGwincVal{Blue.Suspension.Silica123K.K}{0.75}
\DefGwincVal{Blue.Suspension.Silica123K.Alpha}{-5e-07}
\DefGwincVal{Blue.Suspension.Silica123K.dlnEdT}{0.000152}
\DefGwincVal{Blue.Suspension.Silica123K.Y}{70000000000}
\DefGwincVal{Blue.Suspension.Silica123K.Phi}{2e-05}
\DefGwincVal{Blue.Suspension.Silica123K.Dissdepth}{0.015}
\DefGwincVal{Blue.Suspension.C70Steel.Rho}{7800}
\DefGwincVal{Blue.Suspension.C70Steel.C}{486}
\DefGwincVal{Blue.Suspension.C70Steel.K}{49}
\DefGwincVal{Blue.Suspension.C70Steel.Alpha}{1.2e-05}
\DefGwincVal{Blue.Suspension.C70Steel.dlnEdT}{-0.00025}
\DefGwincVal{Blue.Suspension.C70Steel.Phi}{0.0002}
\DefGwincVal{Blue.Suspension.C70Steel.Y}{212000000000}
\DefGwincVal{Blue.Suspension.C70Steel295K.Rho}{7800}
\DefGwincVal{Blue.Suspension.C70Steel295K.C}{486}
\DefGwincVal{Blue.Suspension.C70Steel295K.K}{49}
\DefGwincVal{Blue.Suspension.C70Steel295K.Alpha}{1.2e-05}
\DefGwincVal{Blue.Suspension.C70Steel295K.dlnEdT}{-0.00025}
\DefGwincVal{Blue.Suspension.C70Steel295K.Phi}{0.0002}
\DefGwincVal{Blue.Suspension.C70Steel295K.Y}{212000000000}
\DefGwincVal{Blue.Suspension.C70Steel300K.Rho}{7800}
\DefGwincVal{Blue.Suspension.C70Steel300K.C}{486}
\DefGwincVal{Blue.Suspension.C70Steel300K.K}{49}
\DefGwincVal{Blue.Suspension.C70Steel300K.Alpha}{1.2e-05}
\DefGwincVal{Blue.Suspension.C70Steel300K.dlnEdT}{-0.00025}
\DefGwincVal{Blue.Suspension.C70Steel300K.Phi}{0.0002}
\DefGwincVal{Blue.Suspension.C70Steel300K.Y}{212000000000}
\DefGwincVal{Blue.Suspension.C70Steel123K.Rho}{7800}
\DefGwincVal{Blue.Suspension.C70Steel123K.C}{250}
\DefGwincVal{Blue.Suspension.C70Steel123K.K}{15}
\DefGwincVal{Blue.Suspension.C70Steel123K.Alpha}{8e-06}
\DefGwincVal{Blue.Suspension.C70Steel123K.dlnEdT}{-0.00025}
\DefGwincVal{Blue.Suspension.C70Steel123K.Phi}{0.0002}
\DefGwincVal{Blue.Suspension.C70Steel123K.Y}{212000000000}
\DefGwincVal{Blue.Suspension.MaragingSteel.Rho}{7800}
\DefGwincVal{Blue.Suspension.MaragingSteel.C}{460}
\DefGwincVal{Blue.Suspension.MaragingSteel.K}{20}
\DefGwincVal{Blue.Suspension.MaragingSteel.Alpha}{1.1e-05}
\DefGwincVal{Blue.Suspension.MaragingSteel.dlnEdT}{0}
\DefGwincVal{Blue.Suspension.MaragingSteel.Phi}{0.0001}
\DefGwincVal{Blue.Suspension.MaragingSteel.Y}{187000000000}
\DefGwincVal{Blue.Suspension.MaragingSteel295K.Rho}{7800}
\DefGwincVal{Blue.Suspension.MaragingSteel295K.C}{460}
\DefGwincVal{Blue.Suspension.MaragingSteel295K.K}{20}
\DefGwincVal{Blue.Suspension.MaragingSteel295K.Alpha}{1.1e-05}
\DefGwincVal{Blue.Suspension.MaragingSteel295K.dlnEdT}{0}
\DefGwincVal{Blue.Suspension.MaragingSteel295K.Phi}{0.0001}
\DefGwincVal{Blue.Suspension.MaragingSteel295K.Y}{187000000000}
\DefGwincVal{Blue.Suspension.MaragingSteel300K.Rho}{7800}
\DefGwincVal{Blue.Suspension.MaragingSteel300K.C}{460}
\DefGwincVal{Blue.Suspension.MaragingSteel300K.K}{20}
\DefGwincVal{Blue.Suspension.MaragingSteel300K.Alpha}{1.1e-05}
\DefGwincVal{Blue.Suspension.MaragingSteel300K.dlnEdT}{0}
\DefGwincVal{Blue.Suspension.MaragingSteel300K.Phi}{0.0001}
\DefGwincVal{Blue.Suspension.MaragingSteel300K.Y}{187000000000}
\DefGwincVal{Blue.Suspension.Type}{BQuad}
\DefGwincVal{Blue.Suspension.Stage(1).Length}{0.69685}
\DefGwincVal{Blue.Suspension.Stage(1).Mass}{200}
\DefGwincVal{Blue.Suspension.Stage(1).CummulativeMass}{200}
\DefGwincVal{Blue.Suspension.Stage(1).Dilution}{NaN}
\DefGwincVal{Blue.Suspension.Stage(1).K}{45600}
\DefGwincVal{Blue.Suspension.Stage(1).NWires}{4}
\DefGwincVal{Blue.Suspension.Stage(1).WireRadius}{NaN}
\DefGwincVal{Blue.Suspension.Stage(1).Blade}{0.0043}
\DefGwincVal{Blue.Suspension.Stage(1).WireMaterial}{Silicon}
\DefGwincVal{Blue.Suspension.Stage(1).BladeMaterial}{Silicon}
\DefGwincVal{Blue.Suspension.Stage(2).Length}{0.64515}
\DefGwincVal{Blue.Suspension.Stage(2).Mass}{200}
\DefGwincVal{Blue.Suspension.Stage(2).CummulativeMass}{400}
\DefGwincVal{Blue.Suspension.Stage(2).Dilution}{NaN}
\DefGwincVal{Blue.Suspension.Stage(2).K}{26262.6263}
\DefGwincVal{Blue.Suspension.Stage(2).NWires}{4}
\DefGwincVal{Blue.Suspension.Stage(2).WireRadius}{0.00066776}
\DefGwincVal{Blue.Suspension.Stage(2).Blade}{0.007206}
\DefGwincVal{Blue.Suspension.Stage(2).WireMaterial}{C70Steel}
\DefGwincVal{Blue.Suspension.Stage(2).BladeMaterial}{MaragingSteel}
\DefGwincVal{Blue.Suspension.Stage(3).Length}{0.15}
\DefGwincVal{Blue.Suspension.Stage(3).Mass}{60.8466}
\DefGwincVal{Blue.Suspension.Stage(3).CummulativeMass}{460.8466}
\DefGwincVal{Blue.Suspension.Stage(3).Dilution}{NaN}
\DefGwincVal{Blue.Suspension.Stage(3).K}{17795.0681}
\DefGwincVal{Blue.Suspension.Stage(3).NWires}{4}
\DefGwincVal{Blue.Suspension.Stage(3).WireRadius}{0.00071675}
\DefGwincVal{Blue.Suspension.Stage(3).Blade}{0.0076296}
\DefGwincVal{Blue.Suspension.Stage(3).WireMaterial}{C70Steel}
\DefGwincVal{Blue.Suspension.Stage(3).BladeMaterial}{MaragingSteel}
\DefGwincVal{Blue.Suspension.Stage(4).Length}{0.15}
\DefGwincVal{Blue.Suspension.Stage(4).Mass}{59.1534}
\DefGwincVal{Blue.Suspension.Stage(4).CummulativeMass}{520}
\DefGwincVal{Blue.Suspension.Stage(4).Dilution}{NaN}
\DefGwincVal{Blue.Suspension.Stage(4).K}{114362.3071}
\DefGwincVal{Blue.Suspension.Stage(4).NWires}{2}
\DefGwincVal{Blue.Suspension.Stage(4).WireRadius}{0.0010767}
\DefGwincVal{Blue.Suspension.Stage(4).Blade}{0.01388}
\DefGwincVal{Blue.Suspension.Stage(4).WireMaterial}{C70Steel}
\DefGwincVal{Blue.Suspension.Stage(4).BladeMaterial}{MaragingSteel}
\DefGwincVal{Blue.Suspension.Ribbon.Thickness}{0.00022}
\DefGwincVal{Blue.Suspension.Ribbon.Width}{0.0022}
\DefGwincVal{Blue.Suspension.hForce}{[array]}
\DefGwincVal{Blue.Suspension.vForce}{[array]}
\DefGwincVal{Blue.Suspension.hForce_singlylossy}{[array]}
\DefGwincVal{Blue.Suspension.vForce_singlylossy}{[array]}
\DefGwincVal{Blue.Suspension.hTable}{[array]}
\DefGwincVal{Blue.Suspension.vTable}{[array]}
\DefGwincVal{Blue.Materials.Coating.Yhighn}{60000000000}
\DefGwincVal{Blue.Materials.Coating.Sigmahighn}{0.22}
\DefGwincVal{Blue.Materials.Coating.CVhighn}{1050000}
\DefGwincVal{Blue.Materials.Coating.Alphahighn}{1e-09}
\DefGwincVal{Blue.Materials.Coating.Betahighn}{0.00014}
\DefGwincVal{Blue.Materials.Coating.ThermalDiffusivityhighn}{1.03}
\DefGwincVal{Blue.Materials.Coating.Phihighn}{1e-05}
\DefGwincVal{Blue.Materials.Coating.Indexhighn}{3.5}
\DefGwincVal{Blue.Materials.Coating.Ylown}{72000000000}
\DefGwincVal{Blue.Materials.Coating.Sigmalown}{0.17}
\DefGwincVal{Blue.Materials.Coating.CVlown}{1641200}
\DefGwincVal{Blue.Materials.Coating.Alphalown}{5.1e-07}
\DefGwincVal{Blue.Materials.Coating.Betalown}{8e-06}
\DefGwincVal{Blue.Materials.Coating.ThermalDiffusivitylown}{1.05}
\DefGwincVal{Blue.Materials.Coating.Philown}{0.0001}
\DefGwincVal{Blue.Materials.Coating.Indexlown}{1.436}
\DefGwincVal{Blue.Materials.Substrate.c2}{3e-13}
\DefGwincVal{Blue.Materials.Substrate.MechanicalLossExponent}{1}
\DefGwincVal{Blue.Materials.Substrate.Alphas}{5.2e-12}
\DefGwincVal{Blue.Materials.Substrate.MirrorY}{155800000000}
\DefGwincVal{Blue.Materials.Substrate.MirrorSigma}{0.27}
\DefGwincVal{Blue.Materials.Substrate.MassDensity}{2329}
\DefGwincVal{Blue.Materials.Substrate.MassAlpha}{1e-09}
\DefGwincVal{Blue.Materials.Substrate.MassCM}{300}
\DefGwincVal{Blue.Materials.Substrate.MassKappa}{700}
\DefGwincVal{Blue.Materials.Substrate.RefractiveIndex}{3.5}
\DefGwincVal{Blue.Materials.Substrate.dndT}{0.0001}
\DefGwincVal{Blue.Materials.Substrate.Temp}{123}
\DefGwincVal{Blue.Materials.Substrate.isSemiConductor}{1}
\DefGwincVal{Blue.Materials.Substrate.CarrierDensity}{1e+19}
\DefGwincVal{Blue.Materials.Substrate.ElectronDiffusion}{0.0097}
\DefGwincVal{Blue.Materials.Substrate.HoleDiffusion}{0.0035}
\DefGwincVal{Blue.Materials.Substrate.ElectronEffMass}{9.747e-31}
\DefGwincVal{Blue.Materials.Substrate.HoleEffMass}{8.0163e-31}
\DefGwincVal{Blue.Materials.Substrate.ElectronIndexGamma}{-8.8e-28}
\DefGwincVal{Blue.Materials.Substrate.HoleIndexGamma}{-1.02e-27}
\DefGwincVal{Blue.Materials.MassRadius}{0.225}
\DefGwincVal{Blue.Materials.MassThickness}{0.55}
\DefGwincVal{Blue.Materials.MirrorVolume}{0.087474}
\DefGwincVal{Blue.Materials.MirrorMass}{203.7263}
\DefGwincVal{Blue.Laser.Wavelength}{2e-06}
\DefGwincVal{Blue.Laser.Power}{151.5919}
\DefGwincVal{Blue.Laser.ArmPower}{3050535.1816}
\DefGwincVal{Blue.Optics.Type}{SignalRecycled}
\DefGwincVal{Blue.Optics.SRM.CavityLength}{55}
\DefGwincVal{Blue.Optics.SRM.Transmittance}{0.045773}
\DefGwincVal{Blue.Optics.SRM.Tunephase}{0}
\DefGwincVal{Blue.Optics.PhotoDetectorEfficiency}{0.95}
\DefGwincVal{Blue.Optics.Loss}{1e-05}
\DefGwincVal{Blue.Optics.BSLoss}{0.0005}
\DefGwincVal{Blue.Optics.coupling}{1}
\DefGwincVal{Blue.Optics.Curvature.ITM}{1800}
\DefGwincVal{Blue.Optics.Curvature.ETM}{2500}
\DefGwincVal{Blue.Optics.SubstrateAbsorption}{3e-07}
\DefGwincVal{Blue.Optics.ITM.SubstrateAbsorption}{0.001}
\DefGwincVal{Blue.Optics.ITM.BeamRadius}{0.058504}
\DefGwincVal{Blue.Optics.ITM.CoatingAbsorption}{1e-06}
\DefGwincVal{Blue.Optics.ITM.Transmittance}{0.002}
\DefGwincVal{Blue.Optics.ITM.CoatingThicknessLown}{0.308}
\DefGwincVal{Blue.Optics.ITM.CoatingThicknessCap}{0.5}
\DefGwincVal{Blue.Optics.ITM.CoatLayerOpticalThickness}{[array]}
\DefGwincVal{Blue.Optics.ITM.Thickness}{0.55}
\DefGwincVal{Blue.Optics.ITM.Pabs}{[array]}
\DefGwincVal{Blue.Optics.pcrit}{10}
\DefGwincVal{Blue.Optics.ETM.BeamRadius}{0.083543}
\DefGwincVal{Blue.Optics.ETM.Transmittance}{5e-06}
\DefGwincVal{Blue.Optics.ETM.CoatingThicknessLown}{0.27}
\DefGwincVal{Blue.Optics.ETM.CoatingThicknessCap}{0.5}
\DefGwincVal{Blue.Optics.ETM.CoatLayerOpticalThickness}{[array]}
\DefGwincVal{Blue.Optics.PRM.Transmittance}{0.049202}
\DefGwincVal{Blue.Optics.Quadrature.dc}{1.5832}
\DefGwincVal{Blue.Squeezer.Type}{Freq Dependent}
\DefGwincVal{Blue.Squeezer.AmplitudedB}{10}
\DefGwincVal{Blue.Squeezer.InjectionLoss}{0.05}
\DefGwincVal{Blue.Squeezer.SQZAngle}{0}
\DefGwincVal{Blue.Squeezer.FilterCavity.fdetune}{-25}
\DefGwincVal{Blue.Squeezer.FilterCavity.L}{300}
\DefGwincVal{Blue.Squeezer.FilterCavity.Ti}{0.0006149}
\DefGwincVal{Blue.Squeezer.FilterCavity.Te}{0}
\DefGwincVal{Blue.Squeezer.FilterCavity.Lrt}{1e-05}
\DefGwincVal{Blue.Squeezer.FilterCavity.Rot}{0}
\DefGwincVal{Blue.OutputFilter.Type}{None}
\DefGwincVal{Blue.OutputFilter.FilterCavity.fdetune}{-30}
\DefGwincVal{Blue.OutputFilter.FilterCavity.L}{4000}
\DefGwincVal{Blue.OutputFilter.FilterCavity.Ti}{0.01}
\DefGwincVal{Blue.OutputFilter.FilterCavity.Te}{0}
\DefGwincVal{Blue.OutputFilter.FilterCavity.Lrt}{0.0001}
\DefGwincVal{Blue.OutputFilter.FilterCavity.Rot}{0}
\DefGwincVal{Blue.Model}{IFOModel_123_2000}
\DefGwincVal{Blue.modeSR}{0}
\DefGwincVal{Blue.gwinc.PRfixed}{0}
\DefGwincVal{Blue.gwinc.pbs}{3081.0405}
\DefGwincVal{Blue.gwinc.parm}{3002539.9267}
\DefGwincVal{Blue.gwinc.finesse}{3110.4878}
\DefGwincVal{Blue.gwinc.prfactor}{20.3246}
\DefGwincVal{Blue.gwinc.fSQL}{38.2924}
\DefGwincVal{Blue.gwinc.fGammaIFO}{509.8349}
\DefGwincVal{Blue.gwinc.fGammaArm}{5.9716}
\DefGwincVal{Blue.nse.ResGas}{[array]}
\DefGwincVal{Blue.nse.SuspThermal}{[array]}
\DefGwincVal{Blue.nse.Quantum}{[array]}
\DefGwincVal{Blue.nse.Freq}{[array]}
\DefGwincVal{Blue.nse.Newtonian}{[array]}
\DefGwincVal{Blue.nse.Seismic}{[array]}
\DefGwincVal{Blue.nse.Total}{[array]}
\DefGwincVal{Blue.nse.MirrorThermal.Total}{[array]}
\DefGwincVal{Blue.nse.MirrorThermal.SubBrown}{[array]}
\DefGwincVal{Blue.nse.MirrorThermal.CoatBrown}{[array]}
\DefGwincVal{Blue.nse.MirrorThermal.SubTE}{[array]}
\DefGwincVal{Blue.nse.MirrorThermal.CoatTO}{[array]}
\DefGwincVal{Blue.nse.MirrorThermal.SubTR}{[array]}
\DefGwincVal{Blue.nse.MirrorThermal.SubCD}{[array]}
\DefGwincVal{Blue.Laser.Wavelength_nm}{2000}
\DefGwincVal{Blue.Laser.ArmPower_Watts}{3050535}
\DefGwincVal{Blue.Laser.BSPower_Watts}{3081}
\DefGwincVal{Blue.Laser.IMCOutput}{152}
\DefGwincVal{Blue.Laser.IMCInput}{202}
\DefGwincVal{Blue.Laser.PMCOutput}{212}
\DefGwincVal{Blue.Laser.PMCInput}{222}
\DefGwincVal{Blue.Laser.ITMIntensity}{89127}
\DefGwincVal{Blue.Laser.BSIntensity}{90}
\DefGwincVal{Blue.Materials.MassRadius_cm}{22.5}
\DefGwincVal{Blue.Materials.MassThickness_cm}{55}
\DefGwincVal{Blue.Optics.ITM.BeamRadius_cm}{5.9}
\DefGwincVal{Blue.Optics.ETM.BeamRadius_cm}{8.4}
\DefGwincVal{Blue.Substance}{silicon}
\DefGwincVal{Blue.Suspension.FiberType_str}{ribbon}
\DefGwincVal{Blue.Squeezer.FilterCavity.Lrt_ppm}{10}

\begin{table}[h]
\resizebox{\textwidth}{!}{%
  \centering
  \begin{tabular}{lccc}
    \toprule

    Parameter & Advanced LIGO & Red & Blue \\

    \midrule

    Laser wavelength & \GwincVal{aLIGO.Laser.Wavelength_nm}\,nm &
    \GwincVal{Red.Laser.Wavelength_nm}\,nm &
    \GwincVal{Blue.Laser.Wavelength_nm}\,nm \\

    Laser power & \GwincVal{aLIGO.Laser.Power}\,W &
    \GwincVal{Red.Laser.Power}\,W & \GwincVal{Blue.Laser.Power}\,W \\

    Mirror substrate & \GwincVal{aLIGO.Substance} &
    \GwincVal{Red.Substance} & \GwincVal{Blue.Substance} \\

    Mirror radius & \GwincVal{aLIGO.Materials.MassRadius_cm}\,cm &
    \GwincVal{Red.Materials.MassRadius_cm}\,cm &
    \GwincVal{Blue.Materials.MassRadius_cm}\,cm \\

    Mirror thickness & \GwincVal{aLIGO.Materials.MassThickness_cm}\,cm
    & \GwincVal{Red.Materials.MassThickness_cm}\,cm &
    \GwincVal{Blue.Materials.MassThickness_cm}\,cm \\

    Beam radius on ITM/ETM &
    \GwincVal{aLIGO.Optics.ITM.BeamRadius_cm}/\GwincVal{aLIGO.Optics.ETM.BeamRadius_cm}\,cm
    &
    \GwincVal{Red.Optics.ITM.BeamRadius_cm}/\GwincVal{Red.Optics.ETM.BeamRadius_cm}\,cm~\tablefootnote{The
      Red design projects a factor of 3.2 improvement in all coating
      and substrate noises.  A factor of 1.6 is to be achieved through
      increasing the beam size, and a further factor of 2 through
      coating technology improvements.  For the
      noise budget, this has been modeled as an increase in the beam
      size by the full factor of 3.2.} &
    \GwincVal{Blue.Optics.ITM.BeamRadius_cm}/\GwincVal{Blue.Optics.ETM.BeamRadius_cm}\,cm
    \\

    Mass per stage &
    \GwincVal{aLIGO.Suspension.Stage(4).Mass}/\GwincVal{aLIGO.Suspension.Stage(3).Mass}/\GwincVal{aLIGO.Suspension.Stage(2).Mass}/\GwincVal{aLIGO.Suspension.Stage(1).Mass}\,kg
    &
    \GwincVal{Red.Suspension.Stage(4).Mass}/\GwincVal{Red.Suspension.Stage(3).Mass}/\GwincVal{Red.Suspension.Stage(2).Mass}/\GwincVal{Red.Suspension.Stage(1).Mass}\,kg
    &
    \GwincVal{Blue.Suspension.Stage(4).Mass}/\GwincVal{Blue.Suspension.Stage(3).Mass}/\GwincVal{Blue.Suspension.Stage(2).Mass}/\GwincVal{Blue.Suspension.Stage(1).Mass}\,kg
    \\

    Final stage temperature & \GwincVal{aLIGO.Suspension.Temp}\,K &
    \GwincVal{Red.Suspension.Temp}\,K &
    \GwincVal{Blue.Materials.Substrate.Temp}\,K \\

    Final stage construction & \GwincVal{aLIGO.Substance}
    \GwincVal{aLIGO.Suspension.FiberType_str} &
    \GwincVal{Red.Substance}
    \GwincVal{Red.Suspension.FiberType_str}~\tablefootnote{The noise
      of the optimized suspension fiber in the Red design was roughly
      fitted in GWINC, using the specific heat of fused silica as the
      fit parameter.}  & \GwincVal{Blue.Substance}
    \GwincVal{Blue.Suspension.FiberType_str} \\

    Final stage length &
    \GwincVal{aLIGO.Suspension.Stage(1).Length}\,m &
    \GwincVal{Red.Suspension.Stage(1).Length}\,m &
    \GwincVal{Blue.Suspension.Stage(1).Length}\,m \\

    Newtonian noise suppression & N/A & \GwincVal{Red.Seismic.Omicron}
    & \GwincVal{Blue.Seismic.Omicron} \\

    Squeeze factor & N/A &
    $\GwincVal{Red.Squeezer.AmplitudedB}$\,dB~\tablefootnote{Squeezer
      parameters have been altered to provide an equal ground for
      comparison between the two designs.  T1200005 originally assumed
      20~dB of squeezing.}  &
    $\GwincVal{Blue.Squeezer.AmplitudedB}$\,dB \\

    Squeeze injection loss & N/A &
    \GwincVal{Red.Squeezer.InjectionLoss} &
    \GwincVal{Blue.Squeezer.InjectionLoss} \\

    Squeeze filter cavity length & N/A &
    \GwincVal{Red.Squeezer.FilterCavity.L}\,m &
    \GwincVal{Blue.Squeezer.FilterCavity.L}\,m \\

    Squeeze filter cavity loss & N/A &
    \GwincVal{Red.Squeezer.FilterCavity.Lrt_ppm}\,ppm-rt &
    \GwincVal{Blue.Squeezer.FilterCavity.Lrt_ppm}\,ppm-rt \\

    \bottomrule
  \end{tabular}}
  \captionof{table}{Parameters varied by the Red and Blue designs,
    relative to Advanced LIGO.}
  \label{tab:compare}
\end{table}

\end{appendices}

\bibliographystyle{iopart-num}
\bibliography{bibliography/gw_references,bibliography/GWreferences,bibliography/kalmus_references,bibliography/vladimir,bibliography/cbc_references,bibliography/gw_detector_references,bibliography/TestGR,bibliography/gw_data_analysis_references,bibliography/sn_theory_references,bibliography/bh_formation_references,bibliography/stellarevolution_references,bibliography/grb_references.bib,bibliography/nu_obs_references.bib,bibliography/sn_rates_references,bibliography/pns_cooling_references,bibliography/accretion_disks,bibliography/ilya_mandel,bibliography/vivien,bibliography/postmerger_gws.bib,bibliography/NSNS_NSBH_references,bibliography/sarah}

\providecommand{\newblock}{}
\begin{thebibliography}{100}
\expandafter\ifx\csname url\endcsname\relax
  \def\url#1{{\tt #1}}\fi
\expandafter\ifx\csname urlprefix\endcsname\relax\def\urlprefix{URL }\fi
\providecommand{\eprint}[2][]{\url{#2}}

\bibitem{GW150914}
Abbott B~P, Abbott R, Abbott T~D, Abernathy M~R, Acernese F {\em et~al.\/}
  (LIGO Scientific Collaboration and Virgo Collaboration) 2016 {\em Phys. Rev.
  Lett.\/} {\bf 116}(6) 061102
  \urlprefix\url{http://link.aps.org/doi/10.1103/PhysRevLett.116.061102}

\bibitem{BBH:O1O2}
{The LIGO Scientific Collaboration}, {the Virgo Collaboration}, {Abbott} B~P,
  {Abbott} R, {Abbott} T~D, {Abraham} S, {Acernese} F, {Ackley} K, {Adams} C,
  {Adhikari} R~X and et~al 2018 {\em arXiv e-prints\/} (\textit{Preprint}
  \eprint{1811.12940})

\bibitem{GW170817}
{Abbott} B~P, {Abbott} R, {Abbott} T~D, {Acernese} F, {Ackley} K, {Adams} C,
  {Adams} T, {Addesso} P, {Adhikari} R~X, {Adya} V~B and et~al 2017 {\em
  Physical Review Letters\/} {\bf 119} 161101 (\textit{Preprint}
  \eprint{1710.05832})

\bibitem{Straw}
Adhikari R, Gustafson E~K, Hild S, Ballmer S and Arai K 2012 {Report of the
  Third Generation LIGO Strawman Design Meeting} Tech. rep. LIGO Scientific
  Collaboration
  \urlprefix\url{https://dcc.ligo.org/cgi-bin/private/DocDB/ShowDocument?docid=85610}

\bibitem{ISWP:2016}
Collaboration L~S 2016 {LSC Instrument Science White Paper} Tech. rep. LIGO
  Scientific Collaboration
  \urlprefix\url{https://dcc.ligo.org/LIGO-T1600119/public}

\bibitem{Voyager:Inst}
Adhikari R 2019 A cryogenic interferometer for gravitational-wave detection in
  prep.

\bibitem{3G:Science}
Punturo M, Abernathy M, Acernese F, Allen B, Andersson N and andothers K~A 2010
  {\em Classical and Quantum Gravity\/} {\bf 27} 084007

\bibitem{abbott2017exploring}
Abbott B~P, Abbott R, Abbott T, Abernathy M, Ackley K, Adams C, Addesso P,
  Adhikari R, Adya V, Affeldt C {\em et~al.\/} 2017 {\em Classical and Quantum
  Gravity\/} {\bf 34} 044001

\bibitem{SUS:FEA2009}
Cumming A, Heptonstall A, Kumar R, Cunningham W, Torrie C, Barton M, Strain
  K~A, Hough J and Rowan S 2009 {\em Classical and Quantum Gravity\/} {\bf 26}
  215012 \urlprefix\url{http://stacks.iop.org/0264-9381/26/i=21/a=215012}

\bibitem{SUS:2012}
Cumming A~V, Bell A~S, Barsotti L, Barton M~A, Cagnoli G, Cook D, Cunningham L,
  Evans M, Hammond G~D, Harry G~M, Heptonstall A, Hough J, Jones R, Kumar R,
  Mittleman R, Robertson N~A, Rowan S, Shapiro B, Strain K~A, Tokmakov K,
  Torrie C and van Veggel A~A 2012 {\em Classical and Quantum Gravity\/} {\bf
  29} 035003 \urlprefix\url{http://stacks.iop.org/0264-9381/29/i=3/a=035003}

\bibitem{Den:PRL2016}
Abbott B~P, Abbott R, Abbott T~D, Abernathy M~R, Acernese F {\em et~al.\/}
  (LIGO Scientific Collaboration and Virgo Collaboration) 2016 {\em Phys. Rev.
  Lett.\/} {\bf 116}(13) 131103
  \urlprefix\url{http://link.aps.org/doi/10.1103/PhysRevLett.116.131103}

\bibitem{PhysRevD.91.062005}
Miller J, Barsotti L, Vitale S, Fritschel P, Evans M and Sigg D 2015 {\em Phys.
  Rev. D\/} {\bf 91}(6) 062005
  \urlprefix\url{https://link.aps.org/doi/10.1103/PhysRevD.91.062005}

\bibitem{Harms:LR2015}
{Harms} J 2015 {\em ArXiv e-prints\/} (\textit{Preprint} \eprint{1507.05850})

\bibitem{aLIGO:SEI:2015}
Matichard F, Lantz B, Mittleman R, Mason K, Kissel J, Abbott B {\em et~al.\/}
  2015 {\em Classical and Quantum Gravity\/} {\bf 32} 185003
  \urlprefix\url{http://stacks.iop.org/0264-9381/32/i=18/a=185003}

\bibitem{silicon:SUS}
Cumming A~V, Cunningham L, Hammond G~D, Haughian K, Hough J, Kroker S, Martin
  I~W, Nawrodt R, Rowan S, Schwarz C and van Veggel A~A 2014 {\em Classical and
  Quantum Gravity\/} {\bf 31} 025017
  \urlprefix\url{http://stacks.iop.org/0264-9381/31/i=2/a=025017}

\bibitem{reid2016development}
Reid S and Martin I~W 2016 {\em Coatings\/} {\bf 6} 61

\bibitem{Haixing:CC2012}
Miao H, Yang H, Adhikari R~X and Chen Y 2014 {\em Classical and Quantum
  Gravity\/} {\bf 31} 165010
  \urlprefix\url{http://stacks.iop.org/0264-9381/31/i=16/a=165010}

\bibitem{FC:2013}
Evans M, Barsotti L, Kwee P, Harms J and Miao H 2013 {\em Phys. Rev. D\/} {\bf
  88}(2) 022002
  \urlprefix\url{https://link.aps.org/doi/10.1103/PhysRevD.88.022002}

\bibitem{MandelOShaughnessy:2010}
{Mandel} I and {O'Shaughnessy} R 2010 {\em Classical and Quantum Gravity\/}
  {\bf 27} 114007 (\textit{Preprint} \eprint{0912.1074})

\bibitem{GW150914:astro}
{Abbott} B~P, {Abbott} R, {Abbott} T~D, {Abernathy} M~R, {Acernese} F, {Ackley}
  K, {Adams} C, {Adams} T, {Addesso} P, {Adhikari} R~X and et~al 2016 {\em
  \apjl\/} {\bf 818} L22 (\textit{Preprint} \eprint{1602.03846})

\bibitem{MandelFarmer:2018}
{Mandel} I and {Farmer} A 2018 {\em ArXiv e-prints\/} (\textit{Preprint}
  \eprint{1806.05820})

\bibitem{Mapelli:2018}
{Mapelli} M 2018 {\em arXiv e-prints\/} (\textit{Preprint} \eprint{1809.09130})

\bibitem{2007PhR...442...75K}
{Kalogera} V, {Belczynski} K, {Kim} C, {O'Shaughnessy} R and {Willems} B 2007
  {\em Physics Reports\/} {\bf 442} 75--108 (\textit{Preprint}
  \eprint{astro-ph/0612144})

\bibitem{PostnovYungelson:2014}
{Postnov} K~A and {Yungelson} L~R 2014 {\em Living Reviews in Relativity\/}
  {\bf 17} 3 (\textit{Preprint} \eprint{1403.4754})

\bibitem{Tauris:2017}
{Tauris} T~M, {Kramer} M, {Freire} P~C~C, {Wex} N, {Janka} H~T, {Langer} N,
  {Podsiadlowski} P, {Bozzo} E, {Chaty} S, {Kruckow} M~U, {van den Heuvel}
  E~P~J, {Antoniadis} J, {Breton} R~P and {Champion} D~J 2017 {\em \apj\/} {\bf
  846} 170 (\textit{Preprint} \eprint{1706.09438})

\bibitem{Belczynski:2008}
{Belczynski} K, {Kalogera} V, {Rasio} F~A, {Taam} R~E, {Zezas} A, {Bulik} T,
  {Maccarone} T~J and {Ivanova} N 2008 {\em \apjs\/} {\bf 174} 223--260

\bibitem{Pfahl:2005}
{Pfahl} E, {Podsiadlowski} P and {Rappaport} S 2005 {\em \apj\/} {\bf 628}
  343--352 (\textit{Preprint} \eprint{astro-ph/0502122})

\bibitem{VossTauris:2003}
{Voss} R and {Tauris} T~M 2003 {\em \mnras\/} {\bf 342} 1169--1184
  (\textit{Preprint} \eprint{astro-ph/0303227})

\bibitem{Dominik:2014}
{Dominik} M, {Berti} E, {O'Shaughnessy} R, {Mandel} I, {Belczynski} K, {Fryer}
  C, {Holz} D~E, {Bulik} T and {Pannarale} F 2015 {\em \apj\/} {\bf 806} 263
  (\textit{Preprint} \eprint{1405.7016})

\bibitem{VignaGomez:2018}
{Vigna-G{\'o}mez} A, {Neijssel} C~J, {Stevenson} S, {Barrett} J~W, {Belczynski}
  K, {Justham} S, {de Mink} S~E, {M{\"u}ller} B, {Podsiadlowski} P, {Renzo} M,
  {Sz{\'e}csi} D and {Mandel} I 2018 {\em \mnras\/} {\bf 481} 4009--4029
  (\textit{Preprint} \eprint{1805.07974})

\bibitem{Belczynski:2016}
{Belczynski} K, {Holz} D~E, {Bulik} T and {O'Shaughnessy} R 2016 {\em \nat\/}
  {\bf 534} 512--515 (\textit{Preprint} \eprint{1602.04531})

\bibitem{EldridgeStanway:2016}
{Eldridge} J~J and {Stanway} E~R 2016 {\em ArXiv e-prints\/} (\textit{Preprint}
  \eprint{1602.03790})

\bibitem{Lipunov:2016}
{Lipunov} V~M, {Kornilov} V, {Gorbovskoy} E, {Tiurina} N, {Balanutsa} P and
  {Kuznetsov} A 2016 {\em ArXiv e-prints\/} (\textit{Preprint}
  \eprint{1605.01604})

\bibitem{Stevenson:2017}
{Stevenson} S, {Vigna-G{\'o}mez} A, {Mandel} I, {Barrett} J~W, {Neijssel} C~J,
  {Perkins} D and {de Mink} S~E 2017 {\em Nature Communications\/} {\bf 8}
  14906 (\textit{Preprint} \eprint{1704.01352})

\bibitem{MandeldeMink:2016}
{Mandel} I and {de Mink} S~E 2016 {\em \mnras\/} {\bf 458} 2634--2647
  (\textit{Preprint} \eprint{1601.00007})

\bibitem{Marchant:2016}
{Marchant} P, {Langer} N, {Podsiadlowski} P, {Tauris} T~M and {Moriya} T~J 2016
  {\em \aap\/} {\bf 588} A50 (\textit{Preprint} \eprint{1601.03718})

\bibitem{Bird:2016}
{Bird} S, {Cholis} I, {Mu{\~n}oz} J~B, {Ali-Ha{\"i}moud} Y, {Kamionkowski} M,
  {Kovetz} E~D, {Raccanelli} A and {Riess} A~G 2016 {\em Physical Review
  Letters\/} {\bf 116} 201301 (\textit{Preprint} \eprint{1603.00464})

\bibitem{Sasaki:2016}
Sasaki M, Suyama T, Tanaka T and Yokoyama S 2016 {\em Phys. Rev. Lett.\/} {\bf
  117} 061101 (\textit{Preprint} \eprint{1603.08338})

\bibitem{Rodriguez:2016}
{Rodriguez} C~L, {Haster} C~J, {Chatterjee} S, {Kalogera} V and {Rasio} F~A
  2016 {\em \apjl\/} {\bf 824} L8 (\textit{Preprint} \eprint{1604.04254})

\bibitem{Stone:2016}
{Stone} N~C, {Metzger} B~D and {Haiman} Z 2017 {\em \mnras\/} {\bf 464}
  946--954 (\textit{Preprint} \eprint{1602.04226})

\bibitem{Bartos:2016}
{Bartos} I, {Kocsis} B, {Haiman} Z and {M{\'a}rka} S 2016 {\em ArXiv
  e-prints\/} (\textit{Preprint} \eprint{1602.03831})

\bibitem{ratesdoc}
Abadie J {\em et~al.\/} (LIGO Scientific Collaboration and Virgo Collaboration)
  2010 {\em Classical and Quantum Gravity\/} {\bf 27} 173001--+
  (\textit{Preprint} \eprint{1003.2480})

\bibitem{GW170817:MMA}
{Abbott} B~P, {Abbott} R, {Abbott} T~D, {Acernese} F, {Ackley} K, {Adams} C,
  {Adams} T, {Addesso} P, {Adhikari} R~X, {Adya} V~B and et~al 2017 {\em
  \apjl\/} {\bf 848} L12 (\textit{Preprint} \eprint{1710.05833})

\bibitem{GW150914:PE}
{Abbott} B~P, {Abbott} R, {Abbott} T~D, {Abernathy} M~R, {Acernese} F, {Ackley}
  K, {Adams} C, {Adams} T, {Addesso} P, {Adhikari} R~X and et~al 2016 {\em
  Physical Review Letters\/} {\bf 116} 241102 (\textit{Preprint}
  \eprint{1602.03840})

\bibitem{Veitch:2014}
{Veitch} J, {Raymond} V, {Farr} B, {Farr} W, {Graff} P, {Vitale} S, {Aylott} B,
  {Blackburn} K, {Christensen} N, {Coughlin} M, {Del Pozzo} W, {Feroz} F,
  {Gair} J, {Haster} C~J, {Kalogera} V, {Littenberg} T, {Mandel} I,
  {O'Shaughnessy} R, {Pitkin} M, {Rodriguez} C, {R{\"o}ver} C, {Sidery} T,
  {Smith} R, {Van Der Sluys} M, {Vecchio} A, {Vousden} W and {Wade} L 2015 {\em
  \prd\/} {\bf 91} 042003 (\textit{Preprint} \eprint{1409.7215})

\bibitem{VitaleEvans:2017}
{Vitale} S and {Evans} M 2017 {\em \prd\/} {\bf 95} 064052 (\textit{Preprint}
  \eprint{1610.06917})

\bibitem{Rodriguez:2016big}
{Rodriguez} C~L, {Chatterjee} S and {Rasio} F~A 2016 {\em \prd\/} {\bf 93}
  084029 (\textit{Preprint} \eprint{1602.02444})

\bibitem{Stevenson:2015}
{Stevenson} S, {Ohme} F and {Fairhurst} S 2015 {\em \apj\/} {\bf 810} 58
  (\textit{Preprint} \eprint{1504.07802})

\bibitem{Stevenson:2017spin}
{Stevenson} S, {Berry} C~P~L and {Mandel} I 2017 {\em ArXiv e-prints\/}
  (\textit{Preprint} \eprint{1703.06873})

\bibitem{Barrett:2017FIM}
{Barrett} J~W, {Gaebel} S~M, {Neijssel} C~J, {Vigna-G{\'o}mez} A, {Stevenson}
  S, {Berry} C~P~L, {Farr} W~M and {Mandel} I 2018 {\em \mnras\/} {\bf 477}
  4685--4695 (\textit{Preprint} \eprint{1711.06287})

\bibitem{Mandel:2015}
{Mandel} I, {Haster} C~J, {Dominik} M and {Belczynski} K 2015 {\em \mnras\/}
  {\bf 450} L85--L89 (\textit{Preprint} \eprint{1503.03172})

\bibitem{Vitale:2015}
{Vitale} S, {Lynch} R, {Sturani} R and {Graff} P 2017 {\em Classical and
  Quantum Gravity\/} {\bf 34} 03LT01 (\textit{Preprint} \eprint{1503.04307})

\bibitem{Mandel:2016cluster}
{Mandel} I, {Farr} W~M, {Colonna} A, {Stevenson} S, {Ti{\v n}o} P and {Veitch}
  J 2017 {\em \mnras\/} {\bf 465} 3254--3260 (\textit{Preprint}
  \eprint{1608.08223})

\bibitem{Farr:2017}
{Farr} W~M, {Stevenson} S, {Miller} M~C, {Mandel} I, {Farr} B and {Vecchio} A
  2017 {\em ArXiv e-prints\/} (\textit{Preprint} \eprint{1706.01385})

\bibitem{Read:2009}
{Read} J~S, {Markakis} C, {Shibata} M, {Ury{\= u}} K, {Creighton} J~D~E and
  {Friedman} J~L 2009 {\em \prd\/} {\bf 79} 124033--+ (\textit{Preprint}
  \eprint{arXiv:0901.3258})

\bibitem{GW170817:GRB}
{Abbott} B~P, {Abbott} R, {Abbott} T~D, {Acernese} F, {Ackley} K, {Adams} C,
  {Adams} T, {Addesso} P, {Adhikari} R~X, {Adya} V~B and et~al 2017 {\em
  \apjl\/} {\bf 848} L13 (\textit{Preprint} \eprint{1710.05834})

\bibitem{2012ApJ...748..136C}
{Cannon} K, {Cariou} R, {Chapman} A, {Crispin-Ortuzar} M, {Fotopoulos} N,
  {Frei} M, {Hanna} C, {Kara} E, {Keppel} D, {Liao} L, {Privitera} S, {Searle}
  A, {Singer} L and {Weinstein} A 2012 {\em \apj\/} {\bf 748} 136
  (\textit{Preprint} \eprint{1107.2665})

\bibitem{MillerColbert:2004}
{Miller} M~C and {Colbert} E~J~M 2004 {\em International Journal of Modern
  Physics D\/} {\bf 13} 1--64 (\textit{Preprint}
  \eprint{arXiv:astro-ph/0308402})

\bibitem{Farrell:2009}
{Farrell} S~A, {Webb} N~A, {Barret} D, {Godet} O and {Rodrigues} J~M 2009 {\em
  \nat\/} {\bf 460} 73--75

\bibitem{Collaboration:S5HighMass}
Abadie J {\em et~al.\/} (LIGO Scientific Collaboration and Virgo Collaboration)
  2011 {\em Submitted to PRD\/} (\textit{Preprint} \eprint{arXiv:1102.3781})

\bibitem{Veitch:2015}
{Veitch} J, {P{\"u}rrer} M and {Mandel} I 2015 {\em Physical Review Letters\/}
  {\bf 115} 141101 (\textit{Preprint} \eprint{1503.05953})

\bibitem{Gair:2009ETrev}
{Gair} J~R, {Mandel} I, {Miller} M~C and {Volonteri} M 2011 {\em General
  Relativity and Gravitation\/} {\bf 43} 485--518 (\textit{Preprint}
  \eprint{0907.5450})

\bibitem{Mandel:2008}
{Mandel} I, {Brown} D~A, {Gair} J~R and {Miller} M~C 2008 {\em \apj\/} {\bf
  681} 1431--1447 (\textit{Preprint} \eprint{0705.0285})

\bibitem{Macleod:2016}
{MacLeod} M, {Trenti} M and {Ramirez-Ruiz} E 2016 {\em \apj\/} {\bf 819} 70
  (\textit{Preprint} \eprint{1508.07000})

\bibitem{Haster:2016}
{Haster} C~J, {Antonini} F, {Kalogera} V and {Mandel} I 2016 {\em \apj\/} {\bf
  832} 192 (\textit{Preprint} \eprint{1606.07097})

\bibitem{Fregeau:2006}
{Fregeau} J~M, {Larson} S~L, {Miller} M~C, {O'Shaughnessy} R and {Rasio} F~A
  2006 {\em Astrophysical Journal Letters\/} {\bf 646} L135--L138
  (\textit{Preprint} \eprint{astro-ph/0605732})

\bibitem{AmaroSeoaneSantamaria:2009}
{Amaro-Seoane} P and {Santamar{\'{\i}}a} L 2010 {\em \apj\/} {\bf 722}
  1197--1206 (\textit{Preprint} \eprint{0910.0254})

\bibitem{MadauRees:2001}
{Madau} P and {Rees} M~J 2001 {\em \apjl\/} {\bf 551} L27--L30

\bibitem{Sesana:2009ET}
{Sesana} A, {Gair} J, {Mandel} I and {Vecchio} A 2009 {\em \apjl\/} {\bf 698}
  L129--L132 (\textit{Preprint} \eprint{0903.4177})

\bibitem{Gair:2009ET}
{Gair} J~R, {Mandel} I, {Sesana} A and {Vecchio} A 2009 {\em Classical and
  Quantum Gravity\/} {\bf 26} 204009 (\textit{Preprint} \eprint{0907.3292})

\bibitem{Will:2001}
{Will} C 2001 {\em Living Reviews in Relativity\/} {\bf 4} 4 (\textit{Preprint}
  \eprint{arXiv:gr-qc/0103036})

\bibitem{Yunes:2011}
{Yunes} N 2011 {\em ArXiv e-prints\/} (\textit{Preprint} \eprint{1112.3694})

\bibitem{GW150914:GR}
{Abbott} B~P, {Abbott} R, {Abbott} T~D, {Abernathy} M~R, {Acernese} F, {Ackley}
  K, {Adams} C, {Adams} T, {Addesso} P, {Adhikari} R~X and et~al 2016 {\em
  Physical Review Letters\/} {\bf 116} 221101 (\textit{Preprint}
  \eprint{1602.03841})

\bibitem{lrr-2006-3}
Will C~M 2006 {\em Living Reviews in Relativity\/} {\bf 9}
  \urlprefix\url{http://www.livingreviews.org/lrr-2006-3}

\bibitem{Will98}
Will C~M 1998 {\em Phys. Rev. D\/} {\bf 57} 2061--2068

\bibitem{PhysRevLett.116.221101}
Abbott B~P {\em et~al.\/} (LIGO Scientific and Virgo Collaborations) 2016 {\em
  Phys. Rev. Lett.\/} {\bf 116}(22) 221101
  \urlprefix\url{https://link.aps.org/doi/10.1103/PhysRevLett.116.221101}

\bibitem{Keppel:2010qu}
Keppel D and Ajith P 2010  ArXiv:1004.0284 [gr-qc] (\textit{Preprint}
  \eprint{1004.0284})

\bibitem{Arun:2006a}
{Arun} K~G, {Iyer} B~R, {Qusailah} M~S~S and {Sathyaprakash} B~S 2006 {\em
  Classical and Quantum Gravity\/} {\bf 23} L37--L43 (\textit{Preprint}
  \eprint{arXiv:gr-qc/0604018})

\bibitem{Mishra:2010}
{Mishra} C~K, {Arun} K~G, {Iyer} B~R and {Sathyaprakash} B~S 2010 {\em \prd\/}
  {\bf 82} 064010 (\textit{Preprint} \eprint{1005.0304})

\bibitem{Li:2012}
{Li} T~G~F, {Del Pozzo} W, {Vitale} S, {Van Den Broeck} C, {Agathos} M,
  {Veitch} J, {Grover} K, {Sidery} T, {Sturani} R and {Vecchio} A 2011 {\em
  ArXiv e-prints\/} (\textit{Preprint} \eprint{1110.0530})

\bibitem{Blanchet1993}
Blanchet L and Schafer G 1993 {\em Classical and Quantum Gravity\/} {\bf 10}
  2699--2721
  \urlprefix\url{https://doi.org/10.1088%2F0264-9381%2F10%2F12%2F026}

\bibitem{blanchet2017first}
Blanchet L and Le~Tiec A 2017 {\em Classical and Quantum Gravity\/} {\bf 34}
  164001

\bibitem{Favata:2009}
Favata M 2009 {\em Phys. Rev. D\/} {\bf 80}(2) 024002
  \urlprefix\url{http://link.aps.org/doi/10.1103/PhysRevD.80.024002}

\bibitem{favata:10}
{Favata} M 2010 {\em Classical and Quantum Gravity\/} {\bf 27} 084036
  (\textit{Preprint} \eprint{1003.3486})

\bibitem{Pollney:2011}
Pollney D and Reisswig C 2011 {\em The Astrophysical Journal Letters\/} {\bf
  732} L13 \urlprefix\url{http://stacks.iop.org/2041-8205/732/i=1/a=L13}

\bibitem{PhysRevD.98.064031}
Talbot C, Thrane E, Lasky P~D and Lin F 2018 {\em Phys. Rev. D\/} {\bf 98}(6)
  064031 \urlprefix\url{https://link.aps.org/doi/10.1103/PhysRevD.98.064031}

\bibitem{Lasky2016}
{Lasky} P~D, {Thrane} E, {Levin} Y, {Blackman} J and {Chen} Y 2016 {\em
  Physical Review Letters\/} {\bf 117} 061102 (\textit{Preprint}
  \eprint{1605.01415})

\bibitem{PhysRevLett.121.071102}
Yang H and Martynov D 2018 {\em Phys. Rev. Lett.\/} {\bf 121}(7) 071102
  \urlprefix\url{https://link.aps.org/doi/10.1103/PhysRevLett.121.071102}

\bibitem{Yu2018}
{Yu} H, {Martynov} D, {Vitale} S, {Evans} M, {Shoemaker} D, {Barr} B, {Hammond}
  G, {Hild} S, {Hough} J, {Huttner} S, {Rowan} S, {Sorazu} B, {Carbone} L,
  {Freise} A, {Mow-Lowry} C, {Dooley} K~L, {Fulda} P, {Grote} H and {Sigg} D
  2018 {\em Physical Review Letters\/} {\bf 120} 141102 (\textit{Preprint}
  \eprint{1712.05417})

\bibitem{Ryan:1995}
Ryan F~D 1995 {\em Phys. Rev.\/} {\bf D52} 5707--5718

\bibitem{Meidam:2014}
{Meidam} J, {Agathos} M, {Van Den Broeck} C, {Veitch} J and {Sathyaprakash} B~S
  2014 {\em \prd\/} {\bf 90} 064009 (\textit{Preprint} \eprint{1406.3201})

\bibitem{Brown:2007}
{Brown} D~A, {Brink} J, {Fang} H, {Gair} J~R, {Li} C, {Lovelace} G, {Mandel} I
  and {Thorne} K~S 2007 {\em Phys. Rev. Lett.\/} {\bf 99} 201102--+
  (\textit{Preprint} \eprint{arXiv:gr-qc/0612060})

\bibitem{Rodriguez:2012}
{Rodriguez} C~L, {Mandel} I and {Gair} J~R 2012 {\em \prd\/} {\bf 85} 062002
  (\textit{Preprint} \eprint{1112.1404})

\bibitem{SmithIMRI:2013}
{Smith} R~J~E, {Mandel} I and {Vecchio} A 2013 {\em \prd\/} {\bf 88} 044010
  (\textit{Preprint} \eprint{1302.6049})

\bibitem{Berti:2009}
{Berti} E, {Cardoso} V and {Starinets} A~O 2009 {\em ArXiv e-prints\/}
  (\textit{Preprint} \eprint{0905.2975})

\bibitem{Hughes:2004vw}
Hughes S~A and Menou K 2005 {\em Astrophys.J.\/} {\bf 623} 689--699
  (\textit{Preprint} \eprint{astro-ph/0410148})

\bibitem{Ghosh:2016qgn}
{Ghosh} A, {Ghosh} A, {Johnson-McDaniel} N~K, {Mishra} C~K, {Ajith} P, {Del
  Pozzo} W, {Nichols} D~A, {Chen} Y, {Nielsen} A~B, {Berry} C~P~L and {London}
  L 2016 {\em \prd\/} {\bf 94} 021101 (\textit{Preprint} \eprint{1602.02453})

\bibitem{GW170817:EOS}
{Abbott} B~P, {Abbott} R, {Abbott} T~D, {Acernese} F, {Ackley} K, {Adams} C,
  {Adams} T, {Addesso} P, {Adhikari} R~X, {Adya} V~B and et~al 2018 {\em
  Physical Review Letters\/} {\bf 121} 161101 (\textit{Preprint}
  \eprint{1805.11581})

\bibitem{2009PhRvD..79l4033R}
{Read} J~S, {Markakis} C, {Shibata} M, {Ury{\= u}} K, {Creighton} J~D~E and
  {Friedman} J~L 2009 {\em \prd\/} {\bf 79} 124033 (\textit{Preprint}
  \eprint{0901.3258})

\bibitem{hinderer:10}
{Hinderer} T, {Lackey} B~D, {Lang} R~N and {Read} J~S 2010 {\em \prd\/} {\bf
  81} 123016

\bibitem{hinderer:08}
{Hinderer} T 2008 {\em \apj\/} {\bf 677} 1216

\bibitem{vines:11}
{Vines} J, {Flanagan} {\'E}~{\'E} and {Hinderer} T 2011 {\em \prd\/} {\bf 83}
  084051 (\textit{Preprint} \eprint{1101.1673})

\bibitem{Damour:2012yf}
Damour T, Nagar A and Villain L 2012 {\em Phys. Rev.\/} {\bf D85} 123007
  (\textit{Preprint} \eprint{1203.4352})

\bibitem{read:11c}
{Read} J~S, {Baiotti} L, {Creighton} J~D~E, {Friedman} J~L, {Giacomazzo} B,
  {Kyutoku} K, {Markakis} C, {Rezzolla} L, {Shibata} M and {Taniguchi} K 2013
  {\em \prd\/} {\bf 88} 044042 (\textit{Preprint} \eprint{1306.4065})

\bibitem{2014PhRvL.112j1101F}
{Favata} M 2014 {\em \prl\/} {\bf 112} 101101 (\textit{Preprint}
  \eprint{1310.8288})

\bibitem{2014PhRvD..89j3012W}
{Wade} L, {Creighton} J~D~E, {Ochsner} E, {Lackey} B~D, {Farr} B~F,
  {Littenberg} T~B and {Raymond} V 2014 {\em \prd\/} {\bf 89} 103012
  (\textit{Preprint} \eprint{1402.5156})

\bibitem{Haensel:2002cia}
Haensel P, Zdunik J~L and Douchin F 2002 {\em Astron. Astrophys.\/} {\bf 385}
  301 (\textit{Preprint} \eprint{astro-ph/0201434})

\bibitem{Lattimer:2012nd}
Lattimer J~M 2012 {\em Ann. Rev. Nucl. Part. Sci.\/} {\bf 62} 485--515
  (\textit{Preprint} \eprint{1305.3510})

\bibitem{Perlmutter:1999}
{Perlmutter} S, {Aldering} G, {Goldhaber} G, {Knop} R~A, {Nugent} P, {Castro}
  P~G, {Deustua} S, {Fabbro} S, {Goobar} A, {Groom} D~E, {Hook} I~M, {Kim} A~G,
  {Kim} M~Y, {Lee} J~C, {Nunes} N~J, {Pain} R, {Pennypacker} C~R, {Quimby} R,
  {Lidman} C, {Ellis} R~S, {Irwin} M, {McMahon} R~G, {Ruiz-Lapuente} P,
  {Walton} N, {Schaefer} B, {Boyle} B~J, {Filippenko} A~V, {Matheson} T,
  {Fruchter} A~S, {Panagia} N, {Newberg} H~J~M, {Couch} W~J and {Project} T~S~C
  1999 {\em \apj\/} {\bf 517} 565--586 (\textit{Preprint}
  \eprint{astro-ph/9812133})

\bibitem{2011ApJS..192...18K}
{Komatsu} E, {Smith} K~M, {Dunkley} J, {Bennett} C~L, {Gold} B, {Hinshaw} G,
  {Jarosik} N, {Larson} D, {Nolta} M~R, {Page} L, {Spergel} D~N, {Halpern} M,
  {Hill} R~S, {Kogut} A, {Limon} M, {Meyer} S~S, {Odegard} N, {Tucker} G~S,
  {Weiland} J~L, {Wollack} E and {Wright} E~L 2011 {\em \apjs\/} {\bf 192} 18
  (\textit{Preprint} \eprint{1001.4538})

\bibitem{Planck:2015}
{Planck Collaboration}, {Ade} P~A~R, {Aghanim} N, {Arnaud} M, {Ashdown} M,
  {Aumont} J, {Baccigalupi} C, {Banday} A~J, {Barreiro} R~B, {Bartlett} J~G and
  et~al 2015 {\em ArXiv e-prints\/} (\textit{Preprint} \eprint{1502.01589})

\bibitem{Sathya:Cosmo2010}
{Sathyaprakash} B~S, {Schutz} B~F and {Van Den Broeck} C 2010 {\em Classical
  and Quantum Gravity\/} {\bf 27} 215006 (\textit{Preprint} \eprint{0906.4151})

\bibitem{TaylorGair:2012}
{Taylor} S~R and {Gair} J~R 2012 {\em \prd\/} {\bf 86} 023502
  (\textit{Preprint} \eprint{1204.6739})

\bibitem{Schutz:1986}
{Schutz} B~F 1986 {\em \nat\/} {\bf 323} 310--+

\bibitem{HolzHughes:2005}
{Holz} D~E and {Hughes} S~A 2005 {\em \apj\/} {\bf 629} 15--22
  (\textit{Preprint} \eprint{arXiv:astro-ph/0504616})

\bibitem{GW170817:H0}
{Abbott} B~P, {Abbott} R, {Abbott} T~D, {Acernese} F, {Ackley} K, {Adams} C,
  {Adams} T, {Addesso} P, {Adhikari} R~X, {Adya} V~B and et~al 2017 {\em
  \nat\/} {\bf 551} 85--88 (\textit{Preprint} \eprint{1710.05835})

\bibitem{DelPozzo:2012}
{Del Pozzo} W 2012 {\em \prd\/} {\bf 86} 043011 (\textit{Preprint}
  \eprint{1108.1317})

\bibitem{MessengerRead:2012}
{Messenger} C and {Read} J 2012 {\em Physical Review Letters\/} {\bf 108}
  091101 (\textit{Preprint} \eprint{1107.5725})

\bibitem{DelPozzo:2017}
{Del Pozzo} W, {Li} T~G~F and {Messenger} C 2017 {\em \prd\/} {\bf 95} 043502
  (\textit{Preprint} \eprint{1506.06590})

\bibitem{Taylor:2012}
{Taylor} S~R, {Gair} J~R and {Mandel} I 2012 {\em \prd\/} {\bf 85} 023535
  (\textit{Preprint} \eprint{1108.5161})

\bibitem{Kiziltan:2013}
{Kiziltan} B, {Kottas} A, {De Yoreo} M and {Thorsett} S~E 2013 {\em \apj\/}
  {\bf 778} 66 (\textit{Preprint} \eprint{1011.4291})

\bibitem{Berry:2015}
{Berry} C~P~L, {Mandel} I, {Middleton} H, {Singer} L~P, {Urban} A~L, {Vecchio}
  A, {Vitale} S, {Cannon} K, {Farr} B, {Farr} W~M, {Graff} P~B, {Hanna} C,
  {Haster} C~J, {Mohapatra} S, {Pankow} C, {Price} L~R, {Sidery} T and {Veitch}
  J 2015 {\em \apj\/} {\bf 804} 114 (\textit{Preprint} \eprint{1411.6934})

\bibitem{mtw}
{Misner} C~W, {Thorne} K~S and {Wheeler} J~A 1973 {\em {Gravitation}\/} (San
  Francisco: W.H.~Freeman and Co.)

\bibitem{roberts:2016}
{Roberts} L~F, {Ott} C~D, {Haas} R, {O'Connor} E~P, {Diener} P and {Schnetter}
  E 2016 {\em \apj\/} {\bf 831} 98 (\textit{Preprint} \eprint{1604.07848})

\bibitem{kuroda:2016}
{Kuroda} T, {Kotake} K and {Takiwaki} T 2016 {\em \apjl\/} {\bf 829} L14
  (\textit{Preprint} \eprint{1605.09215})

\bibitem{takiwaki:2016}
{Takiwaki} T, {Kotake} K and {Suwa} Y 2016 {\em \mnras\/} {\bf 461} L112--L116
  (\textit{Preprint} \eprint{1602.06759})

\bibitem{radice:2016}
{Radice} D, {Ott} C~D, {Abdikamalov} E, {Couch} S~M, {Haas} R and {Schnetter} E
  2016 {\em \apj\/} {\bf 820} 76 (\textit{Preprint} \eprint{1510.05022})

\bibitem{abdikamalov:2015}
{Abdikamalov} E, {Ott} C~D, {Radice} D, {Roberts} L~F, {Haas} R, {Reisswig} C,
  {M{\"o}sta} P, {Klion} H and {Schnetter} E 2015 {\em \apj\/} {\bf 808} 70
  (\textit{Preprint} \eprint{1409.7078})

\bibitem{lentz:2015}
{Lentz} E~J, {Bruenn} S~W, {Hix} W~R, {Mezzacappa} A, {Messer} O~E~B, {Endeve}
  E, {Blondin} J~M, {Harris} J~A, {Marronetti} P and {Yakunin} K~N 2015 {\em
  \apjl\/} {\bf 807} L31 (\textit{Preprint} \eprint{1505.05110})

\bibitem{melson:2015-strangeq}
{Melson} T, {Janka} H~T, {Bollig} R, {Hanke} F, {Marek} A and {M{\"u}ller} B
  2015 {\em \apjl\/} {\bf 808} L42 (\textit{Preprint} \eprint{1504.07631})

\bibitem{bmueller:2015}
{M{\"u}ller} B and {Janka} H~T 2015 {\em \mnras\/} {\bf 448} 2141--2174
  (\textit{Preprint} \eprint{1409.4783})

\bibitem{moesta:2015}
{M{\"o}sta} P, {Ott} C~D, {Radice} D, {Roberts} L~F, {Schnetter} E and {Haas} R
  2015 {\em \nat\/} {\bf 528} 376--379 (\textit{Preprint} \eprint{1512.00838})

\bibitem{couch:2014}
{Couch} S~M and {O'Connor} E~P 2014 {\em \apj\/} {\bf 785} 123
  (\textit{Preprint} \eprint{1310.5728})

\bibitem{takiwaki:2014}
{Takiwaki} T, {Kotake} K and {Suwa} Y 2014 {\em \apj\/} {\bf 786} 83
  (\textit{Preprint} \eprint{1308.5755})

\bibitem{murphy:2013}
{Murphy} J~W, {Dolence} J~C and {Burrows} A 2013 {\em \apj\/} {\bf 771} 52
  (\textit{Preprint} \eprint{1205.3491})

\bibitem{hanke:2013}
{Hanke} F, {M{\"u}ller} B, {Wongwathanarat} A, {Marek} A and {Janka} H~T 2013
  {\em \apj\/} {\bf 770} 66 (\textit{Preprint} \eprint{1303.6269})

\bibitem{couch:2013}
{Couch} S~M 2013 {\em \apj\/} {\bf 775} 35 (\textit{Preprint}
  \eprint{1212.0010})

\bibitem{dolence:2013}
{Dolence} J~C, {Burrows} A, {Murphy} J~W and {Nordhaus} J 2013 {\em \apj\/}
  {\bf 765} 110 (\textit{Preprint} \eprint{1210.5241})

\bibitem{ott:2013}
{Ott} C~D, {Abdikamalov} E, {M{\"o}sta} P, {Haas} R, {Drasco} S, {O'Connor}
  E~P, {Reisswig} C, {Meakin} C~A and {Schnetter} E 2013 {\em \apj\/} {\bf 768}
  115 (\textit{Preprint} \eprint{1210.6674})

\bibitem{emueller:2012}
{M{\"u}ller} E, {Janka} H~T and {Wongwathanarat} A 2012 {\em \aap\/} {\bf 537}
  A63 (\textit{Preprint} \eprint{1106.6301})

\bibitem{takiwaki:2012}
{Takiwaki} T, {Kotake} K and {Suwa} Y 2012 {\em \apj\/} {\bf 749} 98
  (\textit{Preprint} \eprint{1108.3989})

\bibitem{hanke:2012}
{Hanke} F, {Marek} A, {M{\"u}ller} B and {Janka} H~T 2012 {\em \apj\/} {\bf
  755} 138 (\textit{Preprint} \eprint{1108.4355})

\bibitem{burrows:2012}
{Burrows} A, {Dolence} J~C and {Murphy} J~W 2012 {\em \apj\/} {\bf 759} 5
  (\textit{Preprint} \eprint{1204.3088})

\bibitem{kotake:2011}
{Kotake} K 2011 {Gravitational-wave signatures in successful vs. failed
  core-collapse supernovae} {\em Journal of Physics Conference Series\/} ({\em
  Journal of Physics Conference Series\/} vol 314) p 012080

\bibitem{scheidegger:2010}
{Scheidegger} S, {K{\"a}ppeli} R, {Whitehouse} S~C, {Fischer} T and
  {Liebend{\"o}rfer} M 2010 {\em \aap\/} {\bf 514} A51 (\textit{Preprint}
  \eprint{1001.1570})

\bibitem{logue:12}
{Logue} J, {Ott} C~D, {Heng} I~S, {Kalmus} P and {Scargill} J 2012 {\em \prd\/}
  {\bf 86} 044023 (\textit{Preprint} \eprint{1202.3256})

\bibitem{powell:2016}
{Powell} J, {Gossan} S~E, {Logue} J and {Heng} I~S 2016 {\em \prd\/} {\bf 94}
  123012 (\textit{Preprint} \eprint{1610.05573})

\bibitem{macfadyen:1999}
{MacFadyen} A~I and {Woosley} S~E 1999 {\em \apj\/} {\bf 524} 262--289
  (\textit{Preprint} \eprint{astro-ph/9810274})

\bibitem{woosley:2006}
Woosley S~E and Bloom J~S 2006 {\em ARAA\/} {\bf 44} 507

\bibitem{yoon:2006}
{Yoon} S~C, {Langer} N and {Norman} C 2006 {\em \aap\/} {\bf 460} 199--208
  (\textit{Preprint} \eprint{astro-ph/0606637})

\bibitem{georgy:2009}
{Georgy} C, {Meynet} G, {Walder} R, {Folini} D and {Maeder} A 2009 {\em \aap\/}
  {\bf 502} 611--622 (\textit{Preprint} \eprint{0906.2284})

\bibitem{faucher-giguere:2006}
{Faucher-Gigu{\`e}re} C~A and {Kaspi} V~M 2006 {\em \apj\/} {\bf 643} 332--355
  (\textit{Preprint} \eprint{astro-ph/0512585})

\bibitem{popov:2010}
{Popov} S~B, {Pons} J~A, {Miralles} J~A, {Boldin} P~A and {Posselt} B 2010 {\em
  \mnras\/} {\bf 401} 2675--2686 (\textit{Preprint} \eprint{0910.2190})

\bibitem{gullon:2014}
{Gull{\'o}n} M, {Miralles} J~A, {Vigan{\`o}} D and {Pons} J~A 2014 {\em
  \mnras\/} {\bf 443} 1891--1899 (\textit{Preprint} \eprint{1406.6794})

\bibitem{fuller:2015}
{Fuller} J, {Cantiello} M, {Lecoanet} D and {Quataert} E 2015 {\em \apj\/} {\bf
  810} 101 (\textit{Preprint} \eprint{1502.07779})

\bibitem{demink:2013}
{de Mink} S~E, {Langer} N, {Izzard} R~G, {Sana} H and {de Koter} A 2013 {\em
  \apj\/} {\bf 764} 166 (\textit{Preprint} \eprint{1211.3742})

\bibitem{Zaldarriaga:2017}
{Zaldarriaga} M, {Kushnir} D and {Kollmeier} J~A 2017 {\em ArXiv e-prints\/}
  (\textit{Preprint} \eprint{1702.00885})

\bibitem{Qin:2019}
{Qin} Y, {Marchant} P, {Fragos} T, {Meynet} G and {Kalogera} V 2019 {\em
  \apjl\/} {\bf 870} L18 (\textit{Preprint} \eprint{1810.13016})

\bibitem{andresen:etal:2017}
{Andresen} H, {M{\"u}ller} B, {M{\"u}ller} E and {Janka} H~T 2017 {\em
  \mnras\/} {\bf 468} 2032--2051 (\textit{Preprint} \eprint{arXiv:1607.05199})

\bibitem{yakunin:2015-gw}
{Yakunin} K~N, {Mezzacappa} A, {Marronetti} P, {Yoshida} S, {Bruenn} S~W, {Hix}
  W~R, {Lentz} E~J, {Bronson Messer} O~E, {Harris} J~A, {Endeve} E, {Blondin}
  J~M and {Lingerfelt} E~J 2015 {\em \prd\/} {\bf 92} 084040 (\textit{Preprint}
  \eprint{1505.05824})

\bibitem{bmueller:2013}
{M{\"u}ller} B, {Janka} H~T and {Marek} A 2013 {\em \apj\/} {\bf 766} 43
  (\textit{Preprint} \eprint{1210.6984})

\bibitem{yakunin:2017}
{Yakunin} K~N, {Mezzacappa} A, {Marronetti} P, {Lentz} E~J, {Bruenn} S~W, {Hix}
  W~R, {Messer} O~E~B, {Endeve} E, {Blondin} J~M and {Harris} J~A 2017 {\em
  ArXiv e-prints\/} (\textit{Preprint} \eprint{1701.07325})

\bibitem{murphy:2009}
{Murphy} J~W, {Ott} C~D and {Burrows} A 2009 {\em \apj\/} {\bf 707} 1173--1190
  (\textit{Preprint} \eprint{0907.4762})

\bibitem{marek:2009}
{Marek} A, {Janka} H~T and {M{\"u}ller} E 2009 {\em \aap\/} {\bf 496} 475--494
  (\textit{Preprint} \eprint{0808.4136})

\bibitem{braginskii:87}
{Braginskii} V~B and {Thorne} K~S 1987 {\em Nature\/} {\bf 327} 123

\bibitem{epstein:78}
{Epstein} R 1978 {\em \apj\/} {\bf 223} 1037

\bibitem{mueller:04}
{M{\" u}ller} E, {Rampp} M, {Buras} R, {Janka} H~T and {Shoemaker} D~H 2004
  {\em \apj\/} {\bf 603} 221

\bibitem{ott:09}
{Ott} C~D 2009 {\em Class. Quantum Grav.\/} {\bf 26} 063001

\bibitem{kotake:11}
{Kotake} K, {Iwakami Nakano} W and {Ohnishi} N 2011 {\em \apj\/} {\bf 736} 124

\bibitem{marek:09b}
{Marek} A, {Janka} H~T and {M{\"u}ller} E 2009 {\em \aap\/} {\bf 496} 475

\bibitem{yakunin:10}
{Yakunin} K~N, {Marronetti} P, {Mezzacappa} A, {Bruenn} S~W, {Lee} C~T,
  {Chertkow} M~A, {Hix} W~R, {Blondin} J~M, {Lentz} E~J, {Bronson Messer} O~E
  and {Yoshida} S 2010 {\em Class. Quantum Grav.\/} {\bf 27} 194005

\bibitem{muellere:12}
{M{\"u}ller} E, {Janka} H~T and {Wongwathanarat} A 2012 {\em \aap\/} {\bf 537}
  A63 (\textit{Preprint} \eprint{1106.6301})

\bibitem{mueller:13gw}
{M{\"u}ller} B, {Janka} H~T and {Marek} A 2013 {\em \apj\/} {\bf 766} 43
  (\textit{Preprint} \eprint{1210.6984})

\bibitem{obergaulinger:06a}
{Obergaulinger} M, {Aloy} M~A and {M{\"u}ller} E 2006 {\em \aap\/} {\bf 450}
  1107

\bibitem{obergaulinger:06b}
{Obergaulinger} M, {Aloy} M~A, {Dimmelmeier} H and {M{\"u}ller} E 2006 {\em
  \aap\/} {\bf 457} 209

\bibitem{takiwaki:11}
{Takiwaki} T and {Kotake} K 2011 {\em \apj\/} {\bf 743} 30 (\textit{Preprint}
  \eprint{1004.2896})

\bibitem{ott:06spin}
{Ott} C~D, {Burrows} A, {Thompson} T~A, {Livne} E and {Walder} R 2006 {\em
  \apjs\/} {\bf 164} 130

\bibitem{woosley:06}
{Woosley} S~E and {Heger} A 2006 {\em Astrophys. J.\/} {\bf 637} 914

\bibitem{heger:05}
{Heger} A, {Woosley} S~E and {Spruit} H~C 2005 {\em \apj\/} {\bf 626} 350

\bibitem{burrows:07b}
{Burrows} A, {Dessart} L, {Livne} E, {Ott} C~D and {Murphy} J 2007 {\em \apj\/}
  {\bf 664} 416

\bibitem{takiwaki:09}
{Takiwaki} T, {Kotake} K and {Sato} K 2009 {\em \apj\/} {\bf 691} 1360

\bibitem{kuroda:2014}
{Kuroda} T, {Takiwaki} T and {Kotake} K 2014 {\em \prd\/} {\bf 89} 044011
  (\textit{Preprint} \eprint{1304.4372})

\bibitem{moesta:14b}
{M{\"o}sta} P, {Richers} S, {Ott} C~D, {Haas} R, {Piro} A~L, {Boydstun} K,
  {Abdikamalov} E, {Reisswig} C and {Schnetter} E 2014 {\em \apjl\/} {\bf 785}
  L29 (\textit{Preprint} \eprint{1403.1230})

\bibitem{masada:2015}
{Masada} Y, {Takiwaki} T and {Kotake} K 2015 {\em \apjl\/} {\bf 798} L22
  (\textit{Preprint} \eprint{1411.6705})

\bibitem{richers:2017}
{Richers} S, {Ott} C~D, {Abdikamalov} E, {O'Connor} E and {Sullivan} C 2017
  {\em ArXiv e-prints\/} (\textit{Preprint} \eprint{1701.02752})

\bibitem{dimmelmeier:08}
{Dimmelmeier} H, {Ott} C~D, {Marek} A and {Janka} H~T 2008 {\em \prd\/} {\bf
  78} 064056

\bibitem{abdikamalov:14}
{Abdikamalov} E, {Gossan} S, {DeMaio} A~M and {Ott} C~D 2014 {\em \prd\/} {\bf
  90} 044001 (\textit{Preprint} \eprint{1311.3678})

\bibitem{scheidegger:10b}
{Scheidegger} S, {K{\"a}ppeli} R, {Whitehouse} S~C, {Fischer} T and
  {Liebend{\"o}rfer} M 2010 {\em \aap\/} {\bf 514} A51

\bibitem{ott:07prl}
{Ott} C~D, {Dimmelmeier} H, {Marek} A, {Janka} H~T, {Hawke} I, {Zink} B and
  {Schnetter} E 2007 {\em \prl\/} {\bf 98} 261101

\bibitem{fu:11}
{Fu} W and {Lai} D 2011 {\em \mnras\/} {\bf 413} 2207 (\textit{Preprint}
  \eprint{1011.4887})

\bibitem{muhlberger:14}
{Muhlberger} C~D, {Nouri} F~H, {Duez} M~D, {Foucart} F, {Kidder} L~E, {Ott}
  C~D, {Scheel} M~A, {Szil{\'a}gyi} B and {Teukolsky} S~A 2014 {\em Submitted
  to \prd. arXiv:1405.2144\/} (\textit{Preprint} \eprint{1405.2144})

\bibitem{hjorth:11}
{Hjorth} J and {Bloom} J~S 2011 {\em arXiv:1104.2274\/}

\bibitem{modjaz:11}
{Modjaz} M 2011 {\em Astron. Nachr.\/} {\bf 332} 434

\bibitem{woosley:93}
{Woosley} S~E 1993 {\em \apj\/} {\bf 405} 273

\bibitem{macfadyen:01}
{MacFadyen} A~I, {Woosley} S~E and {Heger} A 2001 {\em \apj\/} {\bf 550} 410
  (\textit{Preprint} \eprint{arXiv:astro-ph/9910034})

\bibitem{wb:06}
{Woosley} S~E and {Bloom} J~S 2006 {\em Ann. Rev. Astron. Astrophys.\/} {\bf
  44} 507

\bibitem{wheeler:02}
{Wheeler} J~C, {Meier} D~L and {Wilson} J~R 2002 {\em \apj\/} {\bf 568} 807

\bibitem{metzger:11}
{Metzger} B~D, {Giannios} D, {Thompson} T~A, {Bucciantini} N and {Quataert} E
  2011 {\em \mnras\/} {\bf 413} 2031

\bibitem{piro:07}
{Piro} A~L and {Pfahl} E 2007 {\em \apj\/} {\bf 658} 1173

\bibitem{korobkin:11}
{Korobkin} O, {Abdikamalov} E~B, {Schnetter} E, {Stergioulas} N and {Zink} B
  2011 {\em \prd\/} {\bf 83} 043007

\bibitem{kiuchi:11pp}
{Kiuchi} K, {Shibata} M, {Montero} P~J and {Font} J~A 2011 {\em \prl\/} {\bf
  106} 251102 (\textit{Preprint} \eprint{1105.5035})

\bibitem{fryer:02}
Fryer C, Holz D and Hughes S 2002 {\em \apj\/} {\bf 565} 430

\bibitem{ott:11a}
{Ott} C~D, {Reisswig} C, {Schnetter} E, {O'Connor} E, {Sperhake} U,
  {L{\"o}ffler} F, {Diener} P, {Abdikamalov} E, {Hawke} I and {Burrows} A 2011
  {\em Phys. Rev. Lett.\/} {\bf 106} 161103

\bibitem{oconnor:11}
{O'Connor} E and {Ott} C~D 2011 {\em \apj\/} {\bf 730} 70

\bibitem{corsi:09}
{Corsi} A and {M{\'e}sz{\'a}ros} P 2009 {\em \apj\/} {\bf 702} 1171

\bibitem{piro:11}
{Piro} A~L and {Ott} C~D 2011 {\em \apj\/} {\bf 736} 108

\bibitem{ugliano:12}
{Ugliano} M, {Janka} H~T, {Marek} A and {Arcones} A 2012 {\em \apj\/} {\bf 757}
  69 (\textit{Preprint} \eprint{1205.3657})

\bibitem{cerda-duran:2013}
{Cerd{\'a}-Dur{\'a}n} P, {DeBrye} N, {Aloy} M~A, {Font} J~A and {Obergaulinger}
  M 2013 {\em \apjl\/} {\bf 779} L18 (\textit{Preprint} \eprint{1310.8290})

\bibitem{ott:2011}
{Ott} C~D, {Reisswig} C, {Schnetter} E, {O'Connor} E, {Sperhake} U,
  {L{\"o}ffler} F, {Diener} P, {Abdikamalov} E, {Hawke} I and {Burrows} A 2011
  {\em Physical Review Letters\/} {\bf 106} 161103 (\textit{Preprint}
  \eprint{1012.1853})

\bibitem{yoon:05b}
{Yoon} S~C and {Langer} N 2005 {\em \aap\/} {\bf 435} 967

\bibitem{abdikamalov:10}
{Abdikamalov} E~B, {Ott} C~D, {Rezzolla} L, {Dessart} L, {Dimmelmeier} H,
  {Marek} A and {Janka} H 2010 {\em \prd\/} {\bf 81} 044012

\bibitem{vdb:91}
{van den Bergh} S and {Tammann} G~A 1991 {\em \araa\/} {\bf 29} 363

\bibitem{mannucci:05}
{Mannucci} F, {Della Valle} M, {Panagia} N, {Cappellaro} E, {Cresci} G,
  {Maiolino} R, {Petrosian} A and {Turatto} M 2005 {\em Astron. Astrophys.\/}
  {\bf 433} 807 (\textit{Preprint} \eprint{arXiv:astro-ph/0411450})

\bibitem{keane:08}
{Keane} E~F and {Kramer} M 2008 {\em \mnras\/} {\bf 391} 2009
  (\textit{Preprint} \eprint{0810.1512})

\bibitem{ando:05}
{Ando} S, Beacom F and Y\"uksel H 2005 {\em Phys. Rev. Lett.\/} {\bf 95} 171101

\bibitem{kistler:11}
{Kistler} M~D, {Y{\"u}ksel} H, {Ando} S, {Beacom} J~F and {Suzuki} Y 2011 {\em
  \prd\/} {\bf 83} 123008 (\textit{Preprint} \eprint{0810.1959})

\bibitem{abbott:etal:2016:1gSNsearch}
{Abbott} B~P, {Abbott} R, {Abbott} T~D, {Abernathy} M~R, {Acernese} F, {Ackley}
  K, {Adams} C, {Adams} T, {Addesso} P, {Adhikari} R~X and et~al 2016 {\em
  \prd\/} {\bf 94} 102001 (\textit{Preprint} \eprint{1605.01785})

\bibitem{gossan:etal:2016}
{Gossan} S~E, {Sutton} P, {Stuver} A {\em et~al.\/} 2016 {\em \prd\/} {\bf 93}
  042002 (\textit{Preprint} \eprint{arXiv:1511.02836})

\bibitem{hayama:etal:2015}
{Hayama} K, {Kuroda} T, {Kotake} K and {Takiwaki} T 2015 {\em \prd\/} {\bf 92}
  122001 (\textit{Preprint} \eprint{arXiv:1501.00966})

\bibitem{nakamura:etal:2016}
{Nakamura} K, {Horiuchi} S, {Tanaka} M, {Hayama} K, {Takiwaki} T and {Kotake} K
  2016 {\em \mnras\/} {\bf 461} 3296--3313 (\textit{Preprint}
  \eprint{arXiv:1602.03028})

\bibitem{abdikamalov:2014}
{Abdikamalov} E, {Gossan} S, {DeMaio} A~M and {Ott} C~D 2014 {\em \prd\/} {\bf
  90} 044001 (\textit{Preprint} \eprint{1311.3678})

\bibitem{roever:09}
{R{\"o}ver} C, {Bizouard} M, {Christensen} N, {Dimmelmeier} H, {Heng} I~S and
  {Meyer} R 2009 {\em \prd\/} {\bf 80} 102004

\bibitem{engels:2014}
{Engels} W~J, {Frey} R and {Ott} C~D 2014 {\em \prd\/} {\bf 90} 124026
  (\textit{Preprint} \eprint{1406.1164})

\bibitem{edwards:2014}
{Edwards} M~C, {Meyer} R and {Christensen} N 2014 {\em Inverse Problems\/} {\bf
  30} 114008 (\textit{Preprint} \eprint{1407.7549})

\bibitem{yokozawa:2015}
{Yokozawa} T, {Asano} M, {Kayano} T, {Suwa} Y, {Kanda} N, {Koshio} Y and
  {Vagins} M~R 2015 {\em \apj\/} {\bf 811} 86 (\textit{Preprint}
  \eprint{1410.2050})

\bibitem{halzen:09}
{Halzen} F and {Raffelt} G~G 2009 {\em \prd\/} {\bf 80} 087301
  (\textit{Preprint} \eprint{0908.2317})

\bibitem{ikeda:07}
{Ikeda~et~al\ [Super-Kamiokande Collaboration]} M 2007 {\em \apj\/} {\bf 669}
  519

\bibitem{icecube:11sn}
{Abbasi~et~al\ [IceCube Collaboration]} R 2011 {\em \aap\/} {\bf 535} A109

\bibitem{scholberg:11}
{Scholberg} K 2011 {\em J. Phys. Conf. Ser.\/} {\bf 309} 012028

\bibitem{hyperkamiokande:11}
{Abe} K, {Abe} T, {Aihara} H, {Fukuda} Y, {Hayato} Y, {Huang} K, {Ichikawa}
  A~K, {Ikeda} M, {Inoue} K, {Ishino} H, {Itow} Y, {Kajita} T, {Kameda} J,
  {Kishimoto} Y, {Koga} M, {Koshio} Y, {Lee} K~P, {Minamino} A, {Miura} M,
  {Moriyama} S, {Nakahata} M, {Nakamura} K, {Nakaya} T, {Nakayama} S,
  {Nishijima} K, {Nishimura} Y, {Obayashi} Y, {Okumura} K, {Sakuda} M, {Sekiya}
  H, {Shiozawa} M, {Suzuki} A~T, {Suzuki} Y, {Takeda} A, {Takeuchi} Y, {Tanaka}
  H~K~M, {Tasaka} S, {Tomura} T, {Vagins} M~R, {Wang} J and {Yokoyama} M 2011
  {\em arXiv:1109.3262\/}

\bibitem{janka:07}
{Janka} H~T, {Langanke} K, {Marek} A, {Mart{\'{\i}}nez-Pinedo} G and
  {M{\"u}ller} B 2007 {\em \physrep\/} {\bf 442} 38

\bibitem{ott:10dcc}
{Ott} C~D 2010 {GWs from Barmode Instabilities} Tech. Rep. LIGO-T1000553-v2
  LIGO Scientific Collaboration
  \urlprefix\url{https://dcc.ligo.org/LIGO-T1000553-v2}

\bibitem{mueller:12b}
{M{\"u}ller} B, {Janka} H~T and {Heger} A 2012 {\em \apj\/} {\bf 761} 72
  (\textit{Preprint} \eprint{1205.7078})

\bibitem{ott:13a}
{Ott} C~D, {Abdikamalov} E, {M{\"o}sta} P, {Haas} R, {Drasco} S, {O'Connor}
  E~P, {Reisswig} C, {Meakin} C~A and {Schnetter} E 2013 {\em \apj\/} {\bf 768}
  115 (\textit{Preprint} \eprint{1210.6674})

\bibitem{duncan92}
{Duncan} R~C and {Thompson} C 1992 {\em \apjl\/} {\bf 392} L9--L13

\bibitem{mereghetti08}
{Mereghetti} S 2008 {\em \aapr\/} {\bf 15} 225--287

\bibitem{mazets79}
{Mazets} E~P {\em et~al.\/} 1979 {\em Nature\/} {\bf 282} 587--589

\bibitem{terasawa05}
{Terasawa} T, {Tanaka} Y~T, {Takei} Y, {Kawai} N, {Yoshida} A, {Nomoto} K,
  {Yoshikawa} I, {Saito} Y, {Kasaba} Y, {Takashima} T, {Mukai} T, {Noda} H,
  {Murakami} T, {Watanabe} K, {Muraki} Y, {Yokoyama} T and {Hoshino} M 2005
  {\em Nature\/} {\bf 434} 1110--1111

\bibitem{mazets08}
Mazets E~P, Aptekar R~L, Cline T~L, Frederiks D~D, Goldsten J~O, Golenetskii
  S~V, Hurley K, von Kienlin A,  and Pal'shin V~D 2008 {\em The Astrophysical
  Journal\/} {\bf 680} 545--549
  \urlprefix\url{http://stacks.iop.org/0004-637X/680/545}

\bibitem{S5GRB070201}
Abbott B {\em et~al.\/} 2008 {\em \apj\/} {\bf 681} 1419--1430

\bibitem{frederiks07b}
{Frederiks} D~D, {Palshin} V~D, {Aptekar} R~L, {Golenetskii} S~V, {Cline} T~L
  and {Mazets} E~P 2007 {\em Astronomy Letters\/} {\bf 33} 19--24
  (\textit{Preprint} \eprint{arXiv:astro-ph/0609544})

\bibitem{ioka01}
{Ioka} K 2001 {\em \mnras\/} {\bf 327} 639--662 (\textit{Preprint}
  \eprint{astro-ph/0009327})

\bibitem{owen05}
{Owen} B~J 2005 {\em \prl\/} {\bf 95} 211101--+ (\textit{Preprint}
  \eprint{astro-ph/0503399})

\bibitem{horowitz09}
{Horowitz} C~J and {Kadau} K 2009 {\em Phys. Rev. Lett.\/} {\bf 102} 191102
  (\textit{Preprint} \eprint{0904.1986})

\bibitem{corsi11}
{Corsi} A and {Owen} B~J 2011 {\em arXiv:1102.3421\/} (\textit{Preprint}
  \eprint{1102.3421})

\bibitem{kashiyama11}
{Kashiyama} K and {Ioka} K 2011 {\em arXiv:1102.4830\/} (\textit{Preprint}
  \eprint{1102.4830})

\bibitem{levin11}
{Levin} Y and {van Hoven} M 2011 {\em arXiv:astro-ph/1103.0880\/}
  (\textit{Preprint} \eprint{1103.0880})

\bibitem{thompson95}
{Thompson} C and {Duncan} R~C 1995 {\em \mnras\/} {\bf 275} 255--300

\bibitem{benhar04}
{Benhar} O, {Ferrari} V and {Gualtieri} L 2004 {\em \prd\/} {\bf 70} 124015--+
  (\textit{Preprint} \eprint{astro-ph/0407529})

\bibitem{mcdermott88}
{McDermott} P~N, {van Horn} H~M and {Hansen} C~J 1988 {\em \apj\/} {\bf 325}
  725--748

\bibitem{middleditch06}
{Middleditch} J, {Marshall} F~E, {Wang} Q~D, {Gotthelf} E~V and {Zhang} W 2006
  {\em \apj\/} {\bf 652} 1531--1546 (\textit{Preprint}
  \eprint{arXiv:astro-ph/0605007})

\bibitem{anderson75}
{Anderson} P~W and {Itoh} N 1975 {\em \nat\/} {\bf 256} 25--27

\bibitem{Andersson2010ETReview}
Andersson N, Ferrari V, Jones D~I, Kokkotas K~D, Krishnan B, Read J~S, Rezzolla
  L and Zink B 2010 {\em General relativity and gravitation\/} {\bf 43}
  409--436

\bibitem{ushomirsky:00}
{Ushomirsky} G, {Cutler} C and {Bildsten} L 2000 {\em \mnras\/} {\bf 319} 902

\bibitem{Kitiashvili2006ellipticity}
Kitiashvili I~N and Gusev A~V 2008 {\em Astronomy Reports\/} {\bf 52}(1) 61--69
  ISSN 1063-7729 10.1134/S1063772908010071
  \urlprefix\url{http://dx.doi.org/10.1134/S1063772908010071}

\bibitem{colaiuda:08}
{Colaiuda} A, {Ferrari} V, {Gualtieri} L and {Pons} J~A 2008 {\em \mnras\/}
  {\bf 385} 2080

\bibitem{abbott:2010s5pulsar}
{Abbott} B~P, {Abbott} R, {Acernese} F, {Adhikari} R, {Ajith} P, {Allen} B,
  {Allen} G, {Alshourbagy} M, {Amin} R~S, {Anderson} S~B and et~al 2010 {\em
  \apj\/} {\bf 713} 671 (\textit{Preprint} \eprint{0909.3583})

\bibitem{Abbot2008S4IncoherentPaper}
Abbott B, Abbott R, Adhikari R, Agresti J, Ajith P, Allen B {\em et~al.\/}
  (LIGO Scientific Collaboration) 2008 {\em Phys. Rev. D\/} {\bf 77}(2) 022001
  \urlprefix\url{http://link.aps.org/doi/10.1103/PhysRevD.77.022001}

\bibitem{shibata:06bns}
{Shibata} M and {Taniguchi} K 2006 {\em \prd\/} {\bf 73} 064027

\bibitem{giacomazzo:11}
{Giacomazzo} B, {Rezzolla} L and {Baiotti} L 2011 {\em \prd\/} {\bf 83} 044014

\bibitem{hotokezaka:11}
{Hotokezaka} K, {Kyutoku} K, {Okawa} H, {Shibata} M and {Kiuchi} K 2011 {\em
  \prd\/} {\bf 83} 124008

\bibitem{bauswein:12}
{Bauswein} A, {Janka} H~T, {Hebeler} K and {Schwenk} A 2012 {\em \prd\/} {\bf
  86} 063001 (\textit{Preprint} \eprint{1204.1888})

\bibitem{cook:94c}
Cook G~B, Shapiro S~L and Teukolsky S~A 1994 {\em \apj\/} {\bf 424} 823

\bibitem{kaplan:14}
{Kaplan} J~D, {Ott} C~D, {O'Connor} E~P, {Kiuchi} K, {Roberts} L and {Duez} M
  2014 {\em \apj\/} {\bf 790} 19 (\textit{Preprint} \eprint{1306.4034})

\bibitem{stergioulas:11}
{Stergioulas} N, {Bauswein} A, {Zagkouris} K and {Janka} H~T 2011 {\em
  \mnras\/} {\bf 418} 427 (\textit{Preprint} \eprint{1105.0368})

\bibitem{1998PhRvC..58.1804A}
{Akmal} A, {Pandharipande} V~R and {Ravenhall} D~G 1998 {\em \prc\/} {\bf 58}
  1804--1828 (\textit{Preprint} \eprint{nucl-th/9804027})

\bibitem{2010PhRvC..81a5803T}
{Typel} S, {R{\"o}pke} G, {Kl{\"a}hn} T, {Blaschke} D and {Wolter} H~H 2010
  {\em \prc\/} {\bf 81} 015803

\bibitem{2010NuPhA.837..210H}
{Hempel} M and {Schaffner-Bielich} J 2010 {\em Nucl. Phys. A\/} {\bf 837}
  210--254

\bibitem{1998NuPhA.637..435S}
{Shen} H, {Toki} H, {Oyamatsu} K and {Sumiyoshi} K 1998 {\em Nucl. Phys. A\/}
  {\bf 637} 435--450

\bibitem{1997PhRvC..55..540L}
{Lalazissis} G~A, {K{\"o}nig} J and {Ring} P 1997 {\em \prc\/} {\bf 55}
  540--543

\bibitem{hotokezaka:13}
{Hotokezaka} K, {Kiuchi} K, {Kyutoku} K, {Okawa} H, {Sekiguchi} Y~i, {Shibata}
  M and {Taniguchi} K 2013 {\em \prd\/} {\bf 87} 024001 (\textit{Preprint}
  \eprint{1212.0905})

\bibitem{bauswein:14}
{Bauswein} A, {Stergioulas} N and {Janka} H~T 2014 {\em \prd\/} {\bf 90} 023002
  (\textit{Preprint} \eprint{1403.5301})

\bibitem{clark:14}
{Clark} J, {Bauswein} A, {Cadonati} L, {Janka} H~T, {Pankow} C and
  {Stergioulas} N 2014 {\em \prd\/} {\bf 90} 062004 (\textit{Preprint}
  \eprint{1406.5444})

\bibitem{ET:study}
{E T Science Team} 2011 Einstein gravitational wave telescope conceptual design
  study Tech. rep.

\bibitem{Bruce:1988}
Allen B 1988 {\em Phys. Rev. D\/} {\bf 37}(8) 2078--2085
  \urlprefix\url{http://link.aps.org/doi/10.1103/PhysRevD.37.2078}

\bibitem{BoSt2005}
Boyle L~A and Steinhardt P~J 2005

\bibitem{Steinhardt:2011}
Lehners J~L and Steinhardt P~J 2011 {\em Phys. Rev. Lett.\/} {\bf 106}(8)
  081301 \urlprefix\url{http://link.aps.org/doi/10.1103/PhysRevLett.106.081301}

\bibitem{GiblinThrane:2014}
{Giblin} J~T and {Thrane} E 2014 {\em \prd\/} {\bf 90} 107502
  (\textit{Preprint} \eprint{1410.4779})

\bibitem{PhysRevLett.118.121101}
Abbott B~P, Abbott R, Abbott T~D, Abernathy M~R, Acernese F {\em et~al.\/}
  (LIGO Scientific Collaboration and Virgo Collaboration) 2017 {\em Phys. Rev.
  Lett.\/} {\bf 118}(12) 121101
  \urlprefix\url{https://link.aps.org/doi/10.1103/PhysRevLett.118.121101}

\bibitem{Callister:2016}
{Callister} T, {Sammut} L, {Qiu} S, {Mandel} I and {Thrane} E 2016 {\em
  Physical Review X\/} {\bf 6} 031018 (\textit{Preprint} \eprint{1604.02513})

\bibitem{thorne:87}
Thorne K~S 1987 {\em 300 Years of Gravitation\/} ed Hawking S~W and Israel W
  (Cambridge, UK: Cambridge University Press)

\bibitem{flanhughes:98}
{Flanagan} {\'E}~{\'E} and {Hughes} S~A 1998 {\em \prd\/} {\bf 57} 4535

\end{thebibliography}

\end{document}